\documentclass[reprint,superscriptaddress, floatfix]{revtex4-2}
\usepackage[utf8]{inputenc}
\usepackage{amsfonts}
\usepackage{amssymb}
\usepackage{amsmath}
\usepackage{csquotes}
\usepackage{bbm}
\usepackage{verbatim}
\usepackage{graphicx}
\usepackage{xcolor}
\usepackage{hyperref}
\usepackage{listings}
\hypersetup{
    colorlinks = true,
    linkcolor = red,
    anchorcolor = blue,
    citecolor = blue,
    filecolor = blue,
    urlcolor = blue
    }

\usepackage{setspace}

\newcommand{\hmuBI}{\hat{\mu}_B^I}

\begin{document}
\title{%
Contribution to understanding the phase structure of strong interaction matter: Lee-Yang edge singularities from lattice QCD}
\author{P. Dimopoulos}
\affiliation{Dipartimento di Scienze Matematiche, Fisiche e Informatiche, Università di Parma and INFN, Gruppo Collegato di Parma
I-43100 Parma, Italy}
\author{L. Dini}
\affiliation{Universität Bielefeld, Fakultät für Physik, D-33615 Bielefeld, Germnay}
\author{F. Di Renzo}
\affiliation{Dipartimento di Scienze Matematiche, Fisiche e Informatiche, Università di Parma and INFN, Gruppo Collegato di Parma
I-43100 Parma, Italy}
\author{J. Goswami}
\affiliation{Universität Bielefeld, Fakultät für Physik, D-33615 Bielefeld, Germnay}
\author{G.~Nicotra}
\affiliation{Universität Bielefeld, Fakultät für Physik, D-33615 Bielefeld, Germnay}
\author{C. Schmidt}
\affiliation{Universität Bielefeld, Fakultät für Physik, D-33615 Bielefeld, Germnay}
\author{S. Singh}
\affiliation{Dipartimento di Scienze Matematiche, Fisiche e Informatiche, Università di Parma and INFN, Gruppo Collegato di Parma
I-43100 Parma, Italy}
\author{K. Zambello}
\affiliation{Dipartimento di Scienze Matematiche, Fisiche e Informatiche, Università di Parma and INFN, Gruppo Collegato di Parma
I-43100 Parma, Italy}
\author{F. Ziesché}
\affiliation{Universität Bielefeld, Fakultät für Physik, D-33615 Bielefeld, Germnay}
\date{\today}
\begin{abstract}
We present a calculation of the net baryon number density as a function of imaginary baryon number chemical potential, obtained with highly improved staggered quarks at temporal lattice extent of $N_\tau=4,6$. 
We construct various rational function approximations of the lattice data and discuss how poles in the complex plane can be determined from them.
We compare our results of the singularities in the chemical potential plane to the theoretically expected positions of the Lee-Yang edge singularity in the vicinity of the Roberge-Weiss and chiral phase transitions.
We find a temperature scaling that is in accordance with the expected power law behavior.
\end{abstract}
\maketitle

\section{Introduction}
The phase diagram of quantum chromodynamics (QCD) belongs to the most pressing open issues in high energy physics. 
With large scale experimental programs at RHIC and LHC the phase diagram is scanned for hints of a critical point or a first order phase transition. 
In addition, many \textit{ab initio} calculations of lattice QCD are performed to infer on the QCD phase diagram. 

Unfortunately, the notorious sign problem hampers numerical studies of the QCD phase diagram. 
At vanishing baryon chemical potential ($\mu_B\equiv 0)$, lattice QCD calculations rely on Monte Carlo methods for an efficient sampling of the QCD partition sum. 
At nonvanishing baryon chemical potentials ($\mu_B>0$), standard Monte Carlo methods cease working as the fermion determinant becomes genuinely complex.
Hence, the kernel of the QCD partition sum is strongly oscillating with increasing lattice volumes. 

Over the last decades, many methods have been developed which potentially circumvent or solve the QCD sign problem. 
These methods include reweighting \cite{Barbour:1997ej, Fodor:2001au}, Taylor expansions \cite{Allton:2002zi, Allton:2003vx, Gavai:2003mf}, analytic continuation from purely imaginary chemical potentials \cite{deForcrand:2002hgr, DElia:2002tig}, canonical partition functions \cite{Kratochvila:2003rp, Alexandru:2005ix}, strong coupling/dual methods \cite{Karsch:1988zx, deForcrand:2014tha, Gattringer:2016kco, Gagliardi:2019cpa}, the density of states method \cite{Ambjorn:2002pz, Fodor:2007vv, Langfeld:2015fua}, and complex Langevin dynamics \cite{Karsch:1985cb, Aarts:2009uq, Aarts:2012ft, Sexty:2013ica}. Related to the latter, is also the Lefschetz thimble method \cite{Cristoforetti:2012su, Fujii:2013sra, Alexandru:2015sua}, which is based on a deformation of the integration manifold into complex field space.  
Recent developments are reviewed, e.g., in \cite{Alexandru:2020wrj, Attanasio:2020spv, Berger:2019odf}. However, all these methods face severe limitations that restrict their applicability toward the thermodynamic and/or continuum limits.

With this study we systematically investigate singularities of the grand canonical potential in the complex chemical potential plane, which we identify from an (analytically continued) rational approximation of lattice data obtained at purely imaginary chemical potentials.
The rational approximation of the net baryon number density is done in consistency with the second, third and fourth order cumulants of the baryon number density. 
In this sense our method could be seen as a combination of the Taylor expansion approach and the imaginary chemical potential method. 
The position of those singularities provides very valuable information on the QCD phase diagram. 
We find that they are in agreement with the critical scaling of the Lee-Yang edge singularities in the vicinity of the Roberge-Weiss transition and the chiral transition. 
We also discuss the scaling of the Lee-Yang edge singularity in the vicinity of a hypothetical critical end point.
Finally, we point out that the position of the singularities can be used to estimate the radius of convergence of any analytic expansion and to extract nonuniversal parameters that map QCD to the universal scaling function. 
Among the latter might also be the position of the QCD critical end point, which was demonstrated recently in the case of the Gross-Neveu model \cite{Basar:2021hdf}.
Genuine Lee-Yang zeros have been recently also studied in other works \cite{Giordano:2019gev, Mondal:2021jxk}.

This paper is organized as follows. In Sec.~\ref{sec:theory} we introduce the scaling theory of the Lee-Yang edge singularities and apply them to the cases of the Roberge-Weiss transition and the chiral transition of QCD. 
We also discuss the case of the QCD critical endpoint. 
In Sec.~\ref{sec:lattice}, we provide details of our lattice QCD calculations and in Sec.~\ref{sec:rational} we discuss our strategy for determining rational approximations to our lattice data. 
We present our findings of the extracted singularities in Sec.~\ref{sec:results}. 
We summarize and conclude in Sec.~\ref{sec:conclusions}. 
In order to consolidate our findings, we compiled a number of Appendixes~(\ref{sec:ApPADE}-\ref{sec:LastApPADE}), where we discuss numerical issues related to our rational approximations, which are based on a multipoint Padé method. 

\section{Expected singularities \label{sec:theory}}
The grand canonical partition function in lattice QCD, $\mathcal{Z}_{GC}=\mathcal{Z}(V,T,\mu_B)$ has the form of a (high dimensional) polynomial at any finite volume $V$ and is positive for real values of the temperature $T$ and baryon chemical potential $\mu_B$. 
However, following Lee, Yang \cite{PhysRev.87.404,PhysRev.87.410} and Fischer \cite{PhysRevLett.40.1610}, we point out that $\mathcal{Z}_{GC}$ exhibits many zeros in the complex chemical potential plane, which can be used to extract valuable information on the phase transitions that may occur in the system. 
A physical phase transition in the thermodynamic limit can be identified when for $V\to\infty$ one of the complex zeros approaches a point with real parameters $T,\mu_B$.

The grand canonical potential $\log( \mathcal{Z}_{GC})$ diverges when one approaches a zero of $\mathcal{Z}_{GC}$. 
Singularities of the grand canonical potential and its derivatives will limit any analytic expansion performed at zero or purely imaginary chemical potentials. 
The positions of the zeros of $\mathcal{Z}_{GC}$ can thus also be used to estimate the radius of convergence of the Taylor expansion method. 

\begin{figure}
    \centering
    \includegraphics[width=0.48\textwidth]{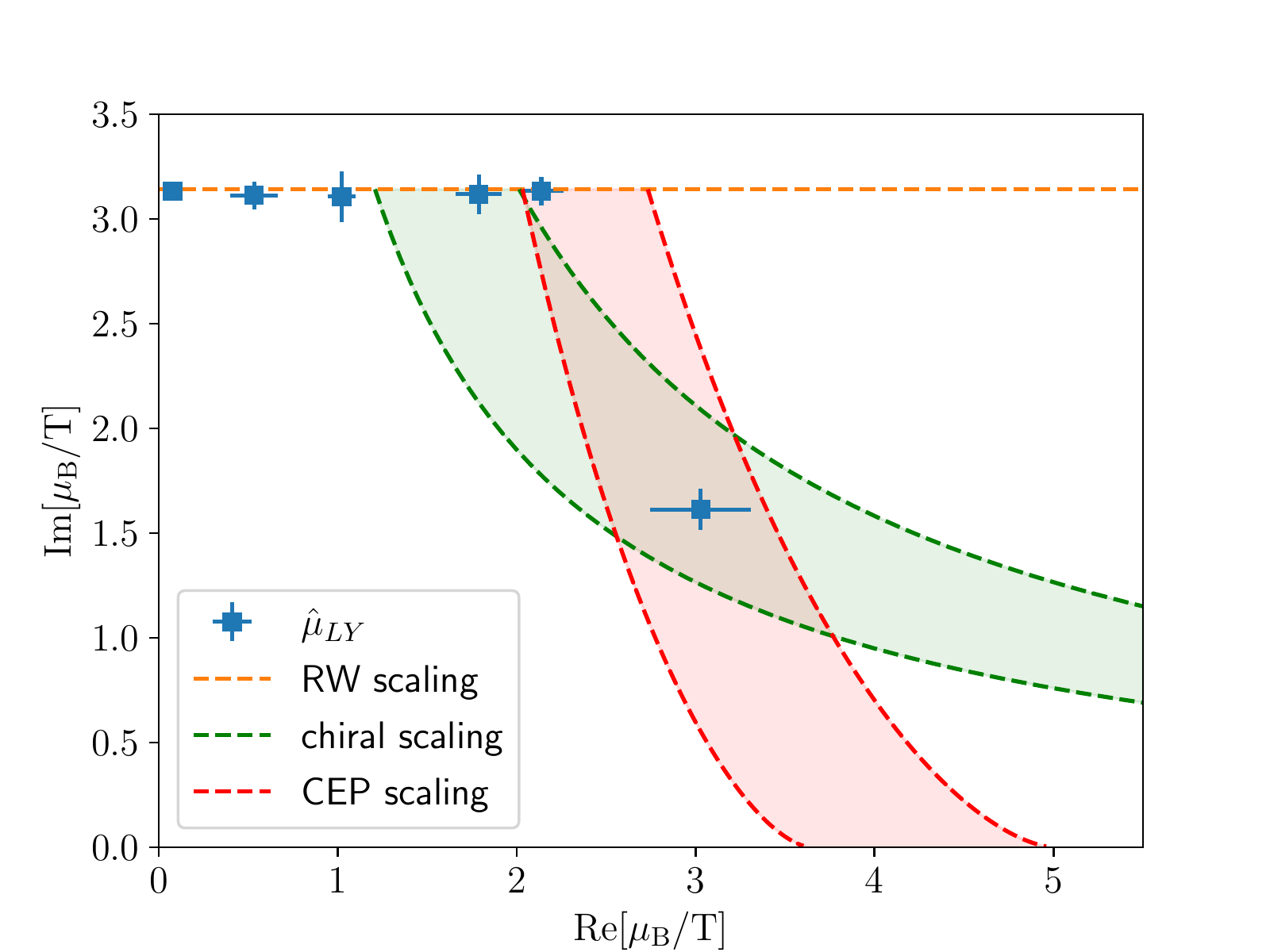}
    \caption{Overview of the expected scaling behavior of the Lee-Yang edge singularities in the complex plane of the baryon chemical potential ($\mu_B/T$). Universal scaling in the vicinity of the Roberge-Weiss transition, the chiral transition, and the critical end point is shown in yellow, green, and red, respectively. The width of the bands indicate uncertainties in the nonuniversal parameters. See the text for a detailed discussion. Data points depict identified Lee-Yang edge singularities (method II) at different temperatures, from lattice QCD calculations on lattices with temporal extent $N_\tau=4,6$.}
    \label{fig:overview}
\end{figure}

In different $(T,\mu_B)$ regions, the QCD partition function can be approximately described by effective theories.
At high temperatures, far above the QCD transition, we might be able to use a free Fermi gas to describe the thermodynamic behavior of the quarks.
Below the QCD crossover temperature $T_{pc}$, the Hadron resonance gas is known to describe the bulk thermodynamics of QCD matter quite well \cite{Bazavov:2017dus, Goswami:2020yez, Bellwied:2021nrt}. 
However, we can also consider universal behavior in the vicinity of the Roberge-Weiss, the chiral transition or even in the vicinity of the QCD critical end point (if existing). 
In particular, we can predict the positions of singularities in the complex $\mu_B$ plane, by applying suitable mappings from the parameter space of QCD to the relevant scaling fields in the vicinity of a critical point. 
In the following, we make extensive use of the fact that the scaling function of the order parameter $f_G(z)$ exhibits a branch cut singularity at $z=z_c$, where the scaling variable $z$ is expressed in terms of the reduced temperature $t$ and symmetry breaking field $h$ as $z=t/|h|^{1/\beta\delta}$. 
The universal position $z_c$ of the universal singularity, known as the Lee-Yang edge singularity, has been recently determined for different universality classes \cite{Connelly:2020gwa}.  
The three distinct scaling approaches are visualized in Fig.~\ref{fig:overview}. 
At this point not all of the nonuniversal normalization constants that fix the above mentioned mappings are known. 
We thus vary some of the parameters to give an impression of the functional dependence, which is discussed in more detail below. 
Also shown are identified Lee-Yang-edge singularities from our lattice QCD calculations on $N_\tau=4,6$ lattices. 
The determination of the data points is discussed in Sec.~\ref{sec:results}.

\subsection{Singularities in the vicinity of the Roberge-Weiss(RW) transition\label{sec:scalingRW}}
The RW critical point ($\mu_B/T=i\pi$) is a remnant of the $Z(3)$ symmetry in the quenched $(m_q\to \infty)$ limit of QCD, where $m_q$ specifies the quark masses, and the QCD partition function has a reflection symmetry around $\mu_B/T=i\pi$ \cite{Roberge:1986mm}. The nature of the RW end point could be either a first order triple point or a second order $Z(2)$ critical point depending on the value of the quark masses \cite{Philipsen:2019ouy}. In $(2+1)$-flavor QCD with a physical value of the light quark masses with improved lattice discretization, which is discussed here, one finds a second order $Z(2)$ critical point \cite{Bonati:2016pwz,Goswami:2018qhc,Goswami:2019exb}.  The order parameter in the vicinity of a second order transition can be written as the sum of a universal and a regular part,
\begin{eqnarray}
M=h^{1/\delta}f_G(z) + M_{\text{reg}} \ ;\ z\equiv t/|h|^{1/\beta \delta},
\end{eqnarray}
where, $t,h$ are scaling fields and $\beta,\delta$ are critical exponents. Also, $f_G$ is the universal scaling function for the order parameter, whereas $M_{\text{reg}}$ accounts for regular contributions.

The relevant scaling fields for a $Z(2)$ symmetric second order RW transition can be defined as
\begin{align}
    t&=t_0^{-1}\left(\frac{T_{RW}-T}{T_{RW}}\right)\,,\\
    h&=h_0^{-1}\left(\frac{\hat{\mu}_B-i\pi}{i\pi}\right)\,,
\end{align}
where $\hat{\mu}_B=\mu_B/T$ and $t_0$, $h_0$, and $T_{RW}$ are nonuniversal parameters. 
$T_{RW}$ is the RW transition temperature, which is known for our particular lattice setup \cite{Goswami:2019exb}. 
We now can solve $t/h^{1/\beta\delta}\equiv z_c = |z_c|e^{i\frac{\pi}{2\beta\delta}}$ for $\hat\mu_B$ to obtain,
\begin{align}
    \hat\mu_{LY}^R&= \pm \pi \left( \frac{z_0}{|z_c|} \right)^{\beta\delta} \left(\frac{T_{RW}-T}{T_{RW}}\right)^{\beta\delta}, 
    \label{eq:RW_scaling_re}\\
    \hat\mu_{LY}^I&= \pm \pi\,,
    \label{eq:RW_scaling_im}
\end{align}
where the normalization constant $z_0$ is defined as $z_0=h_0^{1/\beta\delta}/t_0$. Equations~(\ref{eq:RW_scaling_re}) and (\ref{eq:RW_scaling_im}) thus define the temperature scaling of the Lee-Yang edge singularity, associated with the Roberge-Weiss critical point. 

\subsection{Chiral singularities\label{sec:chiral_sing}}
In the chiral limit of (2+1)-flavor QCD we expect a second order transition in the universality class of the 3d $\rm{O}(4)$-symmetric spin model \cite{Pisarski:1983ms}. 
In recent lattice QCD simulations by the HotQCD Collaboration \cite{Ejiri:2009ac, Ding:2019prx}, consistency with this expected universal behavior could be demonstrated, even though a first order transition at very small pion mass ($m_\pi\lesssim 55~\rm{MeV}$) cannot be excluded with the accuracy of the present data. 
For simulations with staggered fermions on coarse lattices, away from the continuum limit, the universal scaling is expected to be in the universality class of the 3d $\rm{O}(2)$ model, as staggered fermions preserve only a subgroup of the original chiral symmetry. 
The full chiral symmetry is expected to be restored in the continuum limit. 
In the vicinity of the chiral transition the scaling fields can be expressed as 
\begin{align}
\label{chiral_t}
t=& \frac{1}{t_0}\left[\frac{T-T_c}{T_c}+\kappa_2^B\left(\frac{\mu_B}{T}\right)^2\right]\,, \\
h=&\frac{1}{h_0}\frac{m_l}{m_s^{\text{phys}}}\,.
\end{align}
Here the light quark mass $m_l$ in units of the physical strange quark mass $m_s^{\text{phys}}$ takes the role of the symmetry breaking field $(m_l/m_s^{\text{phys}}\propto h)$. 
In addition, this relation involves three nonuniversal parameters $z_0, T_c, \kappa_2^B$. 
The latter two are prominent numbers that quantify the QCD phase diagram and have been determined to quite some precision \cite{Bazavov:2018mes, Ding:2019prx, Borsanyi:2020fev}\footnote{We note that in principle higher order terms in the expansion in Eq.~(\ref{chiral_t}) appear that might become relevant at large $|\mu_B/T|$ and which have been neglected here. Currently the next order in the expansion ($\kappa_4^B$) is found to be zero withing errors \cite{Bazavov:2018mes,Borsanyi:2020fev}. The analysis presented in  Sec.~\ref{sec:scalingChiral} -- albeit subject to some caveats as discussed -- gives evidence that truncating the expansion after $\kappa_2^B$ is indeed verified}.
The normalization constant $z_0$ is known with less precision but it can, in principle, be inferred from scaling fits of QCD observables to the magnetic equation of state.

The solution for $z=z_c$ now reads
\begin{equation}
\hat\mu_{LY}=\left[\frac{1}{\kappa_2^B}\left(\frac{z_c}{z_0}\left(\frac{m_l}{m_s^{\text{phys}}}\right)^{1/\beta\delta}-\frac{T-T_c}{T_c}\right)\right]^{1/2}\,,
\label{eq:chiral_zline}
\end{equation}
where $z_0=h_0^{1/\beta\delta}/t_0$. This solution has also been used in \cite{Mukherjee:2019eou} to derive an estimate of the radius of convergence. 
Equation~(\ref{eq:chiral_zline}) is visualized in Fig.~\ref{fig:overview} as green band, where we chose $m_l/m_s^{\text{phys}}=1/27$ (physical mass ratio) and $T_c=147$ MeV which is our best estimate for the chiral transition temperature for $N_\tau=6$. 
It is however obvious that $T_c$ does not alter the line of constant $z=z_c$ much; it mainly alters the normalization of the temperature behavior. 
The curvature  $\kappa_2^B$ is chosen as $\kappa_2^B=0.012$. 
We vary $z_0$ from 1.5 to 2.5 which generates the width of the green band. Our best estimate for $N_\tau=6$ is $z_0=2.35$, which stems from scaling fits to the magnetic equation of state.

\subsection{The QCD critical point}
The same kind of scaling is expected to hold close to the QCD critical point. 
Unfortunately, the mapping to the universal theory with $Z(2)$ symmetry is unknown. 
A frequently used \textit{Ansatz} for the scaling fields is based on a linear mapping
\begin{align}
t=& \alpha_t(T-T_{\text{cep}})+\beta_t (\mu_B - {\mu_B}_{\text{cep}})\\
h=&\alpha_h(T-T_{\text{cep}})+\beta_h (\mu_B -{\mu_B}_{\text{cep}})\,,
\end{align} 
where the critical point is located at ($T_{cep}, \mu_{cep}$). This \textit{Ansatz} leads to \cite{Stephanov:2006dn}
\begin{equation}
\mu_{LY}=\mu_{\text{cep}}- c_1(T-T_{\text{cep}})+ic_2|z_c|^{-\beta\delta}(T-T_{\text{cep}})^{\beta\delta}\, ,
\end{equation}
where $c_1$ is given by the slope of the transition line at the critical point and $c_2$ is related to the angle between the $(t=0)$ and $(h=0)$ lines. For the red band in Fig.~\ref{fig:overview}, we chose $c_1=-2 \kappa_2^B=-0.024$, assuming that the transition line is a quadratic function all the way down to the critical point.
We vary $\mu_{cep}$ from 500 to 630 MeV, which generates the width of the red band. For $T_{\text{cep}}$, we chose in accordance with the \textit{Ansatz} for the transition line: $T_{cep} = T_c (1- \kappa_2^B ( {\mu_{\text{cep}}}/{T_c})^2)$, with $T_c=156.5$ MeV.
The prefactor $c_2|z_c|^{-\beta\delta}$ was chosen to be 0.5. In principle, we need at least four data points in the scaling region of the critical point to determine all the unknown nonuniversal parameters, including the location of the QCD critical point $T_{\text{cep}}, \mu_{\text{cep}}$. 
For the Gross-Neveu model, it has been recently shown that $T_{\text{cep}}, \mu_{\text{cep}}, c_1, c_2$ can be determined from a fit to the above \textit{Ansatz}~\cite{Basar:2021hdf}. It will be very interesting to apply such scaling fits also to lattice QCD data, which is however currently beyond the scope of this exploratory study.

\subsection{Thermal singularities}
The Lee-Yang edge singularities are not the only singularities in the complex $\mu_B/T$ plane. 
At temperatures above the QCD crossover ($T>T_{pc}$), we expect that quasifree quarks are the relevant degrees of freedom in the system. 
Quarks are distributed according to the Fermi-Dirac distribution $f_p(T,\mu)=1/(\exp{(\varepsilon_p-\mu)/T}+1)$. 
The singularities of this function are located at $\pm i \pi T \pm \varepsilon_p$. 
In particular, the singularities which are closest to the origin are located at $\pm i \pi T \pm \epsilon_0$, where $\varepsilon_0=m$ is the rest mass of the particle. 
The thermal singularities of quasifree quarks are modified by residual interactions as long as we are not considering the Stefan-Boltzmann limit ($T\to \infty$).
We expect that to leading order these modifications are expressed through a substantially larger thermal mass $\tilde{m}(T)\gg m$.

The analytic structure of the Fermi-Dirac distribution function interferes with the scaling of the Lee-Yang edge singularity of the Roberge-Weiss transition, as given in Eqs.~(\ref{eq:RW_scaling_re}) and (\ref{eq:RW_scaling_im}). 
It is \textit{a priori} not clear which type of singularities are closer to the origin/imaginary axis and can thus be found by a Padé/rational approximation of the data. 
As a result of this study, we find that the leading singularities at $\text{Im}[\mu_B/T]=\pm\pi$ follow the RW scaling, see Sec.~\ref{sec:resultsRW}.

\section{Lattice setup \label{sec:lattice} and observables}
The partition function of (2+1)-flavor of highly improved staggered quarks (HISQ) \cite{Follana:2006rc} with imaginary chemical potential can be written as
\begin{align}
    Z=\int\mathcal{D}U\;&\text{det}[M(m_l,i\mu_l^I)]^{2/4} \nonumber\\
    \times\; & \text{det}[M(m_s,i\mu_s^I)]^{1/4}\;e^{-S_G(U)}\,, 
\end{align}
where $M(m,i\mu^I)$ represents the fermion matrix of a HISQ flavor with mass $m$ and chemical potential $\mu=i\mu^I$. The first determinant represents the two degenerate light flavors (up and down quarks). For the gauge part $S_G(U)$, we are using the Symanzik improved Wilson action, which is correct to $\mathcal{O}(a^2)$ in the lattice spacing.
For the gauge field generation, we were using the SIMULATeQCD package \cite{Altenkort:2021fqk} with and implementation of the rational hybrid Monte Carlo algorithm (RHMC) \cite{Clark:2006fx}. 
The lattice bare parameters are used from various publications of HotQCD. The lattice bare quark masses are varied with the lattice coupling such that for each coupling physical meson masses are obtained; i.e., we stay on the line of constant physics (LCP). 
Here, we make use of the parametrization of the LCP (for the physical value of the pion mass, $m_l/m_s^{\text{phys}}=1/27$) obtained and refined in previous works \cite{Bazavov:2011nk, HotQCD:2014kol, Bollweg:2021vqf}. 
The same holds true for the scale setting, where we used the parametrization of the $\beta$ function based on the kaon decay constant. 
For simplicity, we fix the ratio of the explored chemical potential in this study to $\mu_l/\mu_s=1$.

The observables we calculate are the cumulants of the net baryon number density, given as 
\begin{align}
    &\chi_n^B(T,V,\mu_B)=\left(\frac{\partial}{\partial \hat\mu_B}\right)^n\frac{\ln Z(T,V,\mu_l,\mu_s)}{VT^3}\nonumber \\
    &=\left(\frac{1}{3}\frac{\partial}{\partial \hat\mu_l}+\frac{1}{3}\frac{\partial}{\partial \hat\mu_s}\right)^n\frac{\ln Z(T,V,\mu_l,\mu_s)}{VT^3}\, ,
\end{align}
with $\hat\mu_X=\mu_X/T$, $X=B,l,s$. Note that the normalization is done with appropriate powers of the temperature, such that the observables are dimensionless. 
The derivatives generate traces of the type $\text{Tr}[(M^{-1}\partial M/\partial \mu_X)^n]$, which we evaluate with the random noise method, using $\mathcal{O}(500)$ random vectors. 
For more details on the required traces and the method of evaluation, see, e.g., \cite{Allton:2002zi}.

\begin{figure}
    \centering
    \includegraphics[width=0.5\textwidth]{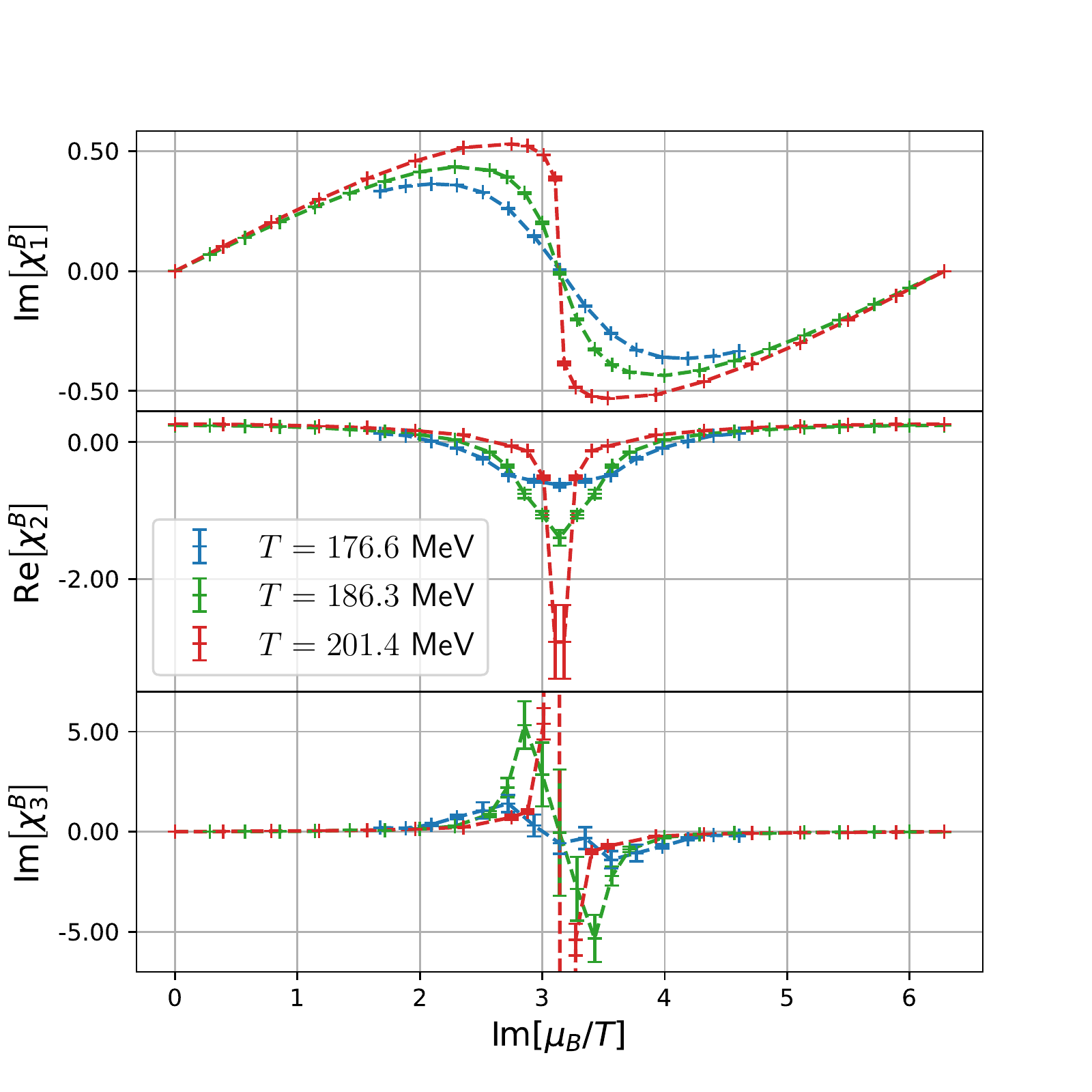}
    \caption{Cumulants of the net baryon number fluctuations as a function of a purely imaginary chemical potential, for three different temperatures, obtained on $24^3 \times 4$ lattices. Shown are Im[$\chi_1^B$] (top), Re[$\chi_2^B$] (middle) and Im[$\chi_3^B$]. Data points are connected by dashed lines to guide the eye.}
    \label{fig:chi-data}
\end{figure}

For obvious reasons (sign problem), we perform our calculations at purely imaginary baryon chemical potential $i\hmuBI$, with $\hmuBI\in\mathbb{R}$.
Exploiting all symmetries, we restrict values for $\hmuBI$ to half the period, i.e., $\hmuBI\in[0,\pi]$. 
The symmetries of the partition function generate specific properties of the observables $\chi_n^B$. 
At imaginary chemical potential, they are imaginary and odd functions of $\hmuBI$ for odd $n$ and real and even functions of $\hmuBI$ for even $n$. 
These properties have been verified by us and can be seen from Fig.~\ref{fig:chi-data}, where we show results for the first three cumulants. 
Preliminary results were presented in \cite{Schmidt:2021pey}. The data are tabulated in Appendix~\ref{appendix_Tables}.

Since this is an exploratory study, we leave the continuum extrapolation for later publications and perform the calculations on rather course lattices, $24^3\times 4$ and $36^3\times 6$. 
We note however that we have deliberately chosen rather large spatial volumes (we have aspect ratio $N_\sigma/N_\tau=6$) in order to minimize finite size effects which are expected to become large in the vicinity of a phase transition. 

\section{Rational function approximation of the lattice data \label{sec:rational}}
\subsection{Pad\'e approximants}
Pad\'e approximants \cite{baker1975} are a popular subject in approximation theory. The main idea is to approximate a given function $f(x)$ with a rational function whose derivatives agree with those of $f(x)$ up to a given order. 
This can be easily rephrased in terms of power series. To set up our notations, we first consider a so-called {\em single point [m/n] Pad\'e}. Suppose the Taylor expansion of $f(x)$ about a single point (in what follows, it is useful to take this point as $x=0$) is known up to a certain order (say L),  
\begin{equation}
  f(x) = \sum\limits_{i=0}^L \, c_i \, x^i + \mathcal{O}(x^{L+1})\,.
\end{equation}
We denote $R^{m}_{n}(x)$ the $[m/n]$ Pad\'e approximant we are looking for, 
\begin{equation}
\label{eq:BasicPade}
R^{m}_{n}(x) = \frac{P_m(x)}{\tilde{Q}_n(x)} = \frac{P_m(x)}{1+Q_n(x)} = \frac{\sum\limits_{i=0}^m \, a_i \, x^i}{1 + \sum\limits_{j=1}^n \, b_j \, x^j}\,,
\end{equation}
which is the ratio of two polynomials ($P_m$ and $\tilde{Q}_n$) of order $m$ and $n$, respectively. 
By discarding a nontrivial $b_0$ and writing $\tilde{Q}_n(x)=1+Q_n(x)$, in Eq.~(\ref{eq:BasicPade}) we have made a definite choice for the coefficients $\{a_i,b_j\}$ our approximant essentially depends on. Given our knowledge of the Taylor expansion for $f(x)$, in principle, our choice can be for any order $[m/n]$ such that $m + n + 1 = L + 1$. Strictly speaking, not all such $[m/n]$ approximants exist.\\
Rational functions are not the only viable solutions in approximation theory; polynomial approximants are very popular as well. 
We stress from the very beginning a main virtue of rational functions we are interested in; they provide a natural handle to probing the singularities structure of $f(x)$. 
Quite trivially, the singularities of $R^{m}_{n}(x)$ are coming from the zeros of $1+Q_n(x)$. To make the latter statement more precise, we need to more precisely state what we mean by a $[m/n]$ approximant. Our attitude is quite pragmatic; we consider a $R^{m}_{n}(x)$ for a given choice of $[m/n]$ and solve for the $\{a_i,b_j\}$ coefficients given that the power series of $f(x)$ is known to a given order $L$ such that $m + n + 1 = L + 1$. Given our input, we are not guaranteed that $P_m$ and $1+Q_n(x)$ are coprime polynomials \footnote{One could argue this is not the most natural definition of the order $[m/n]$; it is for sure the one that gets closer to what one has to do in practice to solve for the unknown coefficients.}. If this is not the case (i.e., there is a nontrivial greatest common divisor), singularities of $R^{m}_{n}(x)$ are coming from those zeros of $1+Q_n(x)$ which either are not zeros of $P_m(x)$ or are zeros of $1+Q_n(x)$ of a higher order than they are of $P_m(x)$. While trivial, this very last statement will be of some relevance in the following. We are also quite concerned with yet another property of $R^{m}_{n}(x)$. Poles are the only singularities we can find in a rational function; still, there are {\em signatures} of other singularities of $f(x)$ (e.g., branch cuts) which we can recognize in $R^{m}_{n}(x)$ once the coefficients of the latter have been determined to reconstruct the power series of $f(x)$. We discuss this topic (which indeed is relevant in the our analysis) in Appendix~\ref{sec:VariousSing}.\\
The most direct way of solving for the unknown coefficients is by {\em approximation through order}; we match coefficients of different powers of $x$ between the (unknown) rational function and the (known) Taylor series by demanding functional independence (up to order $x^{L}$). That means we rewrite 
\begin{equation}
\begin{split}
\sum\limits_{i=0}^m \, a_i \, x^i &= P_m(x) = f(x)\,(1+Q_n(x)) \\ \nonumber
&= (\sum\limits_{i=0}^L \, c_i \, x^i) \, (1+\sum\limits_{j=1}^n \, b_j \, x^j)\,, \nonumber
\end{split} 
\end{equation} 
and match
\begin{equation}
\label{eq:LinearProblem}
\begin{split}
a_0 &= c_0 \\ 
a_1 &= c_1 + b_1 c_0 \\ 
a_2 &= c_2 + b_1 c_1 + b_2 c_0 \\ 
 & \hdots
\end{split}
\end{equation}
Equation.~(\ref{eq:LinearProblem}) defines a set of simultaneous, linear equations, which can be solved by a convenient linear solver. We stress that most often the linear system is not singular but ill-conditioned, which is a warning that determining the solution can be hard. Most importantly, we should keep in mind that our knowledge of the derivatives of $f(x)$ is coming from stochastic evaluation via Monte Carlo simulations. Despite this, we see that we can manage to find the information we are aiming at.\\
Approximation through order somehow hides that the information we are making use of is coming from derivatives of $f(x)$. The solution encoded in Eq.~(\ref{eq:LinearProblem}) can of course also be obtained by evaluating in $x=0$ the tower of relationships
\begin{equation}
\label{eq:LinearProblem2}
\begin{split}
P_m(x) - f(x)Q_n(x) &= f(x) \\ 
P_m'(x) - f'(x)Q_n(x) - f(x)Q_n'(x) &= f'(x) \\ 
P''_m(x) - f''(x)Q_n(x) - f(x)Q_n''(x)\\ - \;2f'(x)Q_n'(x) &= f''(x) \\ 
 & \hdots
\end{split}
\end{equation}
Yet another way of obtaining the unknown ${a_i,b_j}$ is to solve the set of equations 
$\frac{d^k}{dx^k}R^{m}_{n}(x) = f^{(k)}(x)$, i.e.~\footnote{Again, we assume we know the derivatives in $x=0$},
\begin{equation}
\label{eq:Derivatives}
\begin{split}
a_0 &= f(0) \\
a_1 - a_0 b_1 &= f'(0) \\
2 a_2 - 2 a_1 b_1 + a_0 (2 b_1^2 - 2 b_2) &= f''(0) \\ 
 & \hdots
\end{split}
\end{equation}
While Eqs. (\ref{eq:LinearProblem}), (\ref{eq:LinearProblem2}), and (\ref{eq:Derivatives}) are equivalent, the latter is somehow less convenient, not being linear and typically requires the use of computer algebra tools like {\em Mathematica}. In our particular problem, we explicitly showed that the three return the same results (to a very good approximation); Eq.~(\ref{eq:LinearProblem2}) has been to a large extent our preferred choice.\\
There exists a significant amount of literature on single point Padé approximants (about existence, uniqueness, and convergence). 
This is not true for the so-called \textit{multipoint Padé} to the same extent. 
The construction of a multipoint Padé can be extremely useful in situations when Taylor coefficients for a function about a single point are not known to higher orders, but instead either the function values are known or a few Taylor coefficients are known about (possibly) many points. 
Since this is precisely the situation we face in our lattice studies of QCD at finite chemical potential, it is useful to understand how to build rational
approximations from these multipoints.\\
Extending what we saw above to multipoints is straightforward; in particular, we extend the formalism encoded in Eq.~(\ref{eq:LinearProblem2}). A few Taylor coefficients (i.e., derivatives) of a function $f(x)$ known at a collection of points $\{x_i\,|\,i=1 \ldots N\}$ are consistent with the approximation to $f$ provided by Eq.~(\ref{eq:BasicPade}) if they satisfy the set of equations
\begin{equation}
\label{eq:LinearProblem3}
\begin{split}
P_m(x_1) - f(x_1)Q_n(x_1) &= f(x_1) \\ 
P_m'(x_1) - f'(x_1)Q_n(x_1) - f(x_1)Q_n'(x_1) &= f'(x_1) \\ 
 & \hdots \\
P_m(x_2) - f(x_2)Q_n(x_2) &= f(x_2) \\ 
P_m'(x_2) - f'(x_2)Q_n(x_2) - f(x_2)Q_n'(x_2) &= f'(x_2) \\ 
 & \hdots \\
P_m(x_N) - f(x_N)Q_n(x_N) &= f(x_N) \\ 
P_m'(x_N) - f'(x_N)Q_n(x_N) - f(x_N)Q_n'(x_N) &= f'(x_N) \\ 
 & \hdots\,,
\end{split}
\end{equation}
which is once again a linear system in $n+m+1$ unknowns where now $n+m+1=\sum_{i=1}^{N} (L_i+1)$. In the previous formula, the highest order of derivative which we know (i.e., $L_i$) can be different for different points. \\
Multipoints Pad\'e will be our choice for the analysis of this work. We now proceed to discuss a few technical details of our implementation, referring the reader to Appendixes~\ref{sec:ApPADE}-\ref{sec:LastApPADE} for extra remarks and comments.\\

\subsection{Pad\'e approximants for the QCD net baryon number density from imaginary chemical potential}
This is not the first time Pad\'e approximants are applied to the study of finite density (lattice) QCD \cite{Lombardo:2005ks,PhysRevD.78.114503,2011PhLB..696..459G}. A recent paper has, in particular, proposed to join Pad\'e analysis and Bayesian methods, with applications to the study of the crossover line \cite{PhysRevD.103.034511}. In a way that is close to the spirit of this work, in recent times, Pad\'e approximants have been successfully used to probe the singularity structure of simple theories in the context of the Lefschetz thimble method \cite{DiRenzo:2020cgp}. To our knowledge, this work is in a sense the first attempt at a systematic study of the QCD phase diagram and, in particular, of Lee-Yang edge singularities, building on Pad\'e analysis. The function we want to approximate by rational functions is the net baryon number density $\chi_1^B(T,V,\mu_B)$, with the cumulants $\chi_n^B(T,V,\mu_B)$ ($n>1$) entering Eq.~(\ref{eq:LinearProblem3}) as derivatives. Our main goal is to get {\em signatures of singularities} of $\chi_1^B(T,V,\mu_B)$ in the complex-$\mu_B$ ($\mu_B=\mu_B^{I}+i\mu_B^{R}$) plane (at fixed values of $T$ and $V$). Ultimately we aim to understand the phase diagram of the theory.  In particular, we find  clear evidence of the Roberge-Weiss transition in the $\mu_B^{I}-T$ plane. Most importantly, if at some point we could find evidence of singularities eventually pinching the (real) $\mu_B^{R}$ axis, then we would be in the presence of a QCD critical point candidate.\\
We have already seen that Eq.~(\ref{eq:LinearProblem3}) is not the only way to solve for the coefficients entering the Pad\'e approximants (\ref{eq:BasicPade}). Not only, e.g., does the multipoints version of Eq.~(\ref{eq:Derivatives}) works as well, but also other formalisms could be (and actually were) used in our analysis. In the construction of these other formalisms, a key point is 
that the values of the $\chi_n^B$, i.e., the function $f$ and its derivatives in Eq.~(\ref{eq:LinearProblem3}), are known {\em to a limited precision} since they are evaluated by Monte Carlo. \\
Our Pad\'e analysis was performed following three different approaches, aiming at assessing their mutual consistency. 

\begin{enumerate}
    \item The solution of the linear system (\ref{eq:LinearProblem3}) has been worked out in two different ways, namely,
    \begin{itemize}
        \item One can build the system by writing the most general form for $R^{m}_{n}(x)$, i.e., that of 
        (\ref{eq:BasicPade}).
        \item One can instead impose the form 
        \begin{equation}
        \label{eq:EOPade}
        \begin{split}
        R^{m}_{n}(x) = \frac{\sum\limits_{i=0}^{m'} \, a_{2i+1} \, x^{2i+1}}{1 + \sum\limits_{j=1}^{n/2} \, b_{2j} \, x^{2j}}\,,& \\
        (\;m = 2m'+1, \;\, a_1 = \chi_2^B(T,V,0)\;)\,,&
        \end{split}
        \end{equation}
        with the coefficients $\{a_i\}$ and $\{b_j\}$ that turn out to be real.
        This form ensures the following: (a) The function $\chi_1^B(T,V,\mu_B)$ has the right parity (it is an odd function). (b) As a consequence of the coefficients being real valued, for imaginary $\mu_B=\mu_B^{I}$, the odd cumulants $\chi_{2n+1}^B(T,V,\mu_B^I)$ are imaginary valued, while the even $\chi_{2n}^B(T,V,\mu_B^I)$ are real valued, as it must be. (c) When Eq.~(\ref{eq:EOPade}) is computed for real $\mu_B=\mu_B^{R}$, the cumulants are real; i.e., the analytic continuation one is typically interested in is guaranteed to be meaningful.
    \end{itemize}
    Notice that taking into account different functional forms for $R^{m}_{n}$ is not the end of the story. Another alternative which can (and actually was) taken into account is whether one
    \begin{itemize}
        \item performs the Pad\'e analysis in the (original) complex-$\mu_B$ plane or
        \item goes through a {\em conformal map} $\mu_B=\phi(\nu)$ and performs the Pad\'e analysis in the complex-$\nu$ plane.
    \end{itemize}
    This is in the spirit of \cite{Skokov:2010uc,Costin:2021bay,Basar:2021hdf}.
    \item Because of the cumulants being known to finite precision, the minimization of a generalized $\chi^2$ is an obvious alternative to the solution of (\ref{eq:LinearProblem3}). Suppose we want $R^{m}_{n}(x)$ to be a Pad\'e approximant for the function $f(x)$ whose values and derivatives we know at given points $\{x_j\,|\,j=1 \dots N\}$, {\em i.e.} $c_j^{(k)} \equiv \frac{\partial^j f}{\partial x^j}(x_k) \simeq \frac{\partial^j R^{m}_{n}}{\partial x^j}(x_k)$, with the $c_j^{(k)}$ known with errors $\Delta c_j^{(k)}$. Then, the coefficients $\{a_i,b_j\}$ the $R^{m}_{n}$ depends on can be fixed minimizing the generalized $\chi^2$,
    \begin{equation}
    \label{eq:chi2}
        \tilde{\chi}^2 = \sum_{j,k} \frac{|\frac{\partial^j R^{m}_{n}}{\partial x^j}(x_k)-c_j^{(k)}|^2}{|\Delta c_j^{(k)}|^2}.
    \end{equation}
    Of course, all the alternatives that we commented in 1 (namely, different functional forms for $R^{m}_{n}$, use of conformal maps) can be also implemented in this approach.
    \item Both 1 and 2 make use of the knowledge of $f(x)$ (and its derivatives) at given points; i.e., the only information on $f(x)$ we have is at a finite (possibly small) number of points. One could instead compute a smooth interpolation of $f(x)$ before entering the Pad\'e analysis. 
\end{enumerate}

\subsection{Results of Pad\'e analysis of net baryon number density}
\begin{figure} 
        \includegraphics[scale=0.55]{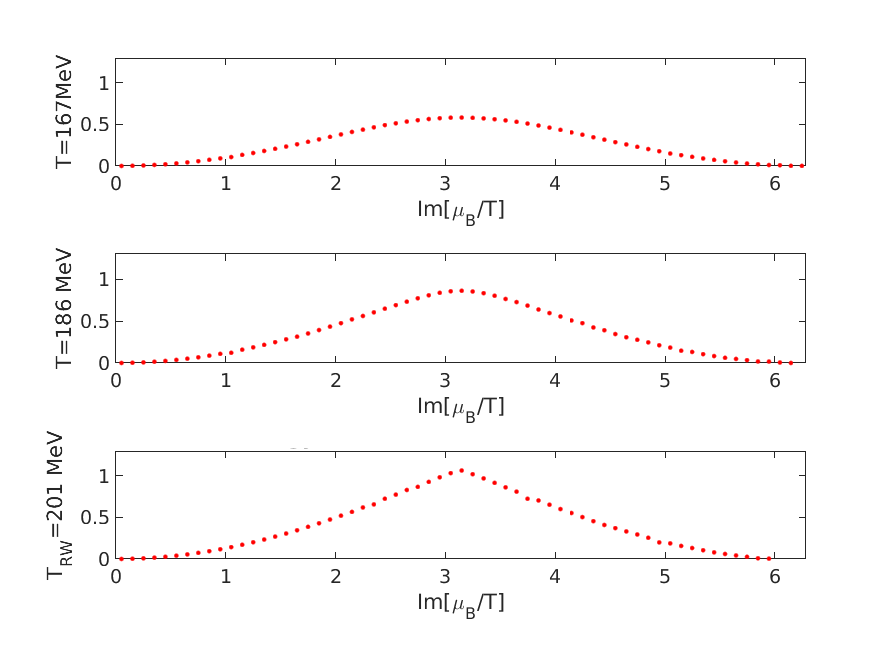}
        \caption{The free energy as a function of $\hat\mu_B^{I}$ at three different temperatures.}
        \label{fig:FreE}
\end{figure}

\begin{figure}[htb] 
        \centering
        \includegraphics[scale=0.5]{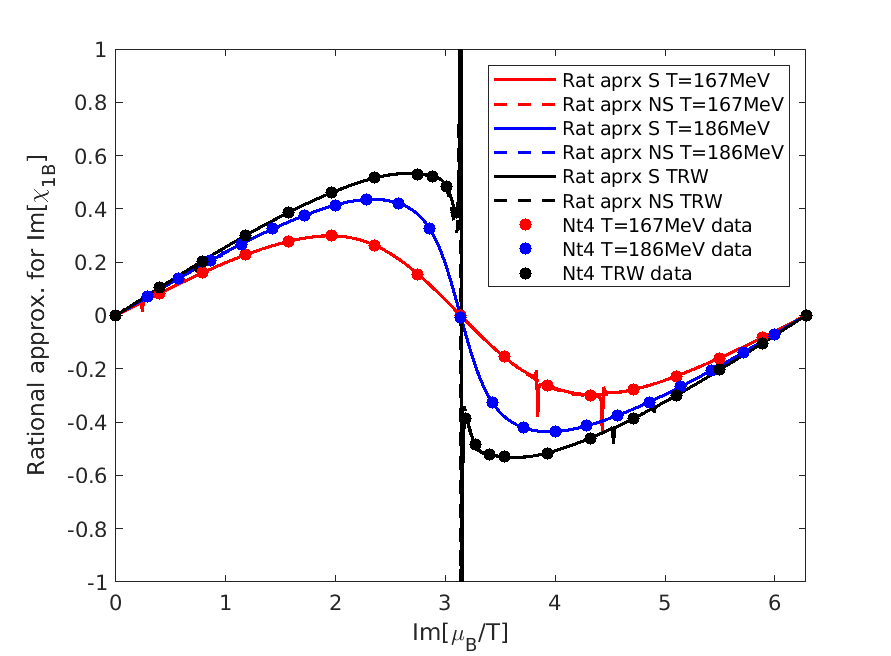}
        \vskip0.5cm
        \includegraphics[scale=0.5]{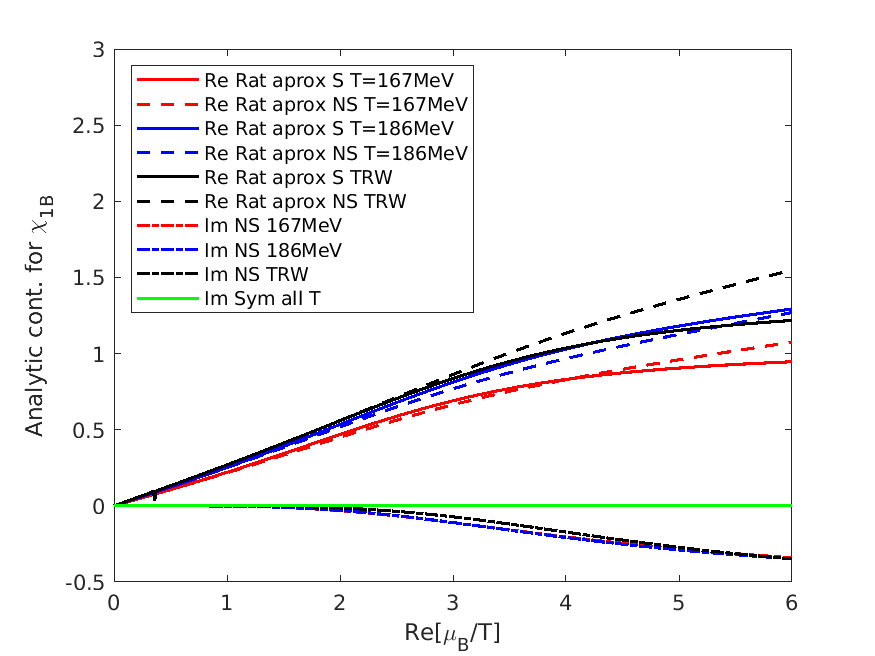}
        \caption{Top: the net baryon number density as a function of $\hat\mu_B^{I}$ at three different temperatures; for each, (\ref{eq:BasicPade}) and (\ref{eq:EOPade}) fall on top of each other. Bottom: the analytic continuation of the baryon number density for three different temperatures. Again, we plot both (\ref{eq:BasicPade}) and (\ref{eq:EOPade}); for the latter, the imaginary part is guaranteed to be zero.}
        \label{fig:ReImMU}
\end{figure}

The focus of our analysis is on singularities of the net baryon number density. Still, before proceeding to this, we make a short digression on a feature which is worth discussing. 
In investigating the phase diagram of QCD in the (imaginary chemical potential-temperature) $\mu_B^I$-$T$ plane, and, in particular, in the study of the Roberge-Weiss transition, a prominent role is played by the free energy as a function of $\hat\mu_B^{I}$ (at given values of the temperature $T$); a cartoon for this quantity is often plotted. Since we have a function $R^{m}_{n}(\hat\mu_{B}^{I})$ approximating the net baryon density, we can obtain the free energy $F(\hat\mu_B^{I})$ by (numerical) integration. In Fig.~\ref{fig:FreE}, we display the free energy $F(\hat\mu_B^{I})$ at three different temperatures; the profile clearly gets closer to a cusp as the temperature gets closer to $T=T_{RW}$ (to the extent that the transition can be detected on a finite volume).\\
We now inspect how well our rational approximants describe the data. On top of that, we are interested in the analytic continuation of results from imaginary to real values of the baryonic chemical potential (this is in the end a key issue in any imaginary-$\mu_B$ study of finite density lattice QCD). Finally, we present the relevant singularity pattern which emerges from our analysis. In Fig.~\ref{fig:ReImMU}, we display what we get both for imaginary and for real baryonic chemical potential.
For three of the temperatures we probed on $N_{\tau}=4$, we plot the results we got from the solution of Eq.~(\ref{eq:LinearProblem3}) for both functional forms~(\ref{eq:BasicPade}) and (\ref{eq:EOPade}). For imaginary values of the baryonic chemical potential, the two solutions are {\em de facto} indistinguishable. For real values (analytic continuation), the real parts are quite close to each other, a significant discrepancy between the two different \textit{Ansätze} being there only at $T=T_{RW}$ and for $\hat\mu_B^{R}>\pi$. As for imaginary parts, Eq.~(\ref{eq:EOPade}) is guaranteed to return zero; it is interesting to notice that also the solution we got for the \textit{Ansatz}~(\ref{eq:BasicPade}) has a quite tiny imaginary part (at least up to $\hat\mu_B^{R} \sim \pi$). All this can be taken as an indication of reasonably tiny systematic effects as far as the dependence on the precise form of the Pad\'e approximants is concerned. All in all, the indeterminations we have to live with when we analytically continue our results to real baryonic chemical potential seem to be competitive when we compare to other methods. This is true despite the fact that, inspecting Fig.~\ref{fig:ReImMU}, a few {\em spikes} are clearly visible; we see in what sense they do not come as a surprise and are in fact harmless. \\
\begin{figure*}[ht] 
        \centering
        \includegraphics[scale=0.41]{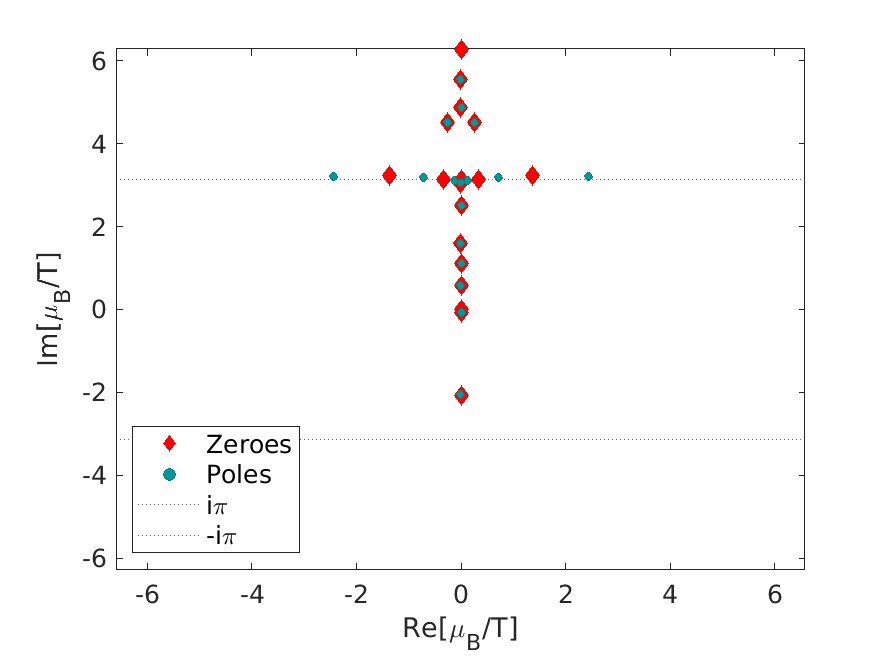}
        \hskip-0.5cm
        \includegraphics[scale=0.41]{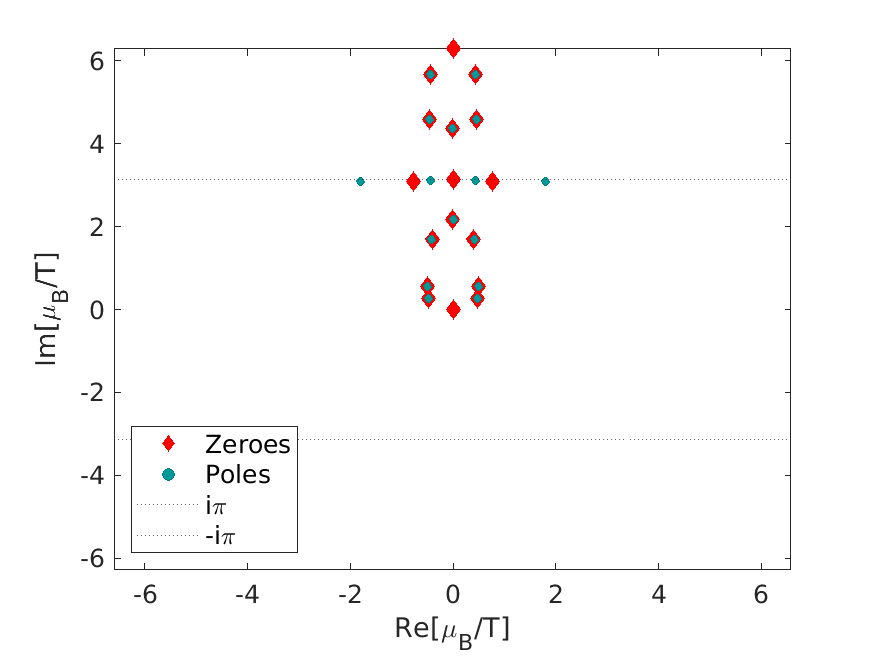}
        \hskip-0.5cm
        \includegraphics[scale=0.41]{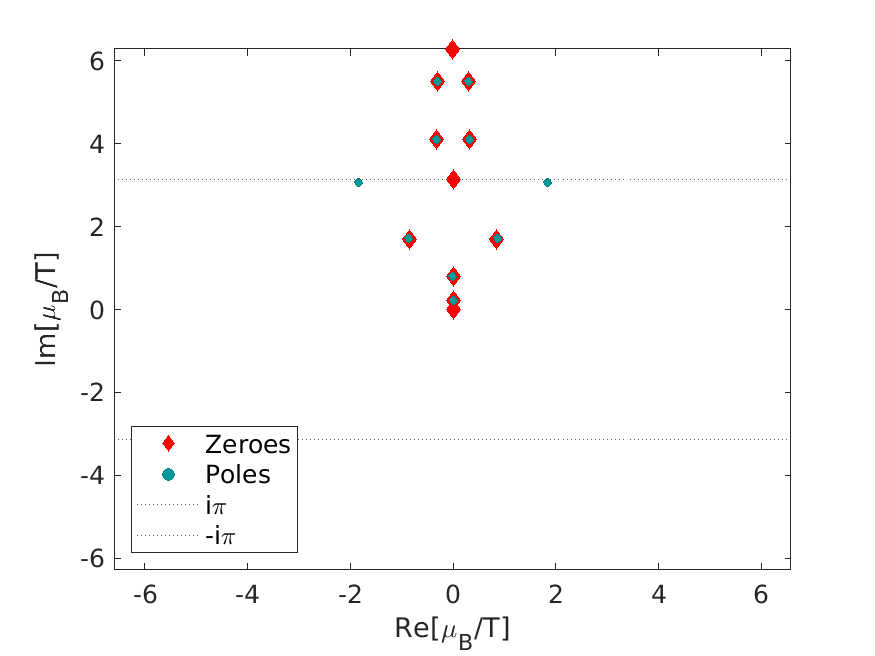}
        \includegraphics[scale=0.41]{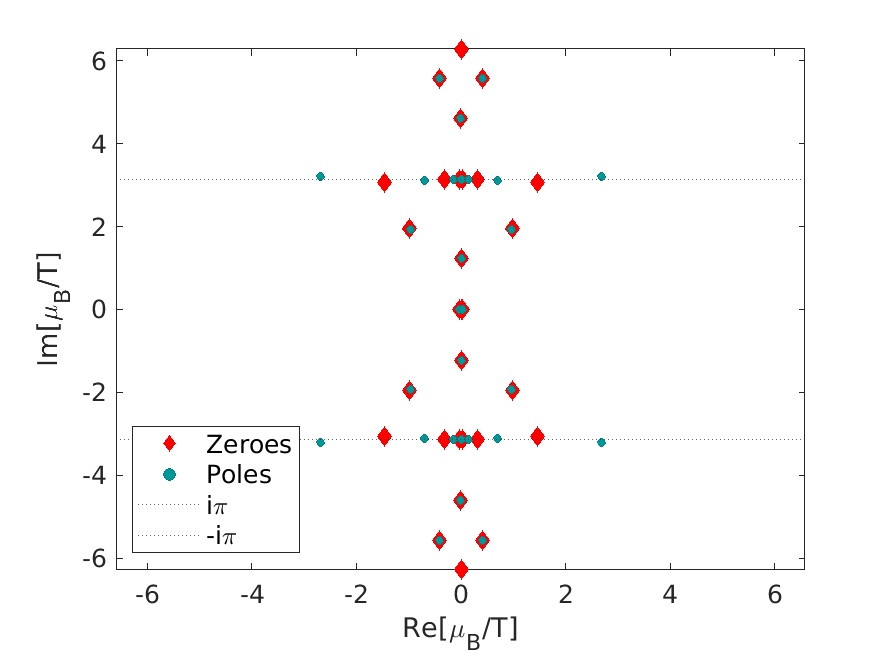}
        \hskip-0.5cm
        \includegraphics[scale=0.42]{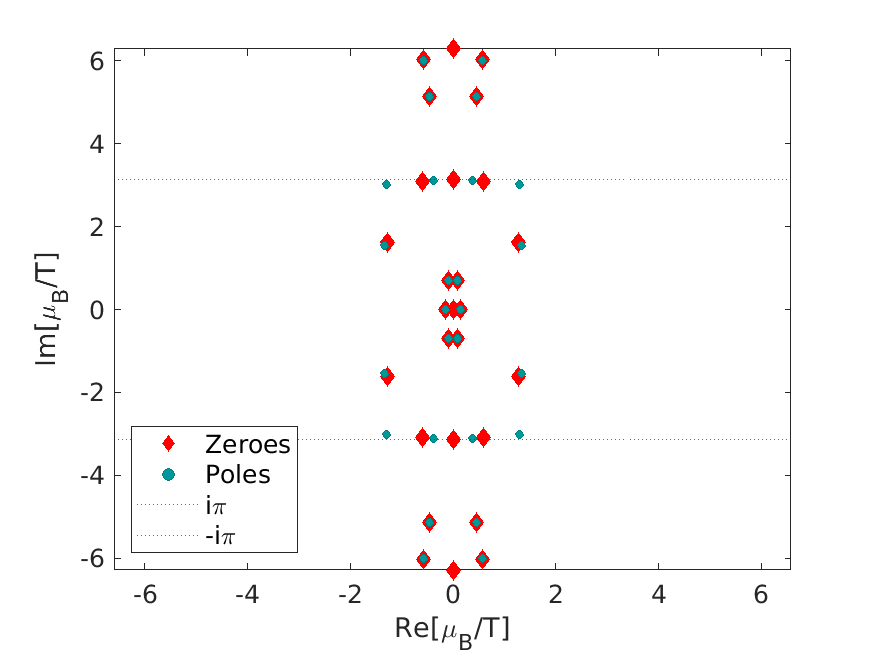}
        \hskip-0.5cm
        \includegraphics[scale=0.42]{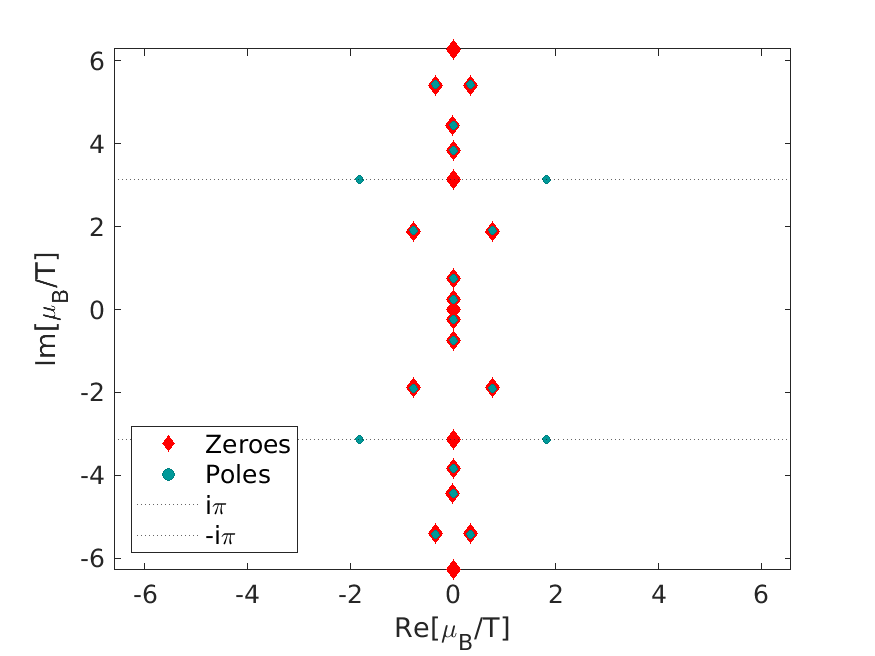}
         \caption{Singularity structure in the $\hat\mu_B$ plane for three different temperatures (from left to right $T=201.4, 186.3, 167.4$). Upper row: \textit{Ansatz}~(\ref{eq:BasicPade}); lower row: \textit{Ansatz}~(\ref{eq:EOPade}). }
         \label{fig:SingMUplane}
\end{figure*}
In Fig.~\ref{fig:SingMUplane}, we plot the singularity pattern we get at three of the temperatures we probed on $N_{\tau}=4$. We once again present results we get for both functional forms~(\ref{eq:BasicPade}) and (\ref{eq:EOPade}). A few remarks are in order, which we invite the reader to consider taking into account the points that we make in Appendixes~\ref{sec:SpuriousP},~\ref{sec:VariousSing}, and,~\ref{sec:LastApPADE}.
\begin{itemize}
    \item Thermal singularities are expected to show up at $\hat\mu_B^{I}=\pi$ and indeed to a very good accuracy they do. 
    \item The signature for a {\em branch cut} is clearly visible at $T=T_{RW}=201.4$ MeV, for both \textit{Ansätze}~(\ref{eq:BasicPade}) (see upper row) and (\ref{eq:EOPade}) (lower row). Notice that the latter is by construction sensitive to all the four replicas of the same singularity, as expected for symmetry reasons: if we find a singularity in $z$, then also $-z$ and $-\bar{z}$ must be singular points. This is a general feature, which is clear at all temperatures; (\ref{eq:BasicPade}) instead only captures singularities in the upper half plane (i.e., we see one of the two symmetries). 
    \item At $T=186.3$ MeV (this is the next to highest temperature that we probed), plots apparently allude to a {\em branch cut} as well, while at $T=167.4$ MeV the singularity shows up as a {\em simple pole}. Much the same happens at the remaining temperatures (i.e., $T=176.6$ MeV and $T=160.4$ MeV are consistent with the appearance of simple poles).
    \item While the pattern of the relevant pieces of information (i.e., {\em true poles and zeros}) is the same for different functional forms, the pattern of other zeros and poles depends on the functional form. Notice that the mechanism of {zero-pole cancellations} is manifest; these cancellations are due to {\em numerical noise}. In a sense, we see {\em fake information}, which would not be there for exact data, but since noise is not that much, this fake information is close to disappearing.
    \item While the zero-pole cancellations seem almost perfect in Fig.~\ref{fig:SingMUplane}, Fig.~\ref{fig:ReImMU} is warning us that this is not really the case, and this is the reason for the {\em spikes} which we see there. We encourage the reader to spot which singular points are responsible for the spikes. Again, the almost perfect cancellations in Fig.~\ref{fig:SingMUplane} reveal that these spikes are harmless.
    \item As should be clear, the big spike at $\hat\mu_B=i \pi$ for $T=T_{RW}$ is a different story: this is indeed the Roberge-Weiss transition showing up.
\end{itemize}
\begin{figure*}[ht] 
        \centering
        \includegraphics[scale=0.22]{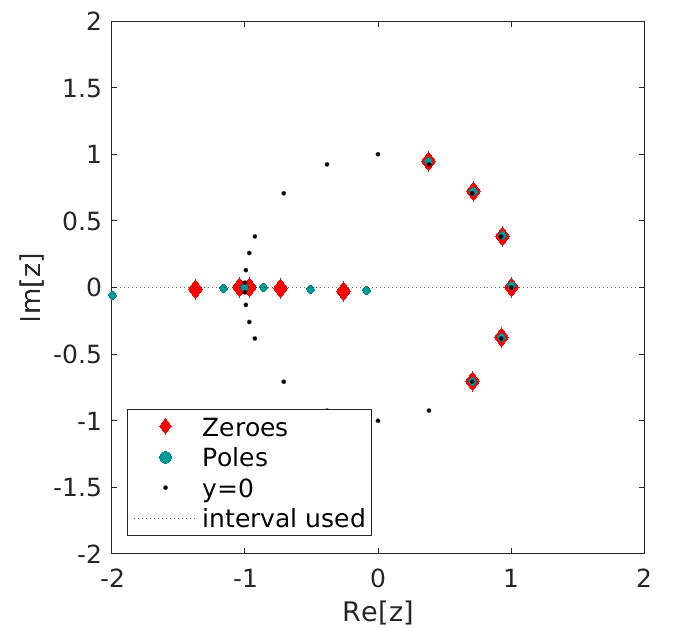}
        \hskip-0.1cm
        \includegraphics[scale=0.22]{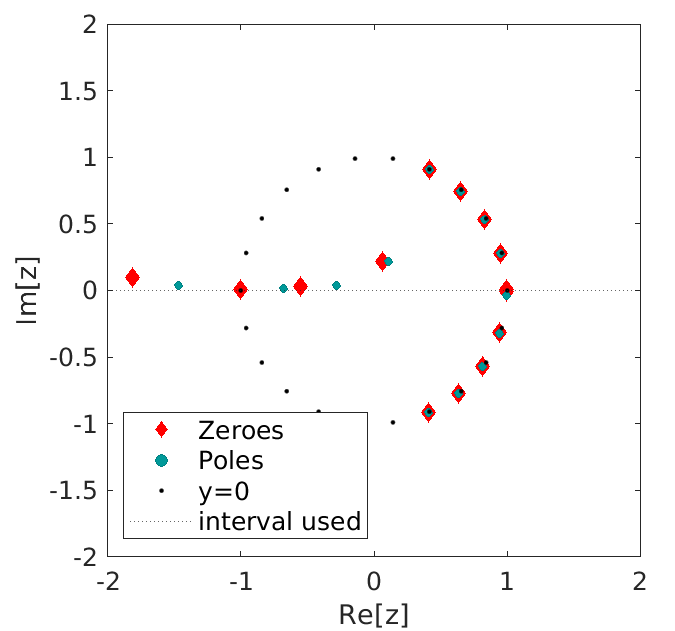}
        \hskip-0.1cm
        \includegraphics[scale=0.22]{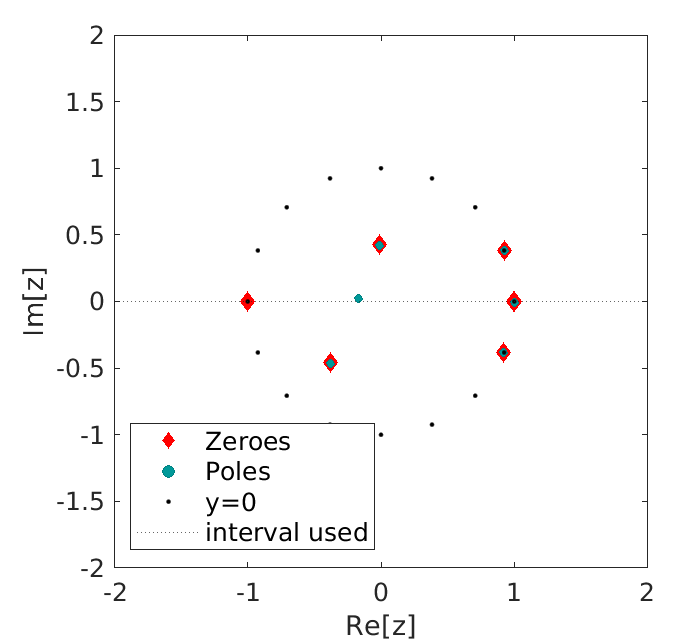}
         \caption{Singularity structure in the fugacity ($z=e^{\frac{\mu_B}{T}}$) plane for three different temperatures (from left to right $T=201.4, 186.3, 167.4$).}
         \label{fig:FUGACITYplane}
\end{figure*}
We repeated our analysis hunting for singular points in the complex-{\em fugacity} plane; after mapping our measurements to this plane, we performed Pad\'e analysis in the $z=e^{\hat\mu_B}$ variable. There are at least two reasons for such an (additional) analysis. First of all, we want to make sure that the information which we get is stable and does not disappear once we change the variable. Also, the linear systems which we have to solve are typically ill-conditioned. Due to the nature of the conformal map, this feature disappears (in a sense, we can trust results to a higher level of confidence). In Fig.~\ref{fig:FUGACITYplane}, we present the singularity pattern in the complex-{\em fugacity} plane.
\begin{itemize}
    \item Since our original data are taken on the imaginary axis ($\mu_B=i \mu_B^{I}$), in the complex fugacity plane, we end up on the unit circle $|z|=1$. In view of the observations that we make in Appendix~\ref{sec:InteDep}, notice that this is a very convenient location with respect to the location of the singularities that we detect.
    \item Since singularities are expected at $\hat\mu_B^{I}=\pi$, in the fugacity plane, they should show up on the real axis, and indeed, they do. Due to the relative positions with respect to the input data ($|z|=1$), we are not sensitive to any (symmetry) replica (that is, one single singularity shows up).  
    \item All the other features (e.g., zero-pole cancellations) show up much the same as they do in the original $\mu_B$ plane.
\end{itemize}
In Table~\ref{tab:nt4LYE}, we collect all the findings that we discussed so far.
\begin{table*}[htp]
    \centering
    \begin{tabular}{|c|c|c|c|c|c|c|c|c|}
    \hline
\multicolumn{1}{|c|}{$T$ (MeV)} & \multicolumn{2}{|c|}{ Method I} & \multicolumn{2}{|c|}{ Method II} & \multicolumn{2}{|c|}{ Method III*} & \multicolumn{2}{|c|}{ Method III} \\ \hline
 & $\hat\mu_{LY}^{R}$ & $\hat\mu_{LY}^{I}$ & $\hat\mu_{LY}^{R}$ & $\hat\mu_{LY}^{I}$ & $\hat\mu_{LY}^{R}$ & $\hat\mu_{LY}^{I}$ & $z^{R}$ & $z^{I}$\\ \hline  
201.4 & 0.11(11) & 3.142(10) & 0.077(45) & 3.133(15) & 0.0541(15) & 3.1294(63) & -0.9472(14) & -0.0116(60) \\
 186.3 & 0.48(14) & 3.118(54) & 0.53(13) & 3.112(66) & 0.397(51) & 3.127(34) & -0.672(34) & 0.010(21) \\
 176.6 & 1.03(10) & 3.112(72) & 1.022(80) & 3.18(12) & 1.040(94) & 3.115(65) & -0.353(33) & -0.010(20) \\
 167.4 & 1.82(11) & 3.125(79) & 1.79(13) & 3.164(95) & 1.694(55) & 3.12(13) & -0.184(12) & 0.004(22) \\
 160.4 & 2.097(90) & 3.147(11) & 2.14(12) & 3.150(70) & 2.07(76) & 3.14(24) & -0.126(70) & 0.000(14) \\ \hline
    \end{tabular}
    \caption{Method I : Linear Solver. Method II : $\chi^2$ fit approach. Method III : Linear solver in fugacity plane.  (Note* : Mapped back values from fugacity plane. We are picking the value in first quadrant given the symmetries of the partition function)}
    \label{tab:nt4LYE}
\end{table*}
In particular, for each temperature that we probed at $N_{\tau}=4$, we list the nearest singularities as obtained (a) from the solution of the linear system (\ref{eq:LinearProblem3}) in the $\mu_B/T$ plane (method I), (b) from the minimisation of the generalised $\chi^2$ (\ref{eq:chi2}) (method II), and (c) from the solution of the linear system (\ref{eq:LinearProblem3}) in fugacity plane, both mapping back results in the original plane and inspecting them in the fugacity plane (method III* and III). The errors are computed out of a bootstrap procedure in which we repeat our Pad\'e analysis letting the input data (i.e., the results of our Monte Carlo measurements) vary within errors. As one can see, results are well consistent.\\
The singularities which we have been discussing so far (and that are listed in Table~\ref{tab:nt4LYE}) are not the only ones on display in Fig.~\ref{fig:overview}. Results obtained on $N_{\tau}=6$ at $T=145$ MeV apparently point at a singular point that could be consistent with a {\em chiral singularity}. While this result is intriguing,  in this case extra care is in order.
\begin{itemize}
    \item In this case, we have a (far) enhanced dependence on the interval our Pad\'e analysis takes into account. In particular, this singularity shows up if we limit our analysis to $\hat\mu_B^{I} \in [0 , \pi]$.
    \item The result which is shown in Fig.~\ref{fig:overview} comes from the minimization of the generalized $\chi^2$ (\ref{eq:chi2}) taking (\ref{eq:BasicPade}) as an \textit{Ansatz}, with $m=n=4$. This choice returns the best $\tilde{\chi}^2$ value.
    \item A consistent result for the singularity is found from other methods if we limit the analysis to the same interval ($\hat\mu_B^{I} \in [0 , \pi]$) (even changing a bit the degree, which thing is easier in this approach). This singularity is not that stable under the variation of the interval. While this is not {\em a priori} that surprising (our multipoint Pad\'e analysis is interval sensitive, see Appendix~\ref{sec:InteDep}), at the same time, it makes the result less solid than what we get on $N_{\tau}=4$ at higher temperatures.
\end{itemize}
\begin{figure*}[htp]
    \centering
    \includegraphics[width=0.8\textwidth]{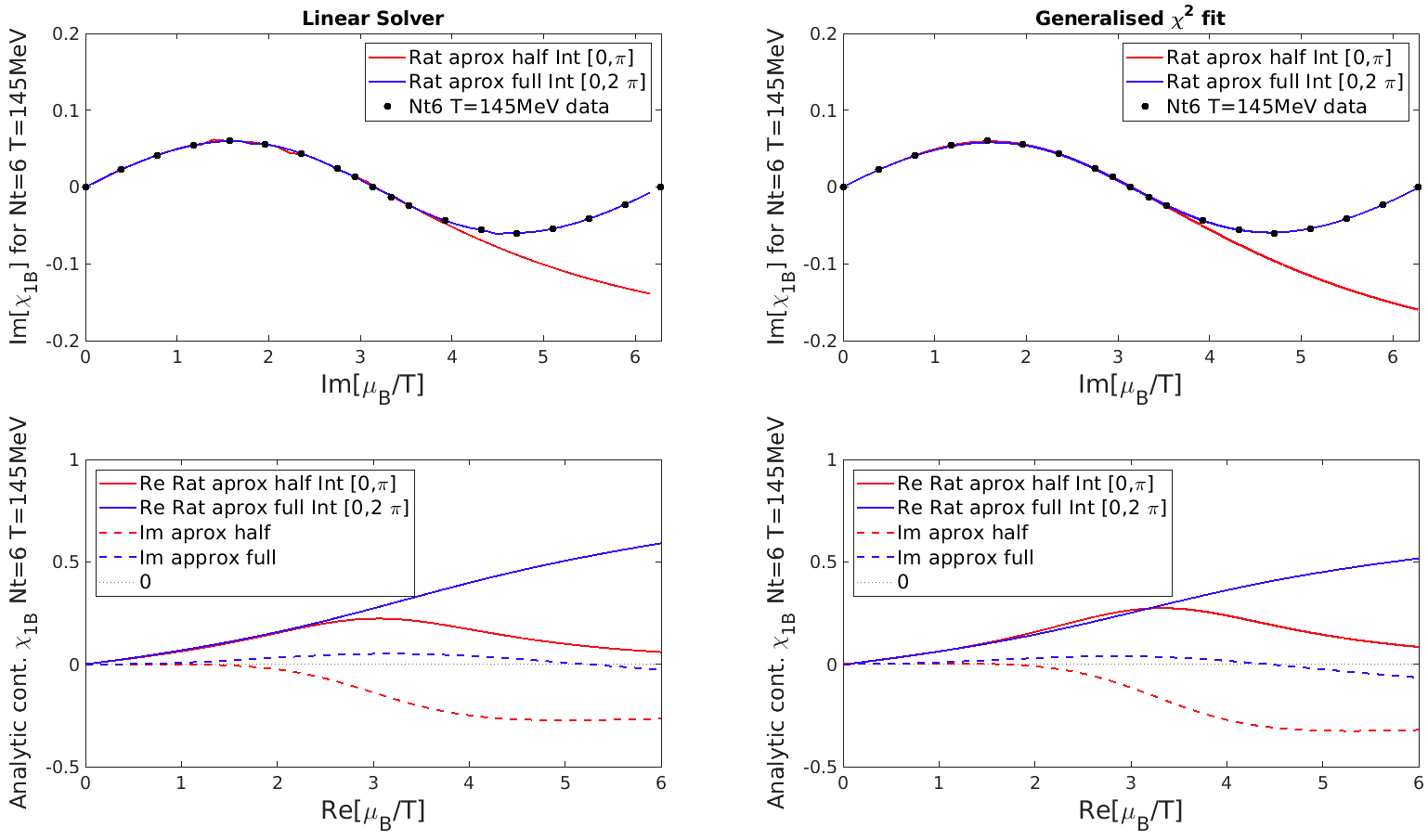}
    \caption{Top: the net baryon number density as a function of $\hat\mu_B^{I}$ on $N_{\tau}=6$ at $T=145$ MeV and the rational approximants obtained by both solving the linear system (left) and minimizing the generalized $\chi^2$ (right); in both approaches, we performed the analysis on both the restricted interval $\hat\mu_B^{I} \in [0 , \pi]$ and the extended one $\hat\mu_B^{I} \in [0 , 2 \pi]$. Bottom: the analytic continuation of the baryon number density for the four different options.}
    \label{fig:NT6}
\end{figure*}
In Fig.~\textcolor{red}{\ref{fig:NT6}}, we display the net baryon number density as obtained on $N_{\tau}=6$ at $T=145$ MeV (notice that the signal is substantially tinier than at higher temperatures). Here we have $2 \times 2$ options displayed: data are compared to rational approximants obtained (a) either from the generalized $\chi^2$ (\ref{eq:chi2}) or from the solution of Eq.~(\ref{eq:LinearProblem3}) for the basic functional form~(\ref{eq:BasicPade}) and (b) taking into account data either for $\hat\mu_B^{I} \in [0 , \pi]$ or for $\hat\mu_B^{I} \in [0 , 2 \pi]$. As one can see, the rational approximants we get from the two methods are (always) substantially equivalent; indeed, there is a difference when it comes to taking into account a larger or smaller $\hat\mu_B^{I}$ interval. Notice, however, that for $\hat\mu_B^{I} \in [0 , \pi]$ every solution is {\em de facto} indistinguishable from any other. We also show analytical continuations; taking into account data from an extended $\hat\mu_B^{I}$ interval results in an imaginary part staying very close to zero in a wider interval of real chemical potential. In Fig.~\textcolor{red}{\ref{fig:NT6_2}}, we display the analytic structure we get from the different $2 \times 2$ options. As one can see, results are again very much consistent whatever method we choose for computing the rational approximants. As anticipated, the singularities we found are indeed different for different input $\hat\mu_B^{I}$ interval. When we take input from the extended $\hat\mu_B^{I} \in [0 , 2 \pi]$ interval, we apparently get what one would regard as a thermal singularity. The outcome is pretty different when input is taken from the restricted $\hat\mu_B^{I} \in [0 , \pi]$ interval.
We see that what we get in this case is a chiral singularity candidate. We stress that this ambiguity shows up only in this case (i.e., for this lowest temperature, which is only probed on $N_{\tau}=6$). We stress once again that our multipoint Pad\'e analysis is interval sensitive and it could well be that in different intervals we have access to different pieces of information. The fact that this is possibly the only piece of information related to chiral symmetry breaking makes all this intriguing and  definitely deserving further investigation. Indeed we are working on this, in particular, repeating our Pad\'e analysis for the chiral condensate, which is the relevant order parameter for chiral symmetry breaking.
\begin{figure}[bt]
    \hskip-0.5cm
    \includegraphics[width=0.51\textwidth]{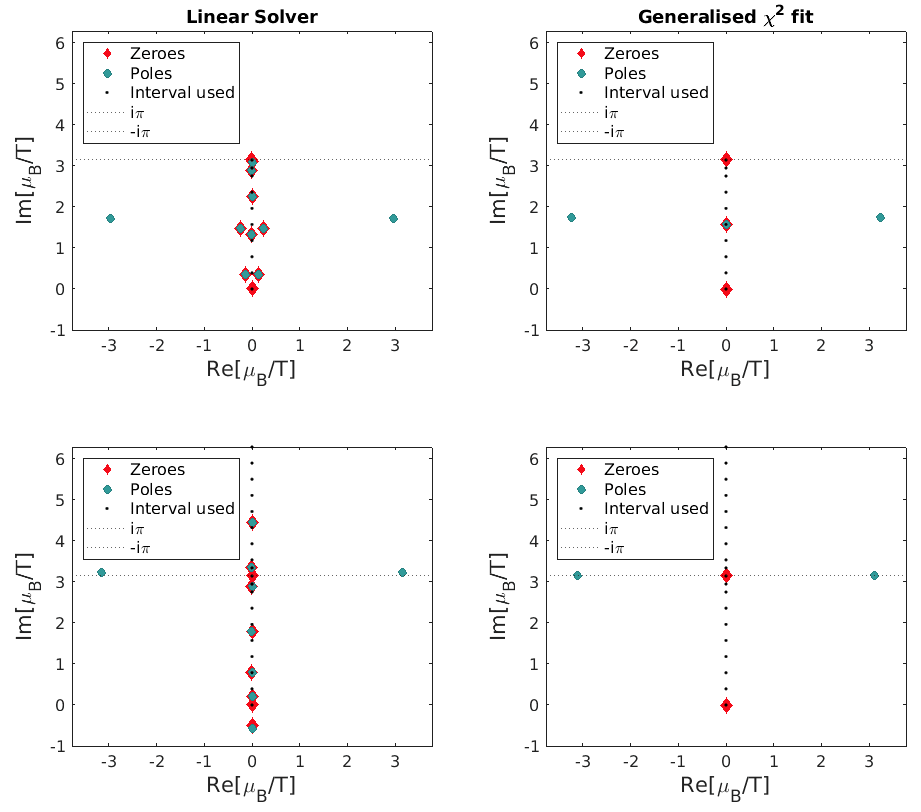}
    \caption{Singularity structure in the $\hat\mu_B$ plane on $N_{\tau}=6$ at $T=145$ MeV. As in Fig.~\ref{fig:NT6}, the method for obtaining the rational approximants can be the solution of the linear system (left) or the minimization of the generalized-$\chi^2$ (right); the input interval for the analysis can be $\hat\mu_B^{I} \in [0 , \pi]$ (top) or $\hat\mu_B^{I} \in [0 , 2 \pi]$ (bottom).}
    \label{fig:NT6_2}
\end{figure}
\section{Scaling Analysis \label{sec:results}}
\subsection{The Roberge-Weiss critical region\label{sec:resultsRW}}
The nearest singularities which we identified in the temperature range $201$~MeV~$<T<160$~MeV from our Padè approximations presented in the last section and which appeared to be stable are listed in Table~\ref{tab:nt4LYE}. 
Despite the fact that we could not observe indications for a branch cut connected to all these singularities, we now demonstrate that they can indeed be identified with Lee-Yang edge singularities of the Roberge-Weiss critical point, i.e., that they scale in accordance with our expectations presented in Sec.~\ref{sec:scalingRW}. 
In particular, it is obvious that we obtain for the imaginary part $\hat\mu_{LY}^I = \text{Im}[\frac{\mu_B}{T}] =\pi$ within errors for all temperatures and methods as demanded by Eq.~(\ref{eq:RW_scaling_im}). 
In order to show that the real part scales in accordance with Eq.~(\ref{eq:RW_scaling_re}) we perform fits to the data listed in Table~\ref{tab:nt4LYE} with the \textit{Ansatz} 
\begin{equation}
    \hat\mu_{LY}^R = a 
    \left( \frac{ T_{RW}-T }{ T_{RW} } \right)^{\beta\delta}+b\,,
    \label{eq:RW_fit}
\end{equation}
with fit-parameter $a,b$. 
For the Roberge-Weiss critical temperature, we set $T_{RW}=201.4$~MeV in accordance with \cite{Goswami:2018qhc}. 
We fixed the critical exponents to that of the Ising universality class; i.e., we have $\beta\delta\approx 1.5635$.
The parameter $b$ is added to capture the leading order finite size effects. 
Since our calculations are done in a finite volume, we expect that the Lee-Yang edge singularities will not reach the real $h$ axis, which is here the $\hat\mu_B^I$ axis. 
Or with other words, there is no phase transition in a finite volume. A more elaborate finite size analysis will be left for future publications. 

\begin{figure}[htp]
    \centering
    \includegraphics[width=0.45\textwidth]{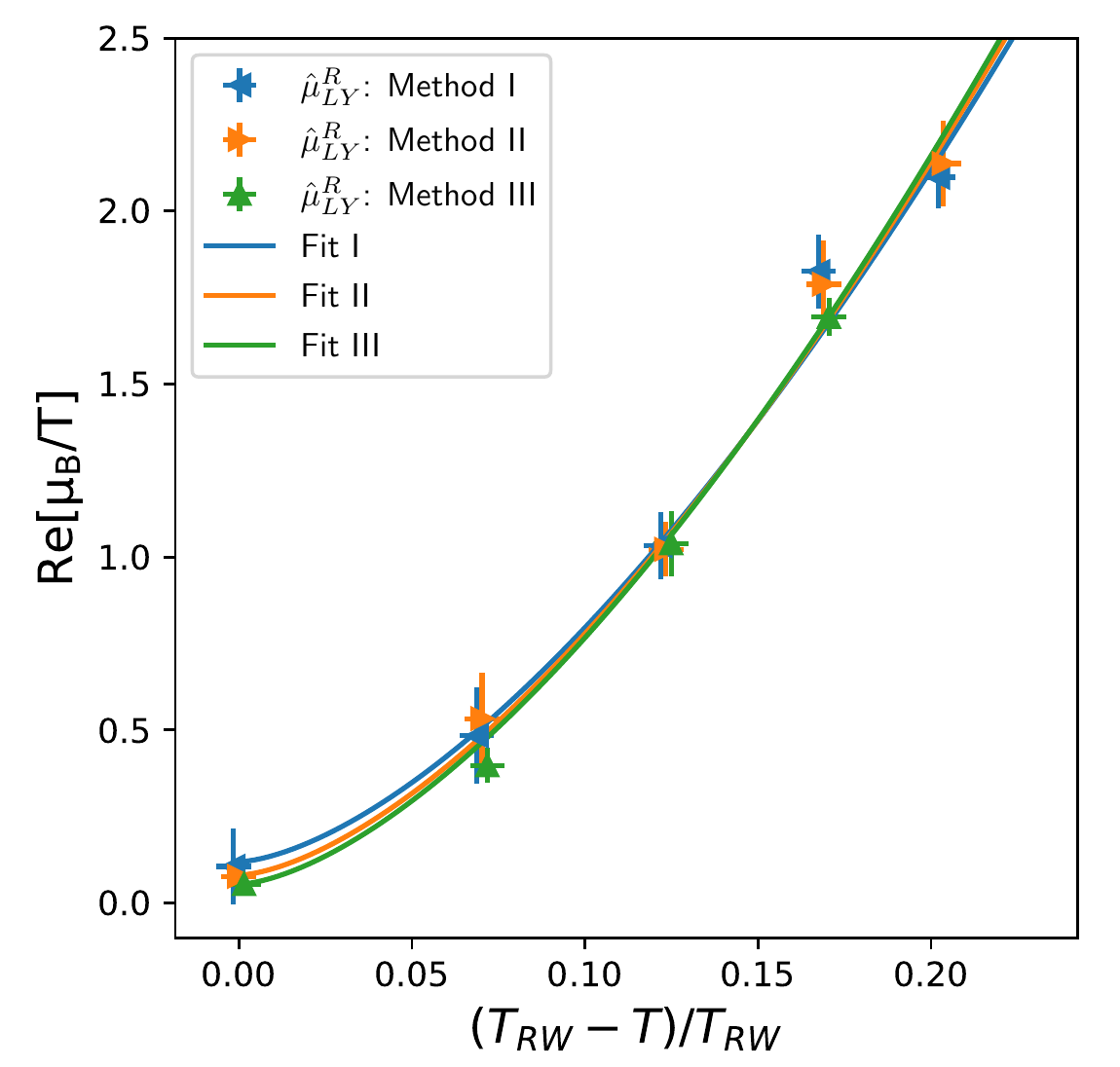}
    \caption{Scaling fit to the Lee-Yang edge singularities in the vicinity of the Roberge-Weiss transition to the \textit{Ansatz}~(\ref{eq:RW_fit}). Shown are three distinct data sets for the real parts of the $\hat\mu_B$ (imaginary parts of $h$) as a function of the reduced temperature $(T_{RW}-T)/T_{RW}$, as obtained from methods I-III.}
    \label{fig:RW_fit}
\end{figure}

The fits work quite well \footnote{The rooting procedure applied to staggered fermions is discussed controversially. This is in particular true for calculations that involve the evaluation of eigenvalues of the (reduced) fermion matrix at nonzero baryon number density \cite{Golterman:2006rw}. It has been pointed out that genuine Lee-Yang zeros that are obtained on the basis of these eigenvalues are plagued by phase ambiguities \cite{Giordano:2019slo, Giordano:2019gev}. To what extent this issue affects also the Taylor expansion method or calculations at imaginary chemical potentials in unclear to us. We take the fact that we find the expected critical scaling of the Lee-Yang edge singularities as a hint that our approach is not or not significantly affected by those artefacts.} and are shown in Fig.~\ref{fig:RW_fit}. Results for the fit parameter and reduced $\chi^2$ values are given in Table~\ref{tab:fit_param}. 

\begin{table}[htb]
    \centering
    \begin{tabular}{|c|c|c|c|c|} \hline
         Method & $a$         & $b$       & $\chi^2$ & $z_0$    \\ \hline
         { I           }& 
         { 24.77 (2.68)}& 
         { 0.1192(80)  }& 
         { 1.14        }& 
         { 9.18(99)    } \\
         { II          }& 
         { 25.54 (79)  }& 
         { 0.0806(9)   }& 
         { 0.49        }& 
         { 9.37(29)    } \\
         III   & 26.08 (63)  & 0.0541(1) & 0.96     & 9.49(23) \\ \hline
    \end{tabular} 
    \caption{Fit parameter $a,b$, obtained from a scaling fit to the Lee-Yang edge singularities in the vicinity of the Roberge-Weiss transition. Also given are the reduced $\chi^2$ and the deduced values for the nonuniversal constant $z_0$ for the data sets obtained from methods I-III, respectively.}
    \label{tab:fit_param}
\end{table}

Besides the demonstration for scaling, we can relate our results for the fit parameter $a$ to the nonuniversal constant $z_0$. From Eq.~(\ref{eq:RW_scaling_re}), we obtain
\begin{equation}
    z_0=|z_c|\left(\frac{a}{\pi}\right)^{\frac{1}{\beta\delta}}\,.
\end{equation}
Using the value $|z_c|=2.452$ for the 3d-Ising universality class \cite{Connelly:2020gwa}, we obtain $z_0\approx 9.2$ -- $9.5$. 
The specific values for our three data sets are given in Table~\ref{tab:fit_param}. 
To our knowledge, that is the first determination of $z_0$ for the Roberge-Weiss transition. 
Note, however, that the value is obtained on course lattices ($N_\tau=4$) with no proper continuum extrapolation yet. 

In principle, $z_0$ could also be determined from a fit to the magnetic equation of state (EoS). 
Here we anticipate more severe corrections from the finite size, as well as large contributions from regular terms. In particular, more data close to the RW transition are needed to obtain a reliable fit. 
A determination of $z_0$ from a fit to the magnetic EoS is thus beyond the scope of this work.  

\subsection{The chiral critical region \label{sec:scalingChiral}}
We have probed one additional temperature below the pseudocritical phase transition temperature of $(2+1)$-flavor QCD, namely, $T=145$~MeV. 
For this low temperature, the calculations have been done on $36^3\times 6$ lattices. 
To put the temperature value into perspective, we recall the continuum extrapolated numbers for the pseudocritical temperature $T_{pc}=(156.5 \pm 1.5)$~MeV \cite{Bazavov:2018mes} and the chiral critical temperature $T_c=132^{+3}_{-6}$~MeV \cite{Ding:2019prx}. 
We also note that the corresponding $N_\tau=6$ results are 10--15 MeV higher. 
In conclusion, the probed temperature of $T=145$~MeV is compatible with the chiral critical temperature, and we thus expected it to be sensitive to chiral scaling. 

\begin{figure}
    \centering
    \includegraphics[width=0.49\textwidth]{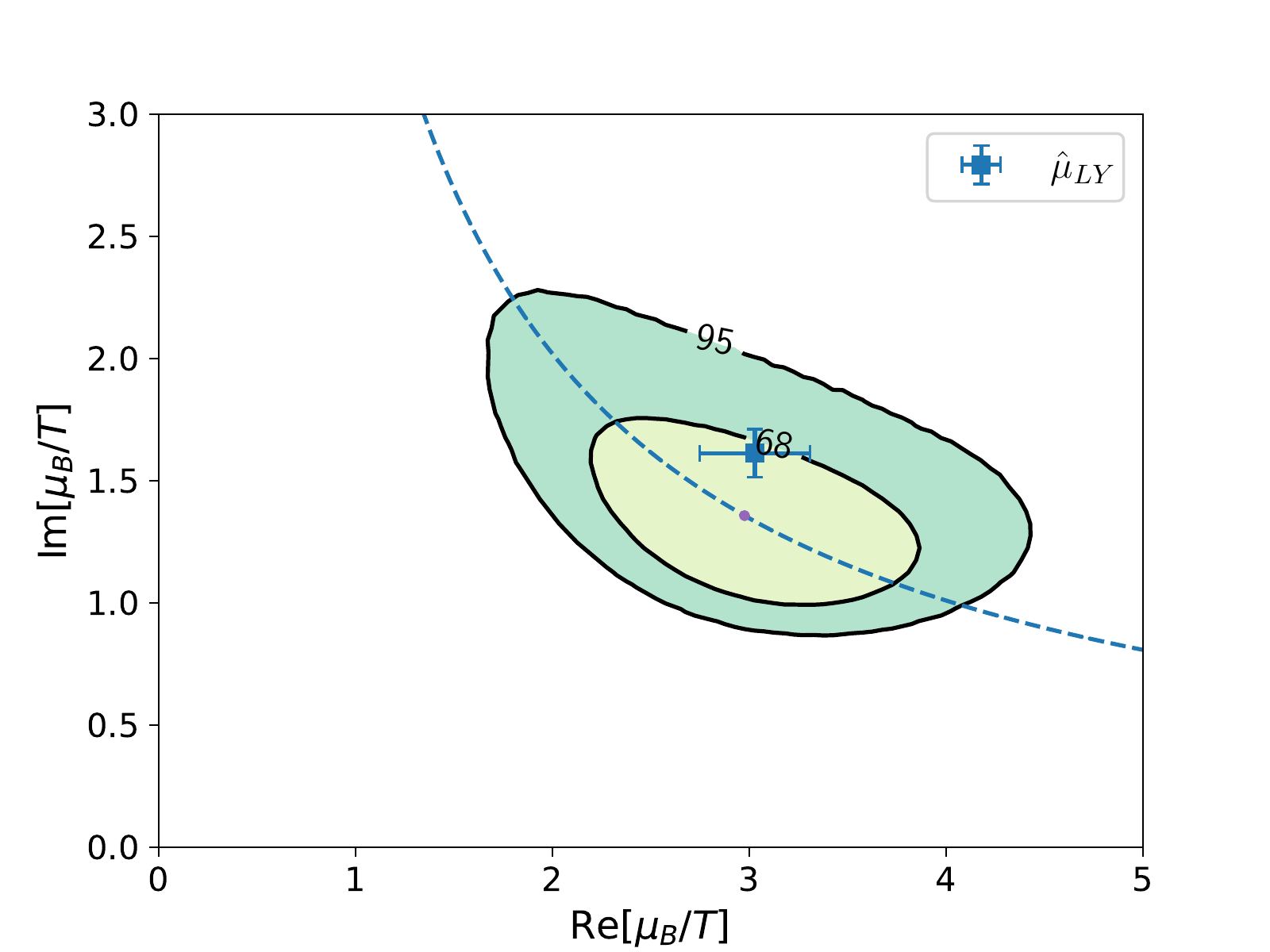}
    \caption{Comparison of the expected Lee-Yang edge singularity at $T=145$~MeV ($N_\tau=6$) from previously estimated nonuniversal parameters (68\% and 95\% confidence areas), with the singularity obtained from our multipoint Padé analysis (data point). The dashed line indicates the predicted temperature dependence of the Lee-Yang edge singularity.}
    \label{fig:chiral_sing}
\end{figure}

We now compare the position of the singularity we find for this temperature with the expected position of the Lee-Yang edge singularity, governed by $O(2)$ critical behavior\footnote{We chose here the $O(2)$-, rather than the $O(4)$-universality class, as staggered fermions break chiral symmetry in a way that only a $O(2)$-symmetric subgroup is preserved. Full $O(4)$-symmetry is expected to be restored in the continuum limit of the massless theory.}. Hence, we fix the critical exponents to $\beta\delta=1.6682$.
The chiral transition has been subject to various lattice QCD studies in the past, the nonuniversal parameters that appear in Eq.~(\ref{eq:chiral_zline}) are known to some extent, as discussed already in Sec.~\ref{sec:chiral_sing}. 
In Fig.~\ref{fig:chiral_sing}, we calculate the 68\% and 95\% confidence areas of the expected Lee-Yang edge singularity when we vary the nonuniversal parameter under the assumption of Gaussian distributed errors. 
In particular, we chose for the $N_\tau=6$ specific values and errors~\footnote{The values are based on scaling studies of the HotQCD collaboration.}, 
\begin{eqnarray}
T_c        &=& (147 \pm 6)\;\text{MeV}\,, \nonumber \\
z_0        &=& 2.35 \pm 0.2\,, \nonumber \\
\kappa_2^B &=& 0.012\pm 0.002 \,,
\end{eqnarray}
and in addition, we take $|z_c|=2.032$ \cite{Connelly:2020gwa}. 
As can be seen from Fig.~\ref{fig:chiral_sing}, the results from the rational approximation to our data (method II), $(\hat\mu_B^R,\hat\mu_B^I)=(3.03(28),1.61(10))$, lie within the 68\% confidence area of this prediction.

\section{Summary and conclusions\label{sec:conclusions}}
We computed cumulants of the net baryon number density as a function of the imaginary baryon number chemical potential in lattice QCD, the fermionic regularization being that of highly improved staggered quarks (HISQ). The results were the input for a multipoint Pad\'e analysis by which rational approximations were calculated, with various choices of both the functional forms of the latter and of the methods by which the approximants were determined. The results have been shown to be stable, in particular, also if we repeat our analysis in the fugacity plane (i.e., after a conformal map). Our rational approximations not only describe very well the data but appear to be quite well under control when we analytically continue them to real values of the baryonic chemical potential. \\
The main focus of our analysis has been on the singularity structure that we can infer from the complex poles of our rational approximations. By comparing the latter with the theoretically expected Lee-Yang edge singularities in the vicinity of the Roberge-Weiss phase transition, we found a quite good agreement. In particular, the temperature scaling of the singularities is consistent with the expected power law behavior. We also found a preliminary evidence of a singular point consistent with the phase transition which is expected in the chiral limit of (2+1)-flavor QCD in a staggered regularization. All our findings (and, in particular, the last that we mentioned) deserve further investigation by getting more precise measurements and probing less coarse lattices, a target that we are aiming at in the near future. An interesting task that is also in front of us is that of comparing our methodology and results with other recent approaches to the study of the analytical structure of finite density QCD, e.g. that of \cite{Mondal:2021jxk, Mukherjee:2021tyg}. All data from our calculations, presented in the figures of this papers can be found in \cite{datapub}.

\section*{Acknowledgements}
Our work is dedicated to our late colleague and friend E.~Laermann. J.G. and C.S. would like to thank F.~Karsch for stimulating discussions. C.S. would also like to thank S.~Mukerherjee and V.~Skokov for discussions and A.~Lahiri for providing data on the cutoff dependence of $z_0,T_c$.
This work was supported by (i) the European Union’s Horizon 2020 research and innovation program under the Marie Skłodowska-Curie Grant Agreement No. H2020-MSCAITN-2018-813942 (EuroPLEx), (ii) The Deutsche Forschungsgemeinschaft (DFG, German Research Foundation), Project Number 315477589-TRR 211, and (iii) I.N.F.N. under the research project {\em i.s.} QCDLAT.
This research used computing resources made available through (i) the Gauss Centre for Supercomputing on the Juwels GPU nodes at the  Jülich Supercomputing Centre, (ii) Bielefeld University on the Bielefeld GPU-Cluster, (iii) CINECA on Marconi100 under both the I.N.F.N.-CINECA agreement and the ISCRA C program (HP10CWD9YA project), and (v) the University of Parma on the UNIPR HPC facility.

\appendix
\section{LATTICE DATA OF NET BARYON NUMBER CUMULANTS\label{appendix_Tables}}
The gauge fields have been generated with a rational hybrid Monte Carlo algorithm (RHMC). In Tables~\ref{tab:data1} and~\ref{tab:data2} we list results from calculation on the $24^3\times4$ and $36^3\times 6$ lattices, respectively. Also listed are the number of configurations on which we have measured the observables and which are separated by 10 RHMC trajectories of length $0.5$-$1.0$. 
\begin{table*}[ht]
    \centering
    \begin{tabular}{|c|c|c|c|c||c|c|c|c|c|} 
    \hline
    $\hmuBI$ & $\text{Im}\left[\chi_1^B\right]$ & $\text{Re}\left[\chi_2^B\right]$ & $\text{Im}\left[\chi_3^B\right]$ & \#conf. & 
    $\hmuBI$ & $\text{Im}\left[\chi_1^B\right]$ & $\text{Re}\left[\chi_2^B\right]$ & $\text{Im}\left[\chi_3^B\right]$ & \#conf. \\ \hline
\multicolumn{5}{|c||}{$T=201.4$ [MeV]}  & \multicolumn{5}{|c|}{ $T=176.6$ [MeV]} \\ \hline
 0.000 & -0.00002(18) & 0.26421(52) & 0.0009(21) &   4800 & 0.000 & 0.00062(28) & 0.2288(10) & -0.0018(43) & 1600 \\
0.393 & 0.10319(17) & 0.26066(46) & 0.0155(27) &   4800 & 0.209 & 0.04840(25) & 0.22800(83) & 0.01876(43) & 1600 \\
 0.785 & 0.20388(21) & 0.25134(72) & 0.0367(30) &   4800 & 0.419 & 0.09556(42) & 0.22487(11) & 0.0279(52) & 1600 \\
 1.178 & 0.29940(27) & 0.2344(11) & 0.0478(47) &   4800 & 0.628 & 0.14253(36) & 0.2178(10) & 0.0305(74) & 1600 \\
 1.571 & 0.38637(20) & 0.2107(11) & 0.0762(68) &   4800 & 0.838 & 0.18703(38) & 0.2084(15) & 0.0420(93) & 1600 \\
1.963 & 0.46004(44) & 0.1675(22) & 0.132(15) &   5400 & 1.047 & 0.22836(50) & 0.1951(18) & 0.0706(87) & 1600 \\
2.356 & 0.51602(51) & 0.1049(28) & 0.224(22) &   5400 & 1.257 & 0.26744(63) & 0.1761(23) & 0.092(14) & 1600 \\
2.749 & 0.53076(91) & -0.0589(73) & 0.72(13) &   5400 & 1.466 & 030325(65) & 0.1599(29) & 0096(20) & 1600 \\

2.880 & 0.52279(99) & -0.1291(79) & 1.00(11) &  10800 & 1.676 & 0.33369(37) & 0.1253(17) & 0.201(16) &     1600 \\
3.011 & 0.4851(21) & -0.516(34) & 5.39(79) &  10800 &1.885 & 0.35465(57) & 0.0961(35) & 0.175(18) &     1600 \\
3.105 & 0.3859(78) & -2.92(54) & 141(48) &  11000 & 2.094 & 0.36401(57) & 0.0147(62) & 0.360(72) &     1600 \\
\cline{1-5}
\multicolumn{5}{|c||}{$T=186.3$ [MeV]} & 2.304 & 0.3594(11) & -0.086(13) & 0.73(12) &     1600 \\ \cline{1-5}
0.000 & 0.00025(12) & 0.24537(35) & 0.0021(20) &   4100 & 2.513 & 0.32881(51) & -0.235(18) & 1.06(42) &     1600 \\
0.286 & 0.06986(20) & 0.24361(51) & 0.0062(17) &   4100 & 2.723 & 0.2608(24) & -0.482(17) & 1.40(44) &     1600 \\
0.571 & 0.13876(20) & 0.23687(52) & 0.0127(23) &   4100 & 2.932 & 0.1454(32) & -0.567(29) & 0.31(55) &     1600 \\
0.857 & 0.20486(26) & 0.22681(75) & 0.0181(29) &   4100 & 3.142 & 0.0055(30) & -0.626(33) & -0.58(53) &     1600 \\ \cline{6-10}
1.142 & 0.26792(27) & 0.21248(91) & 0.0271(46) &   4100 & \multicolumn{5}{|c|}{$T=167.4$ [MeV]} \\ \cline{6-10}
1.428 & 0.32485(30) & 0.1850(16) & 0.0796(97) &   4100 & 0.000 & -0.00029(23) & 0.21093(75) & -0.0070(53) &   6000 \\
1.714 & 0.37469(41) & 0.1588(15) & 0.068(12) &   4100 & 0.393 & 0.08176(27) & 0.2059(13) & 0.0283(67) &   6000 \\
1.999 & 0.41448(59) & 0.1088(38) & 0.140(22) &   4100 & 0.785 & 0.15952(31) & 0.1860(12) & 0.0668(70) &   6000  \\
2.285 & 0.43543(90) & 0.0350(73) & 0.271(48) &   4100 & 1.178 & 0.22720(45) & 0.1551(21) & 0.1017(85) &   6000 \\
2.570 & 0.4213(15) & -0.145(11) & 0.88(15) &   4100 & 1.571 & 0.27782(81) & 0.0982(36) & 0.181(33) &   6000 \\
2.713 & 0.3918(24) & -0.349(25) & 2.21(49) &   4000 & 1.963 & 0.29885(84) & -0.0083(59) & 0.372(68) &   6000 \\
2.856 & 0.3262(34) & -0.757(61) & 5.3(1.2) &   4100 & 2.356 & 0.2630(16) & -0.179(15) & 0.57(12) &  12000 \\
2.999 & 0.2020(41) & -1.061(54) & 2.9(1.6) &   4000 & 2.749 & 0.1554(16) & -0.343(19) & 0.23(25) &  12000 \\
3.142 & -0.0069(61) & -1.40(12) & -0.0(3.2) &   4100 & 3.142 & -0.0015(16) & -0.421(13) & 0.07(27) &  12000 \\
\hline
 \multicolumn{5}{|c||}{$T=160.4$ [MeV]} & \multicolumn{5}{|c|}{$T=160.4$ [MeV]} \\ \cline{1-5}\cline{6-10}
0.000 & -0.00027(25) & 0.1919(12) & -0.0000(82) &   5550  & 1.571 & 0.23653(81) & 0.0584(60) & 0.190(39) &   5550 \\
0.393 & 0.07427(36) & 0.1865(14) & 0.0287(84) &   5550 & 1.963 & 0.2391(13) & -0.0476(83) & 0.262(89) &   5550 \\
0.785 & 0.14273(32) & 0.1637(18) & 0.071(14) &   5550  & 2.356 & 0.1990(16) & -0.1580(88) & 0.16(16) &   5550 \\
1.178 & 0.20104(40) & 0.1262(27) & 0.109(17) &   5550 & 2.749 & 0.1061(13) & -0.271(17) & 0.13(23) &   5550  \\ \hline
    \end{tabular}
    \caption{Mean values and statistical errors of net baryon number cumulants from $24^3\times 4$ lattices. Also indicated is the number of measured configurations.}
    \label{tab:data1}
\end{table*}

\begin{table*}[ht]
    \centering
    \begin{tabular}{|c|c|c|c|c||c|c|c|c|c|} 
    \hline
    $\hmuBI$ & $\text{Im}\left[\chi_1^B\right]$ & $\text{Re}\left[\chi_2^B\right]$ & $\text{Im}\left[\chi_3^B\right]$ & \#conf. & 
    $\hmuBI$ & $\text{Im}\left[\chi_1^B\right]$ & $\text{Re}\left[\chi_2^B\right]$ & $\text{Im}\left[\chi_3^B\right]$ & \#conf. \\ \hline
    \multicolumn{5}{|c||}{$T=145.1$ (MeV)} & \multicolumn{5}{|c|}{$T=145.1$ (MeV)} \\ \hline
    0.000 & 0.00024(51) & 0.0579(24) & 0.001(20) &   5280 & 1.963 & 0.05578(61) & -0.0217(39) & 0.074(36) &   5280 \\
    0.393 & 0.02276(42) & 0.0526(27) & 0.026(24) &   5280 & 2.356 & 0.04384(73) & -0.0467(49) & 0.099(46) &   5280 \\
    0.785 & 0.04142(56) & 0.0426(22) & 0.057(19) &   5280 & 2.749 & 0.02391(81) & -0.0569(44) & 0.024(37) &   5280 \\
    1.178 & 0.05436(54) & 0.0176(27) & 0.055(27) &   5280 & 2.945 & 0.01315(95) & -0.0663(47) & 0.008(46) &   5280 \\
    1.571 & 0.05995(76) & -0.0042(24) & 0.075(25) &   5280 & 3.142 & 0.00024(80) & -0.0538(41) & 0.071(39) &   5280 \\ \hline
    \end{tabular}
    \caption{Mean values and statistical errors of net baryon number cumulants from $36^3\times 6$ lattices. Also indicated is the number of measured configurations.}
    \label{tab:data2}
\end{table*}

\section{MULTIPOINT VS SINGLE-POINT PADÉ}
\label{sec:ApPADE}
\noindent As mentioned before, most of the literature that exists on existence, uniqueness and convergence \cite{baker1975,NUTTALL1970147} of Padé sequences exists mainly for single-point Padé expansions wherein the rational approximation is constructed from a single Taylor expansion with arbitrarily many Taylor coefficients \cite{Masjuan:2010ck}. 
On the other hand, we may be presented with a situation in which we have low-order Taylor data but at arbitrarily many points.
This is known in the literature as multipoint Padé -- but most commonly only values at other points are used. 
In our work we also use higher Taylor coefficients at other points. Since not a lot of literature on multipoint Padé exists, we will validate our findings with \emph{numerical experiments} conducted on a number of test functions.
Based on our numerical experiments it is also shown that there are situations in which a multipoint Padé does better while there are other situations in which a single point may be better. \\

\begin{figure}[!ht] 
        \centering
        \includegraphics[scale=0.49]{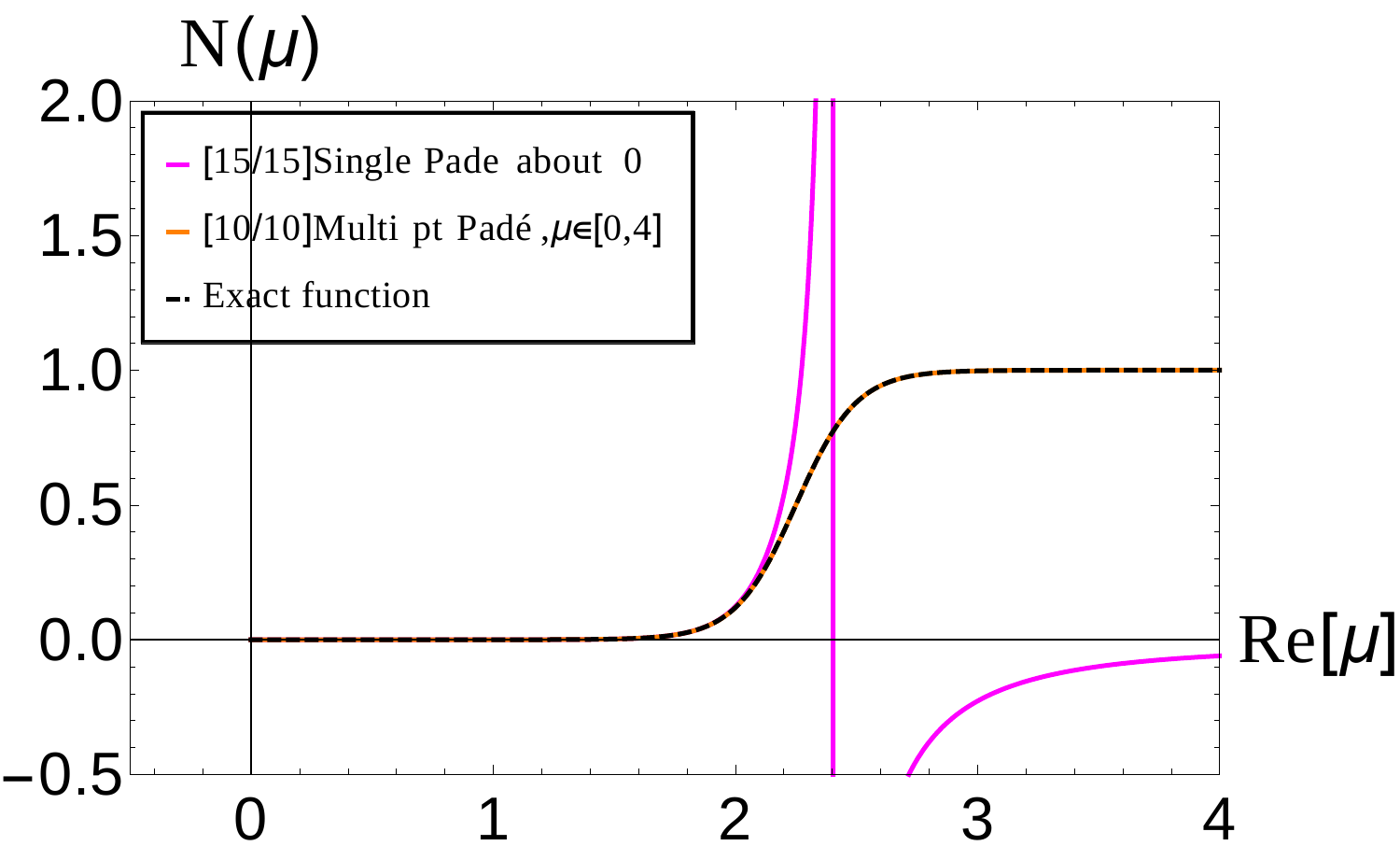}
        \includegraphics[scale=0.49]{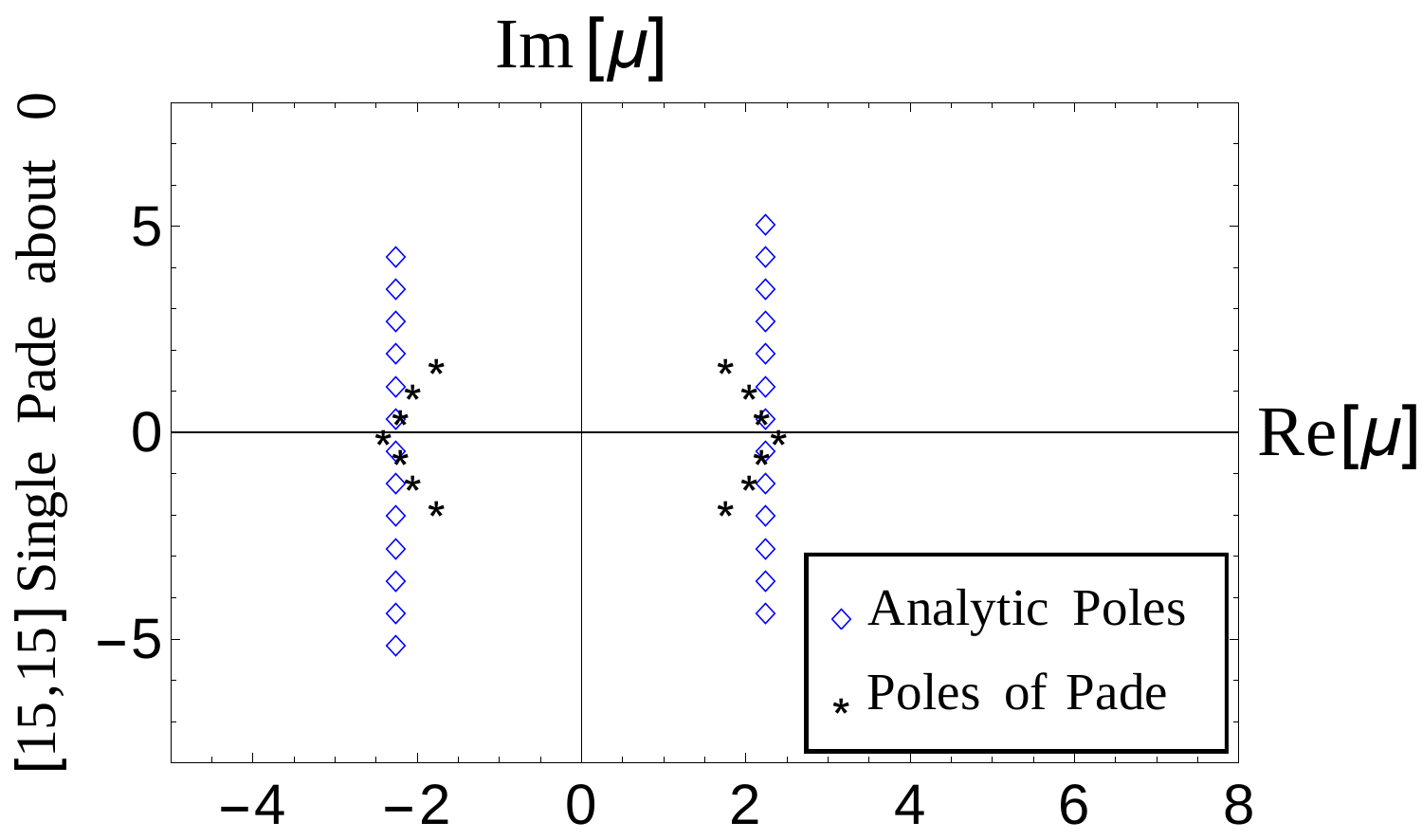}
        \includegraphics[scale=0.49]{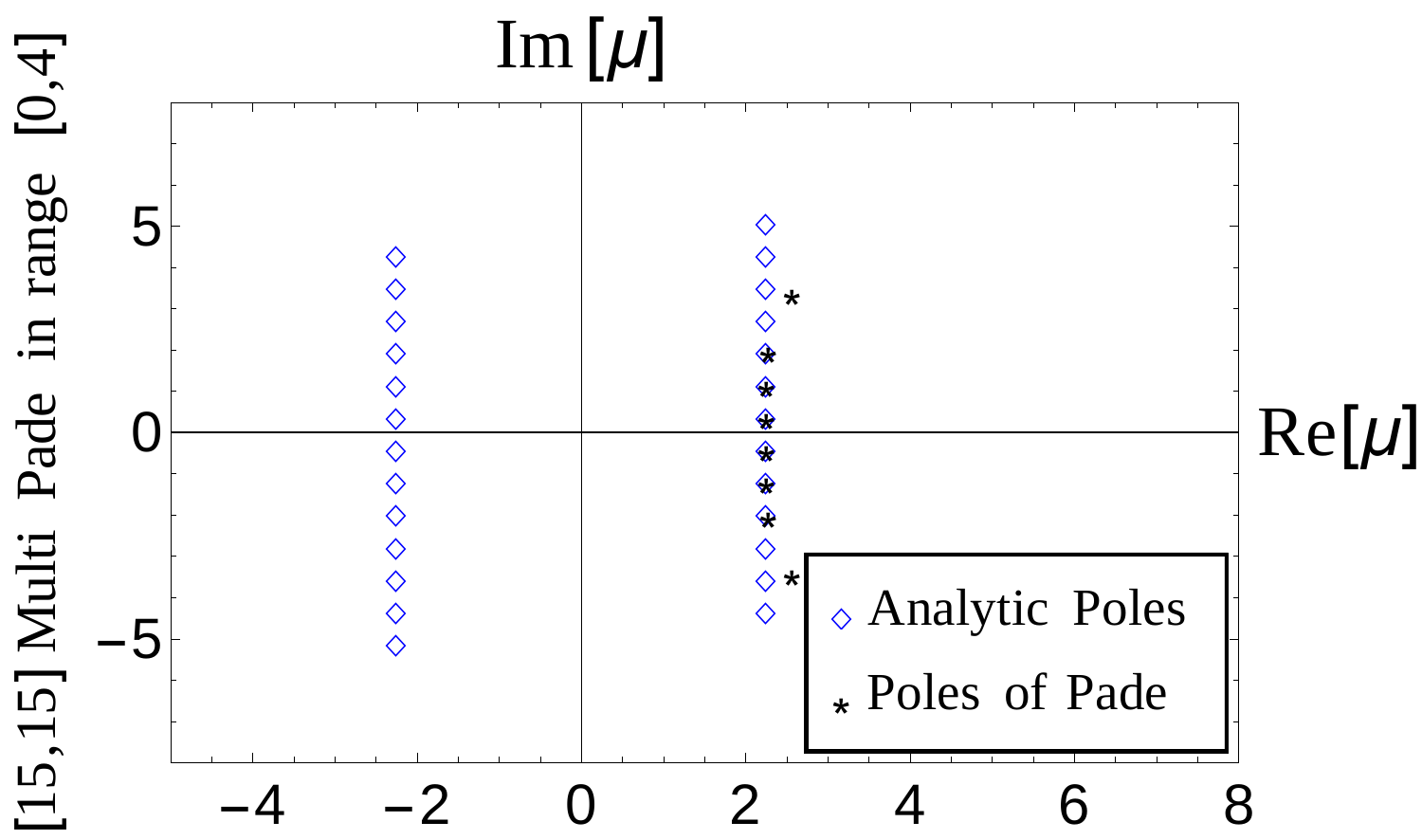}
         \caption{Thirring 1D : (Top) Comparison between the approximation of a [15/15] order single point Padé about 0 and a [10/10] order multipoint Padé constructed in the interval [0,4] with only up-to first derivatives. (Middle and Bottom) Depiction of the poles as seen by the single- and multipoint Padé respectively.}
         \label{fig:appendixA}
\end{figure}

\noindent Most of our numerical experiments are based on the 1D Thirring model (a model that was studied in \cite{DiRenzo:2020cgp}). This model is chosen because its partition function has a known analytical solution. For the purposes of these experiments we will simulate the number density of the 1D Thirring model:
\begin{equation}
N = \frac{I_{1}(\beta)^L \sinh(L \mu)}{I_{1}(\beta)^L \cosh(L\mu) + I_{0}(\beta)^L cosh(L sinh^{-1}(m))}   
\end{equation}
where because we know the exact location of poles of the number density - it is easy to validate/invalidate our approximation. Shown in Fig.~\ref{fig:appendixA} are the approximations and singularity structure of the number density simulated at $\beta=1$, $L=8$ and $m=2$ (same parameters used in all figures depicting the 1D Thirring model).

\section{MANIFESTATION OF SPURIOUS POLES}
\label{sec:SpuriousP}
\begin{figure}[!ht]
        \centering
        \includegraphics[scale=0.44]{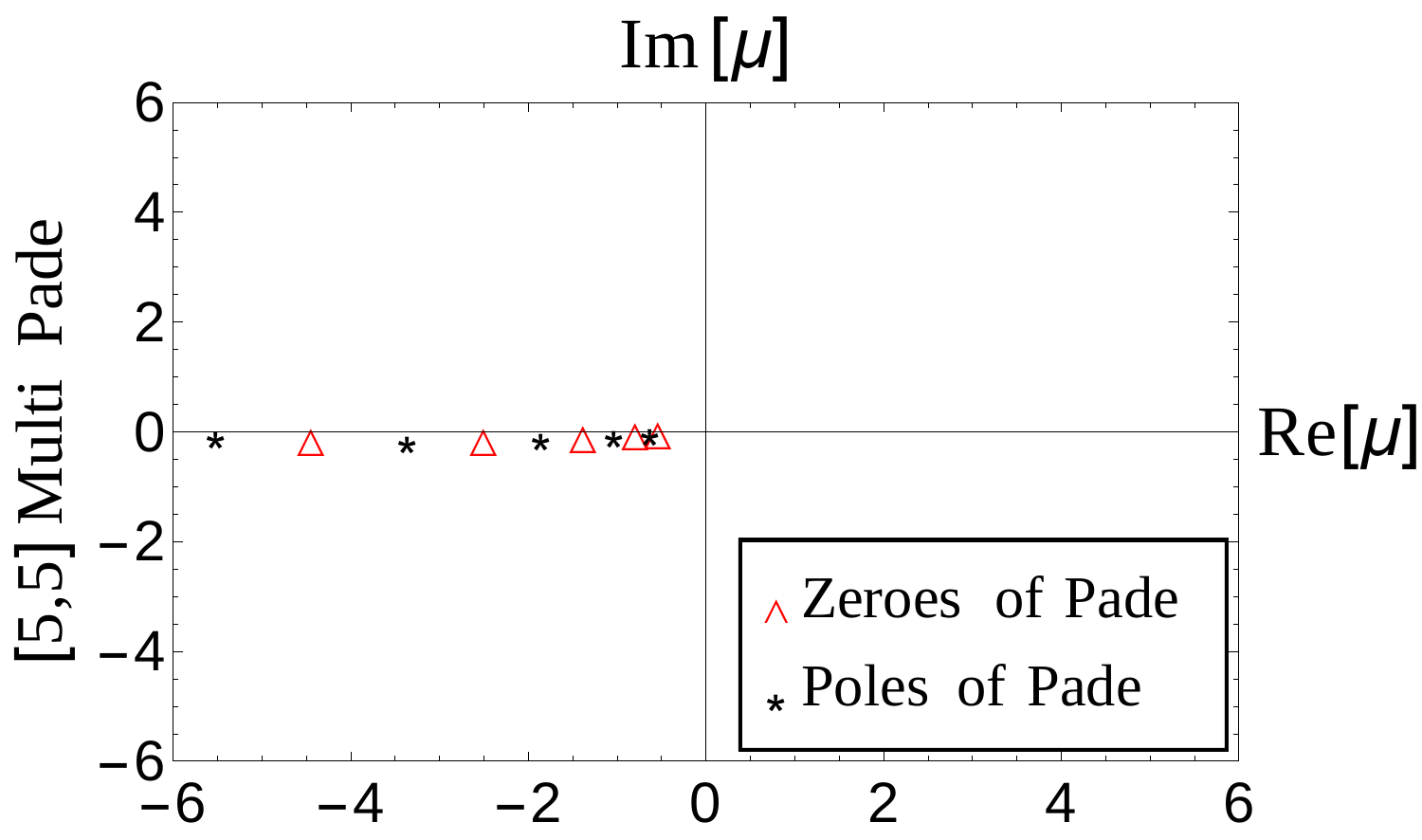}
        \includegraphics[scale=0.44]{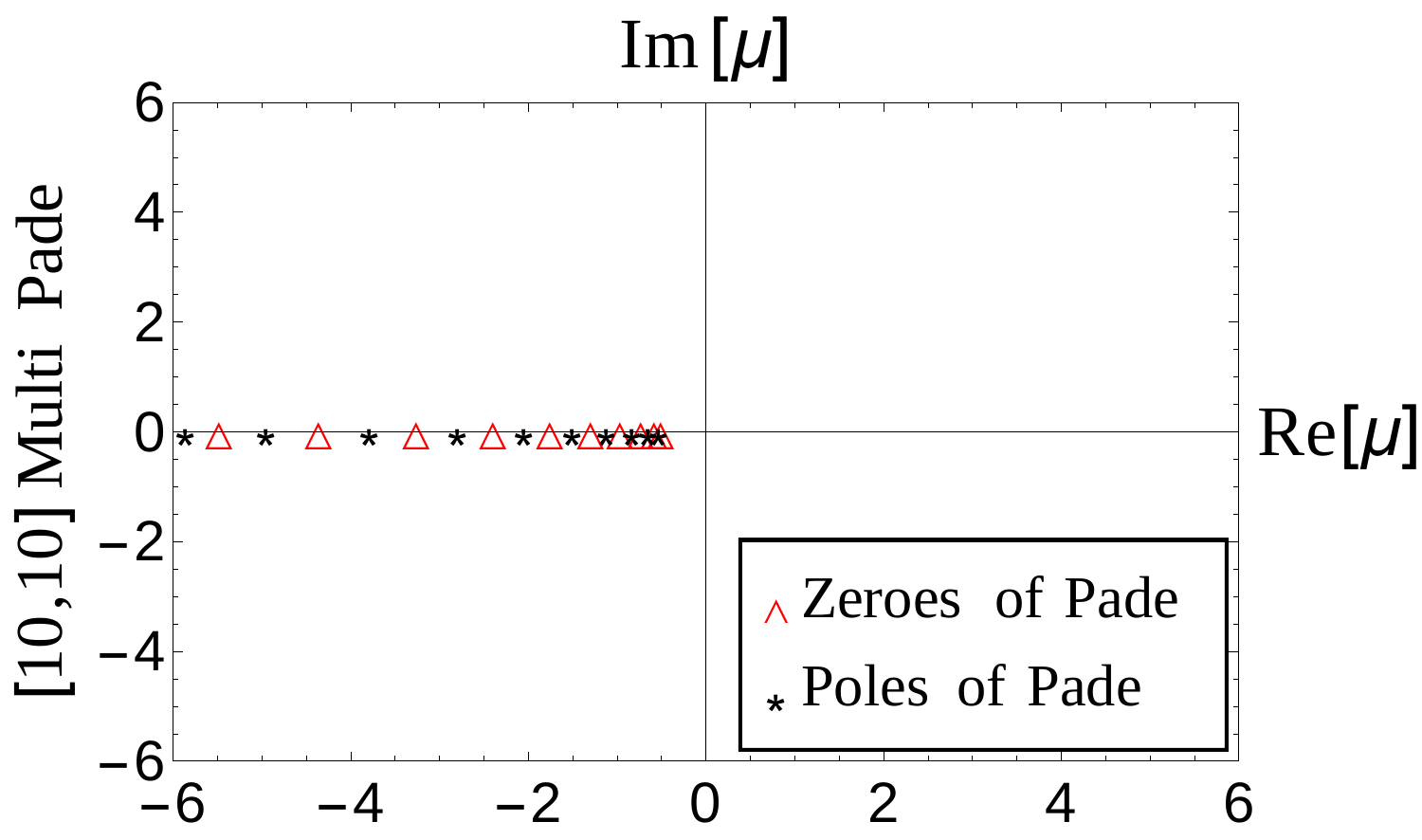}
        \includegraphics[scale=0.44]{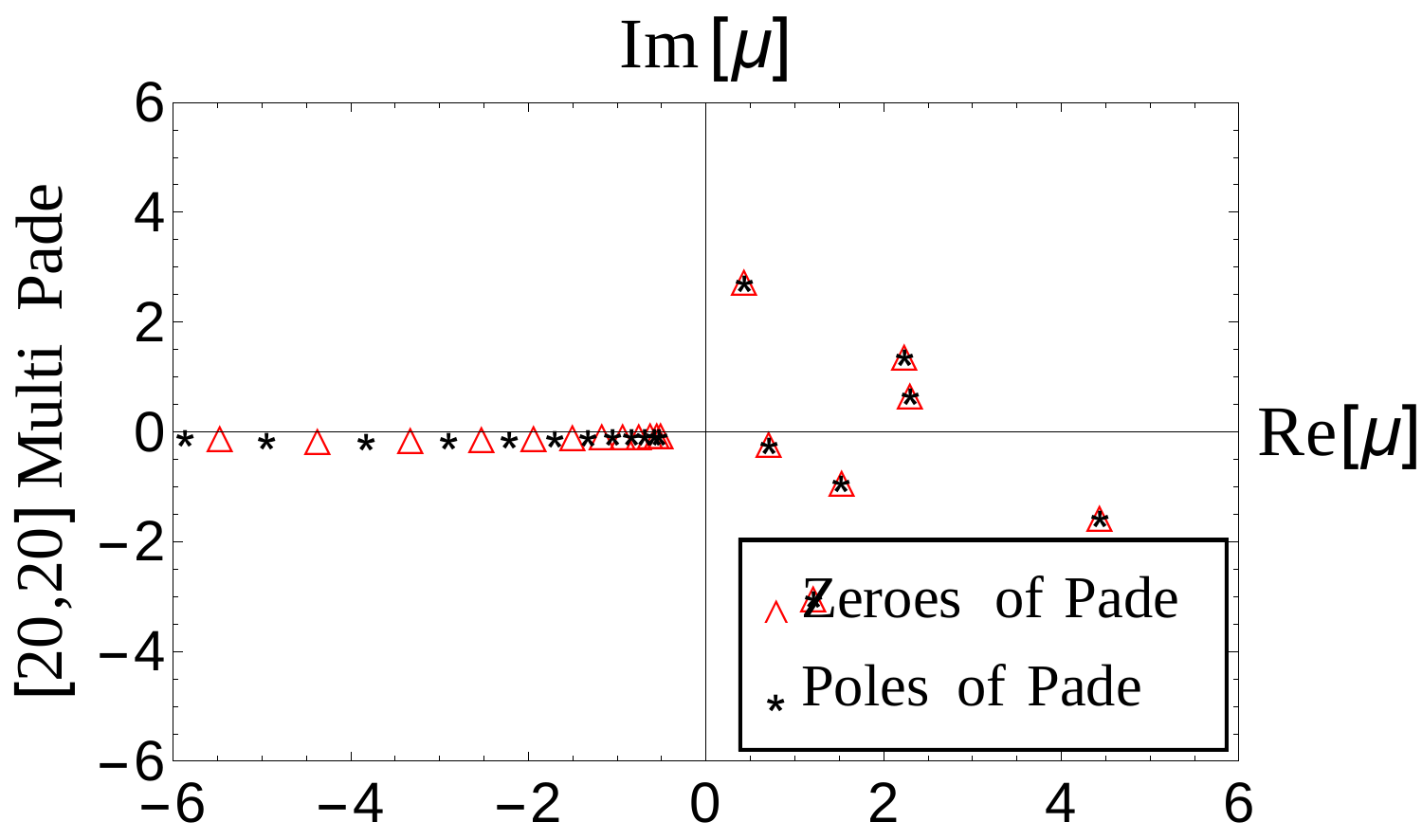}
         \caption{The figure depicts zeroes and poles of the function: $\sqrt{\frac{2\mu+1}{\mu+6}}$ with increasing the order of the Padé approximation. The message to be conveyed is the appearance of \enquote{spurious} poles and zeroes exactly canceling each other on right half plane when we go very high in the order.}
         \label{fig:appendixB_spur}
\end{figure}

\begin{figure}[htp] 
        \centering
        \includegraphics[scale=0.36]{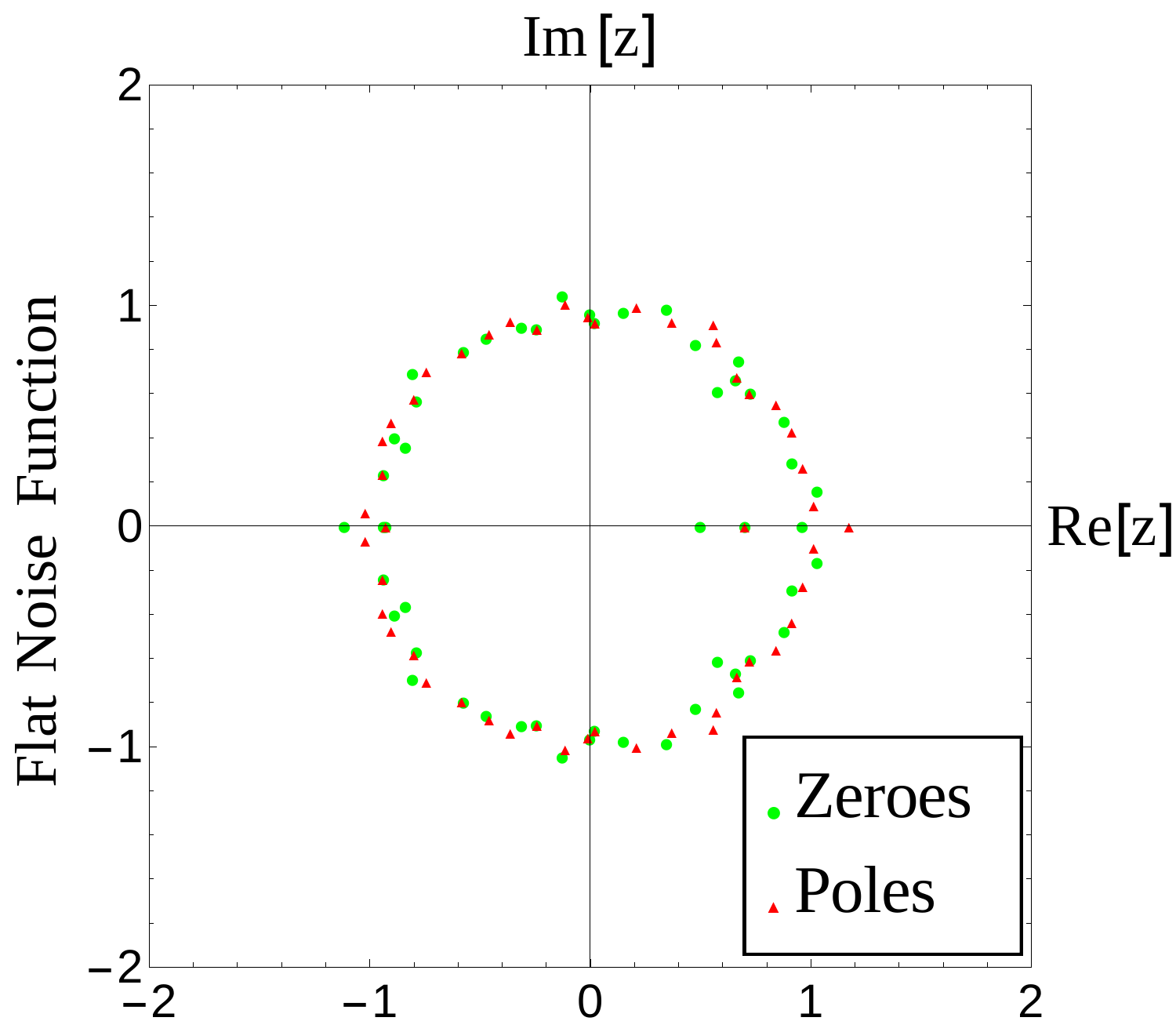}
        \hskip1cm
        \includegraphics[scale=0.36]{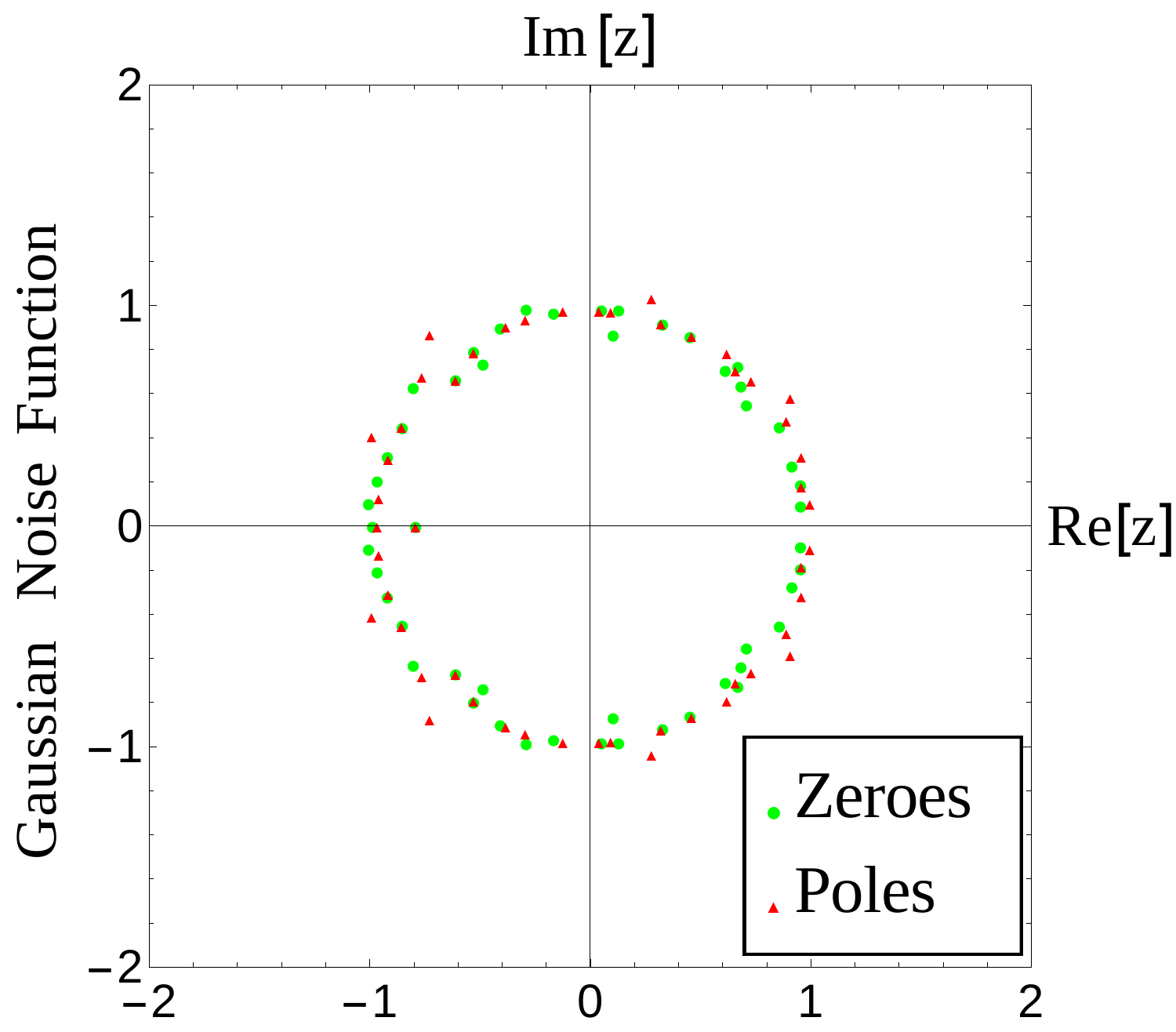}
         \caption{Froissart doublets for (Top) Flat noise and (Bottom) Gaussian noise.}
         \label{fig:appendixB_Froiss}
\end{figure}

As will be shown below, \enquote{spurious} poles can enter our analysis in two ways:
\begin{enumerate}
\item Firstly, we can (will) get spurious poles in noisy data - we will discuss this in the last section. But the important message is that if our function has a genuine pole, we will find a quasi-stable pole from our approximation even in noisy data and the effect of decreasing (increasing) noise will be that the pole becomes more (less) stable - eventually converging to (diverging from) the correct value in the absence of noise.
\item Secondly, even in the absence of noise - when we simulate a test function with its clean data - we will find that after a certain (optimal) order, our Padé will start spitting out spurious poles which will be exactly cancelled by corresponding zeroes. This is a clear result of demanding a very high order of approximation. This happens both for single point Padé and multipoint Padé. This effect can be seen in Fig.~\ref{fig:appendixB_spur}.
\end{enumerate}
A note on Froissart doublets \cite{GILEWICZ2003235}: these appear as zero-pole doublets in a unit circle in a Padé approximation to a series perturbed by noise. The separation between such pairs is proportional to the scale of the noise present. Fig. \ref{fig:appendixB_Froiss} is an example of Froissart doublets in case of simulating pure noise.
\section{INTERVAL DEPENDENCE OF MULTIPOINT PADÉ}
\label{sec:InteDep}
\begin{figure*}[h!]
        \centering
        \includegraphics[scale=0.34]{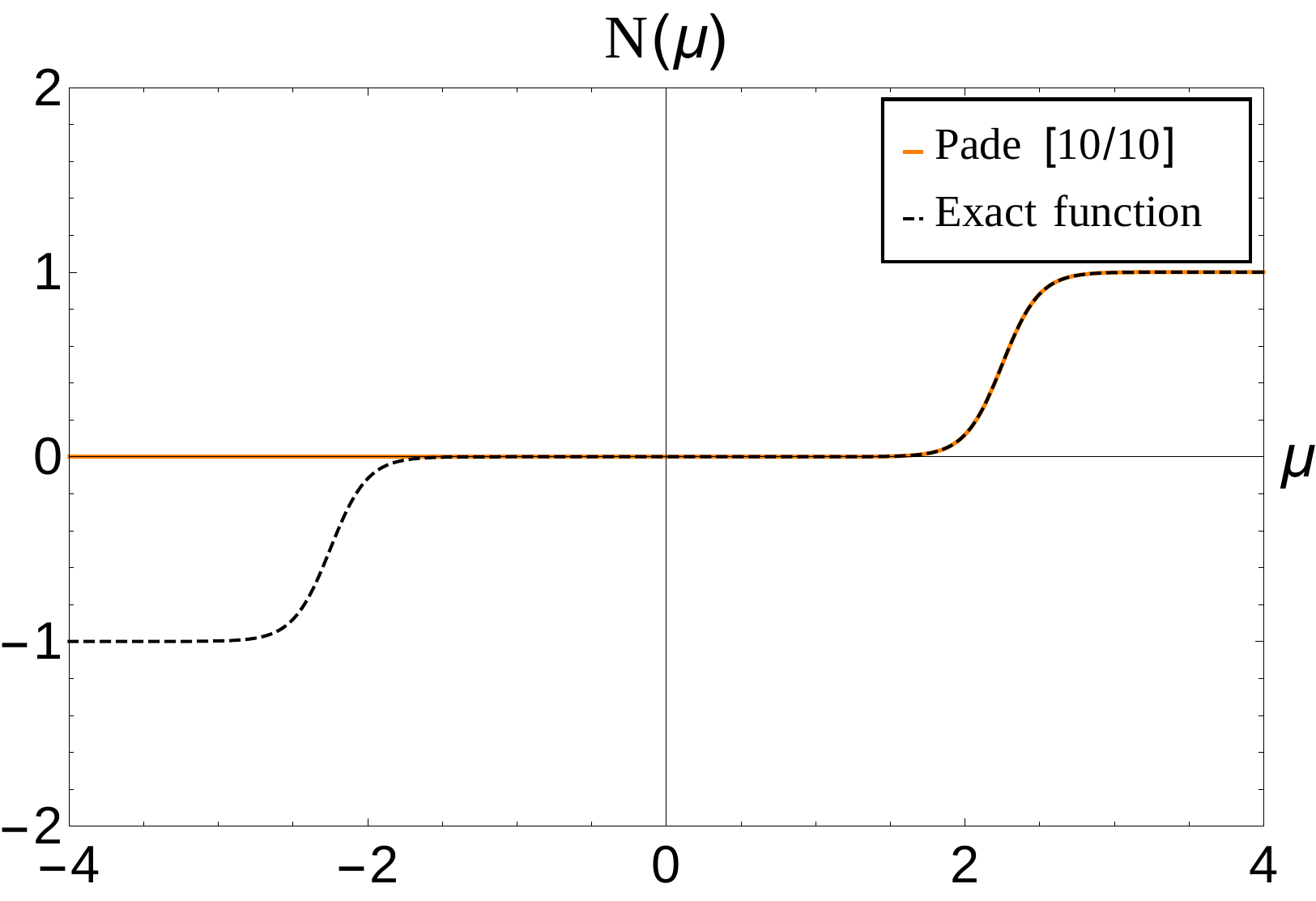}
        \hfill
        \includegraphics[scale=0.40]{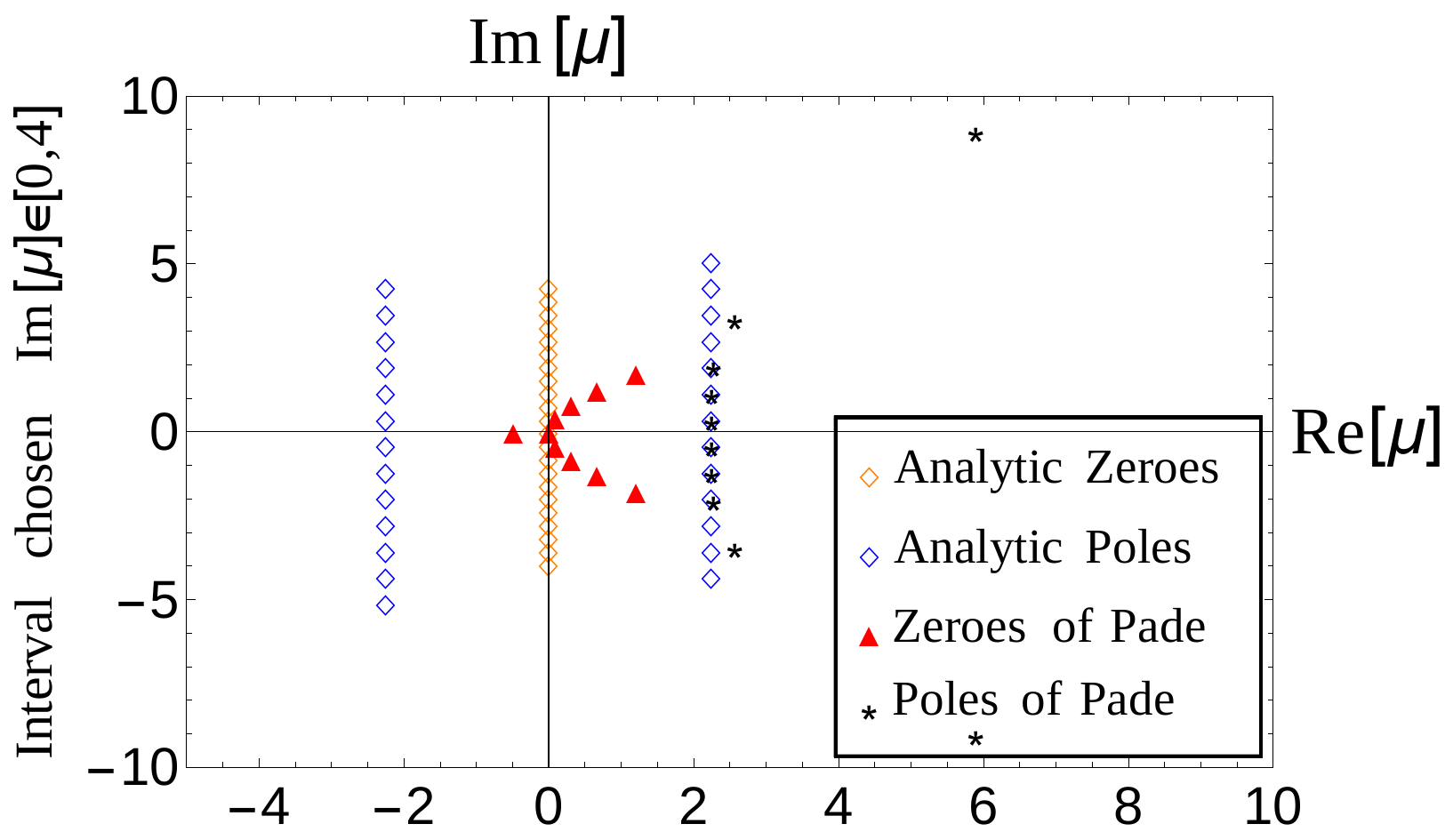}
        \includegraphics[scale=0.34]{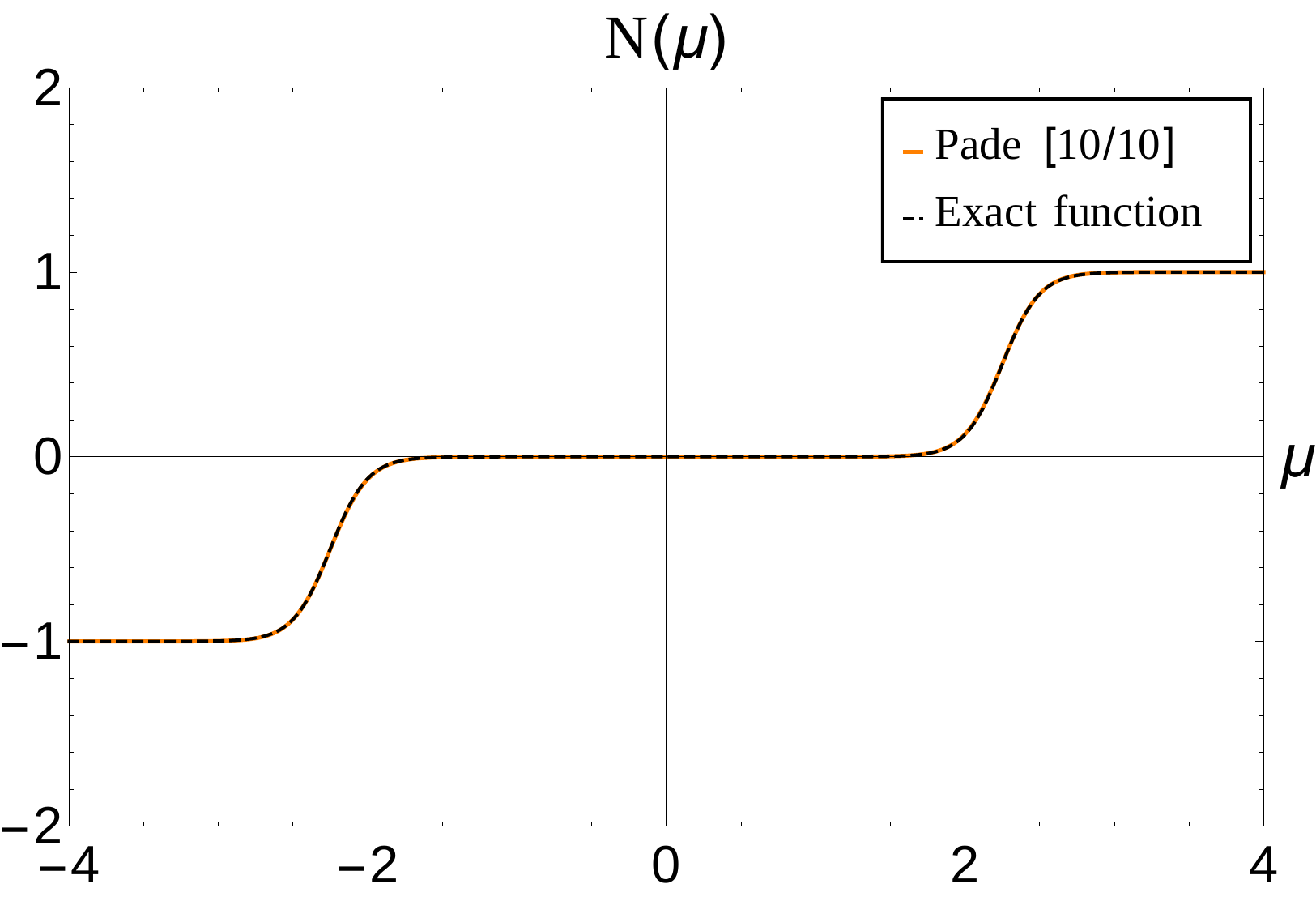}
        \hfill
        \includegraphics[scale=0.40]{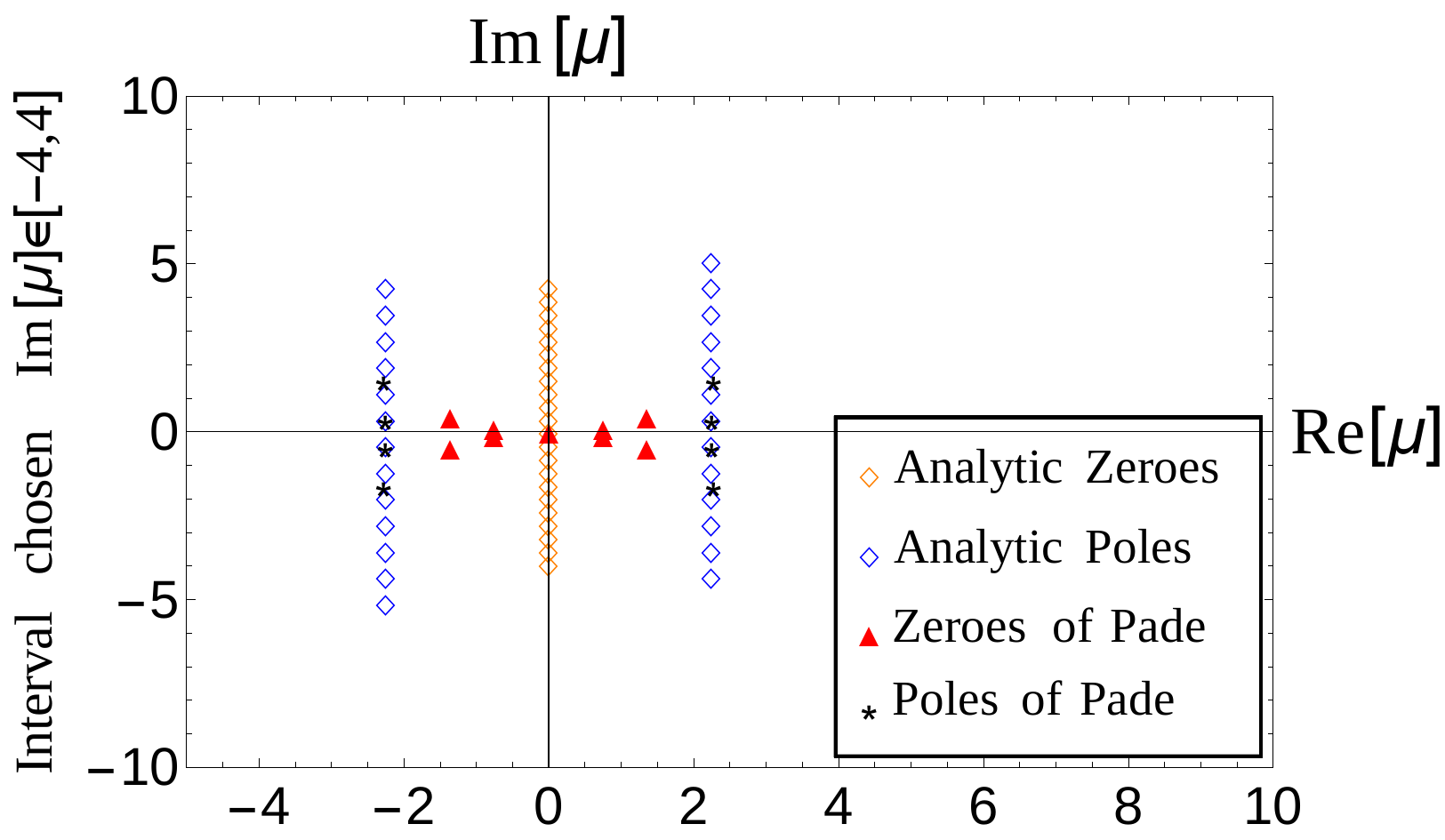}
        \includegraphics[scale=0.34]{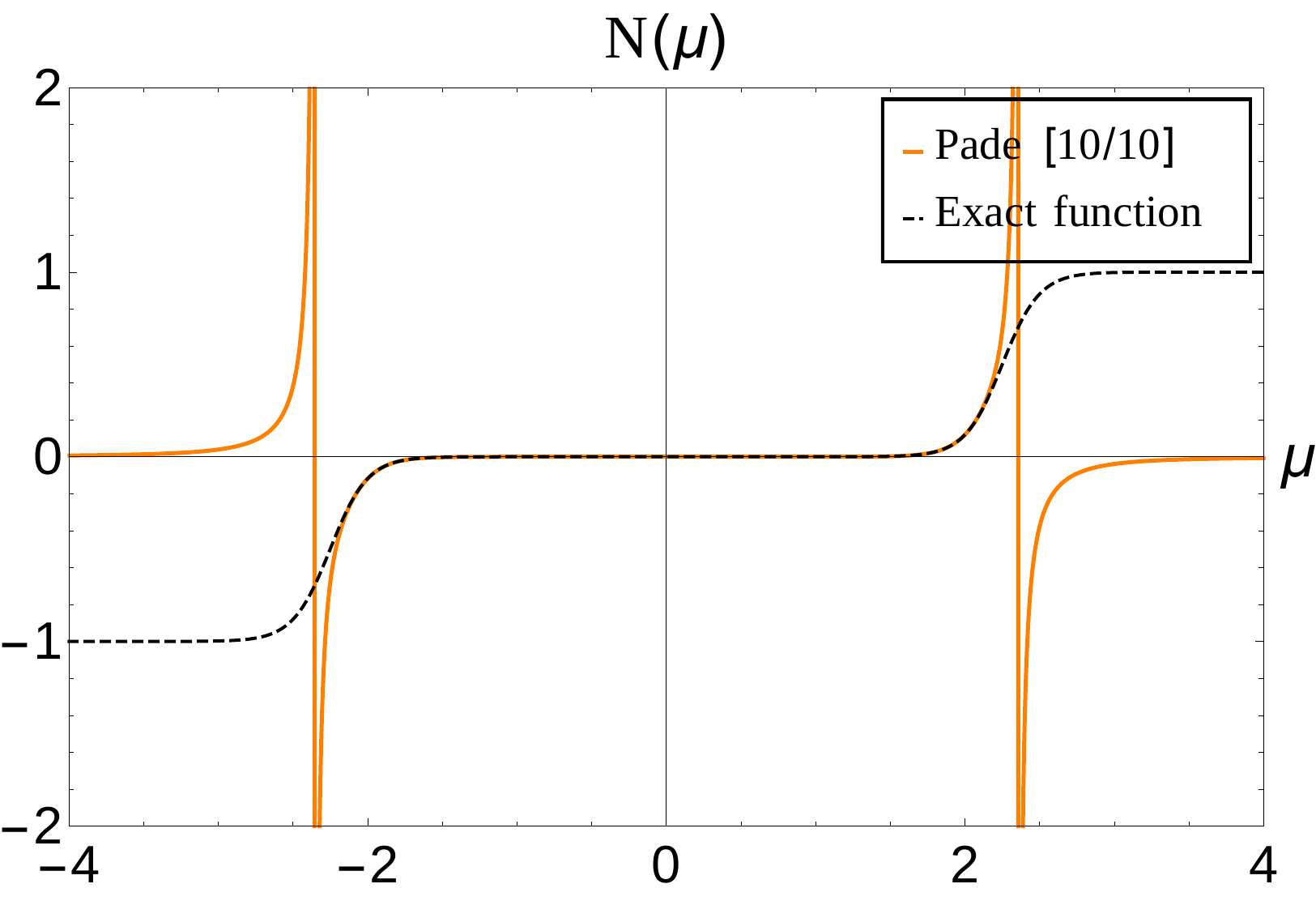}
        \hfill
        \includegraphics[scale=0.40]{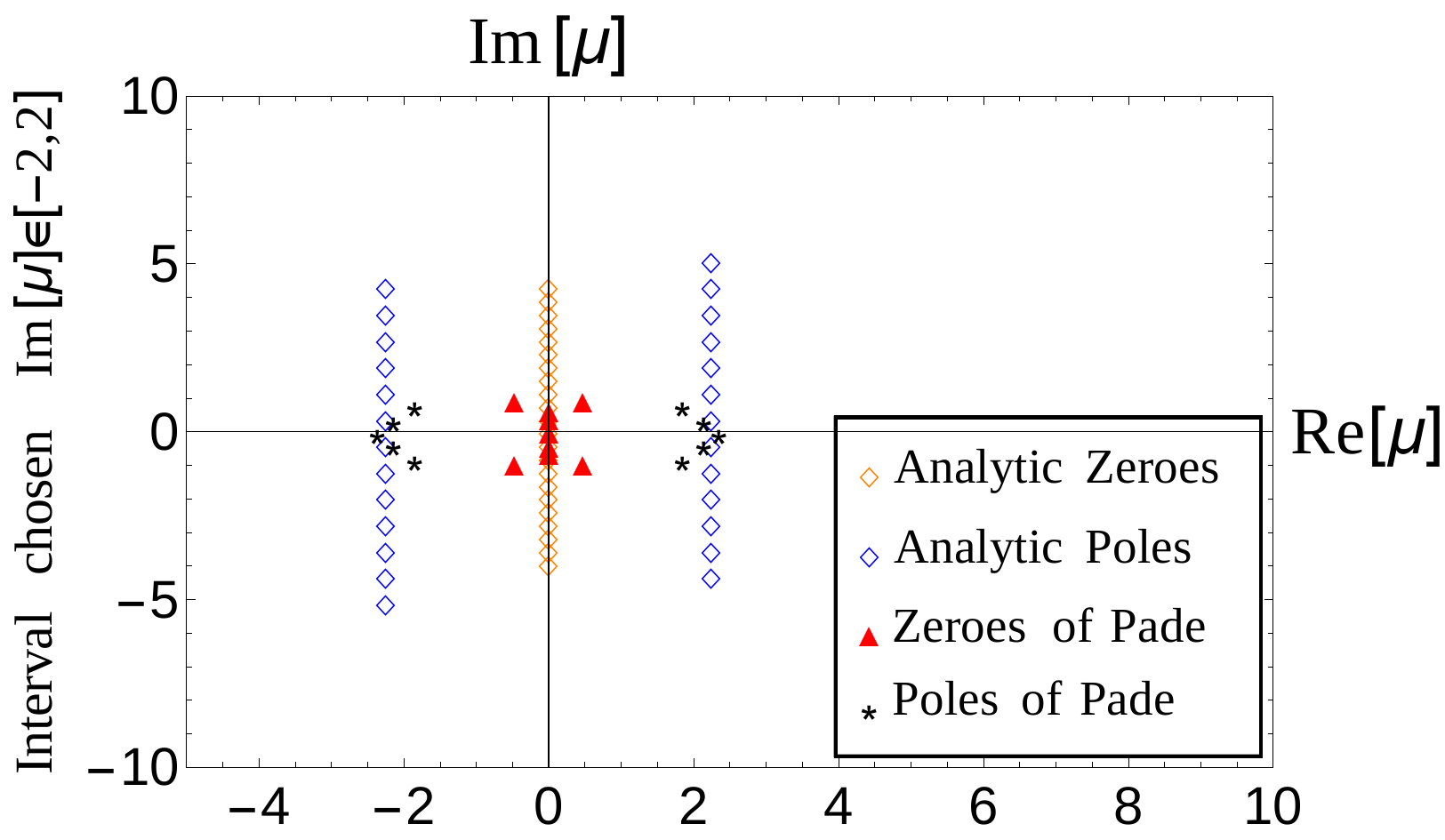}
         \caption{1D Thirring model: Functional form of the rational approximation(left) and sensitivity to different sets of poles(right) when sampled in different intervals: [0,4] (top), [-4,4] (middle), [-2,2] (bottom).}
         \label{fig:appendixC_func}
\end{figure*}

\noindent Based on our numerical experiments, it was observed that the approximation obtained from the multipoint Padé approach is sensitive to the interval sampled. By this we mean that for some functions the signature of the singularity might be missed if the interval is not chosen appropriately. \\ \newline
\noindent In Fig.~\ref{fig:appendixC_func}, this sensitivity is demonstrated with the help of the 1D Thirring model as our test case, since we know the positions of its singularities. The singularities which we detect from rational approximations obtained in different intervals are shown and compared with the analytic (exact) positions. \\
\begin{figure*}[htp] 
        \centering
        \includegraphics[scale=0.385]{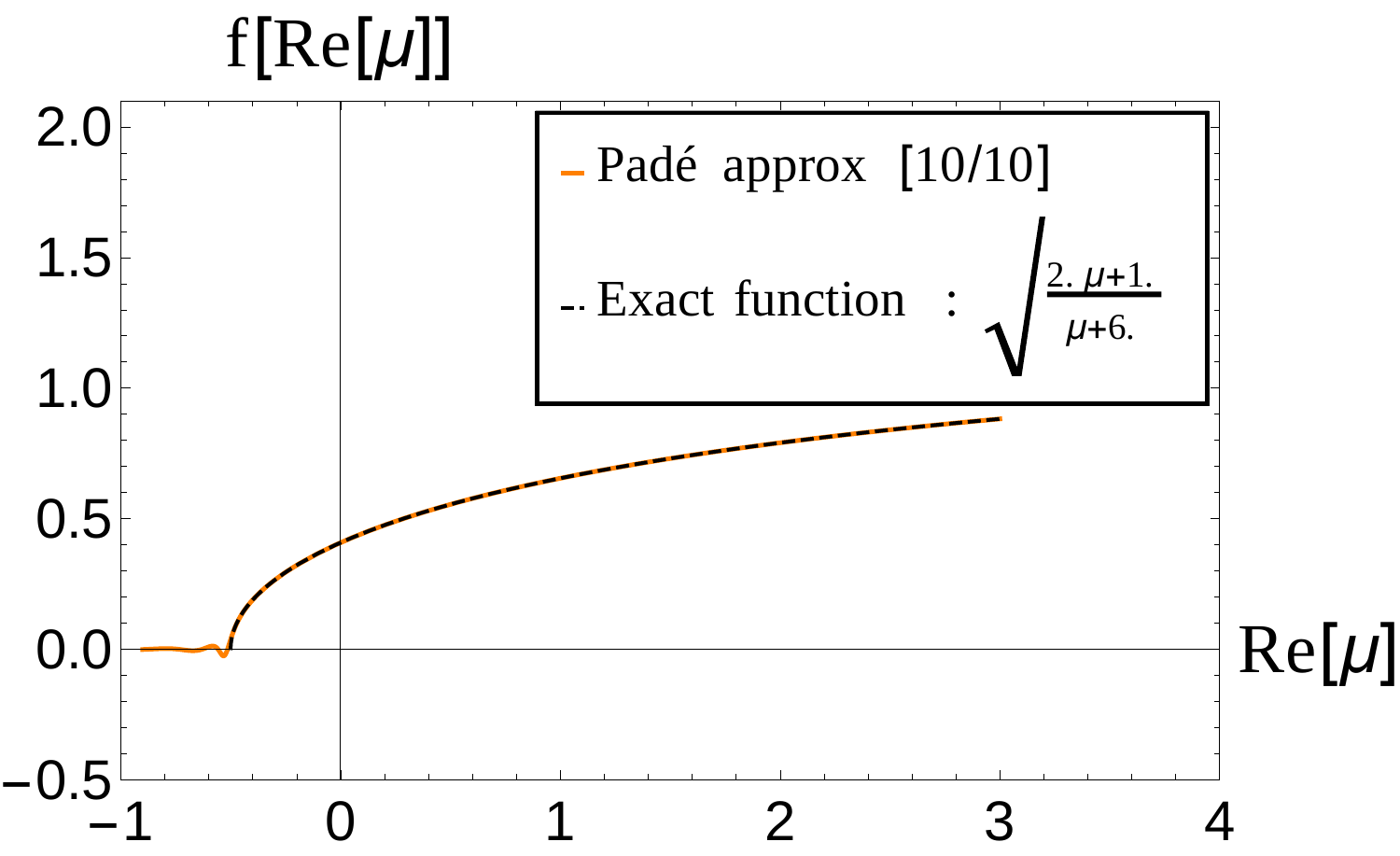}
        \hfill
        \includegraphics[scale=0.385]{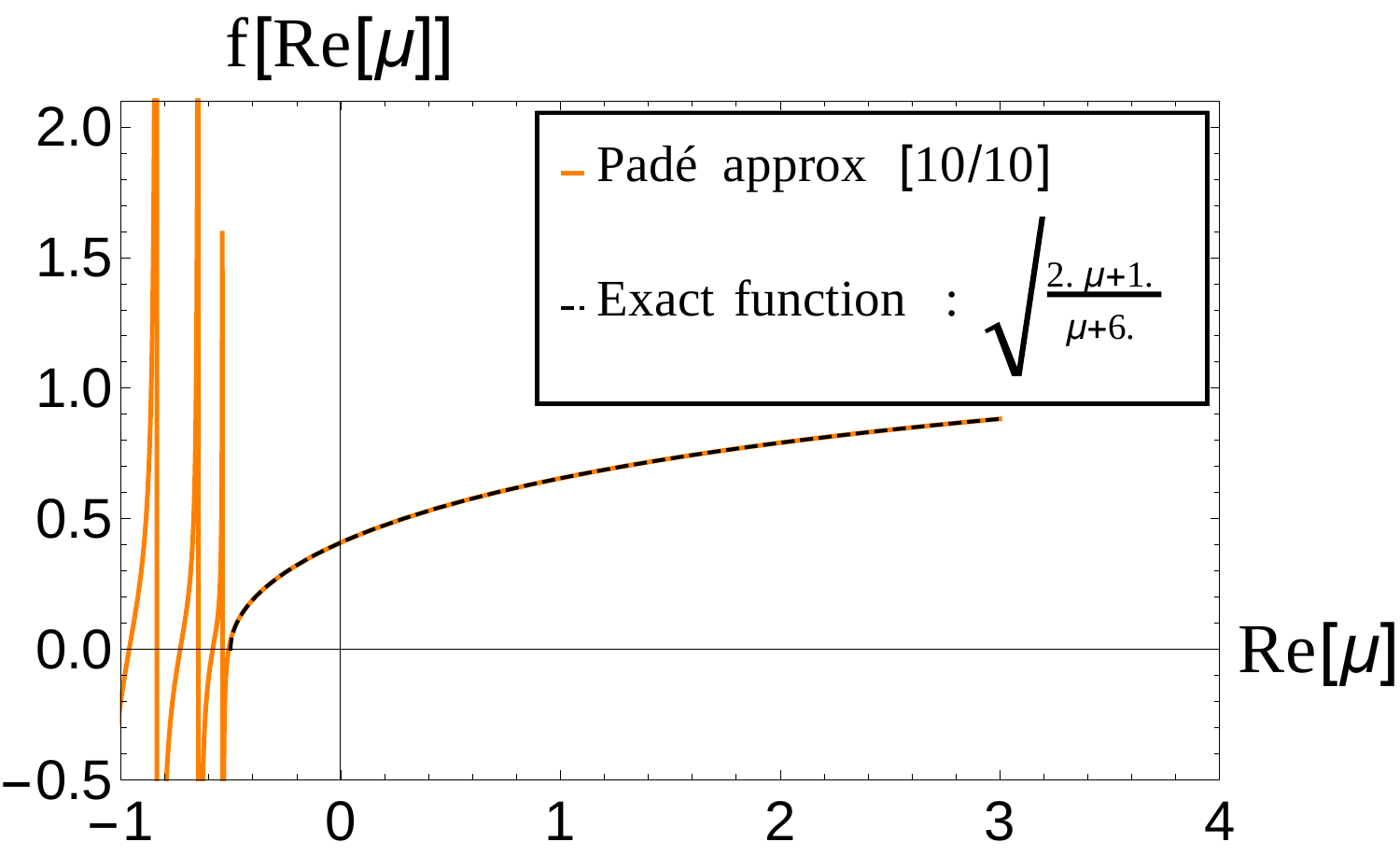}
        \includegraphics[scale=0.40]{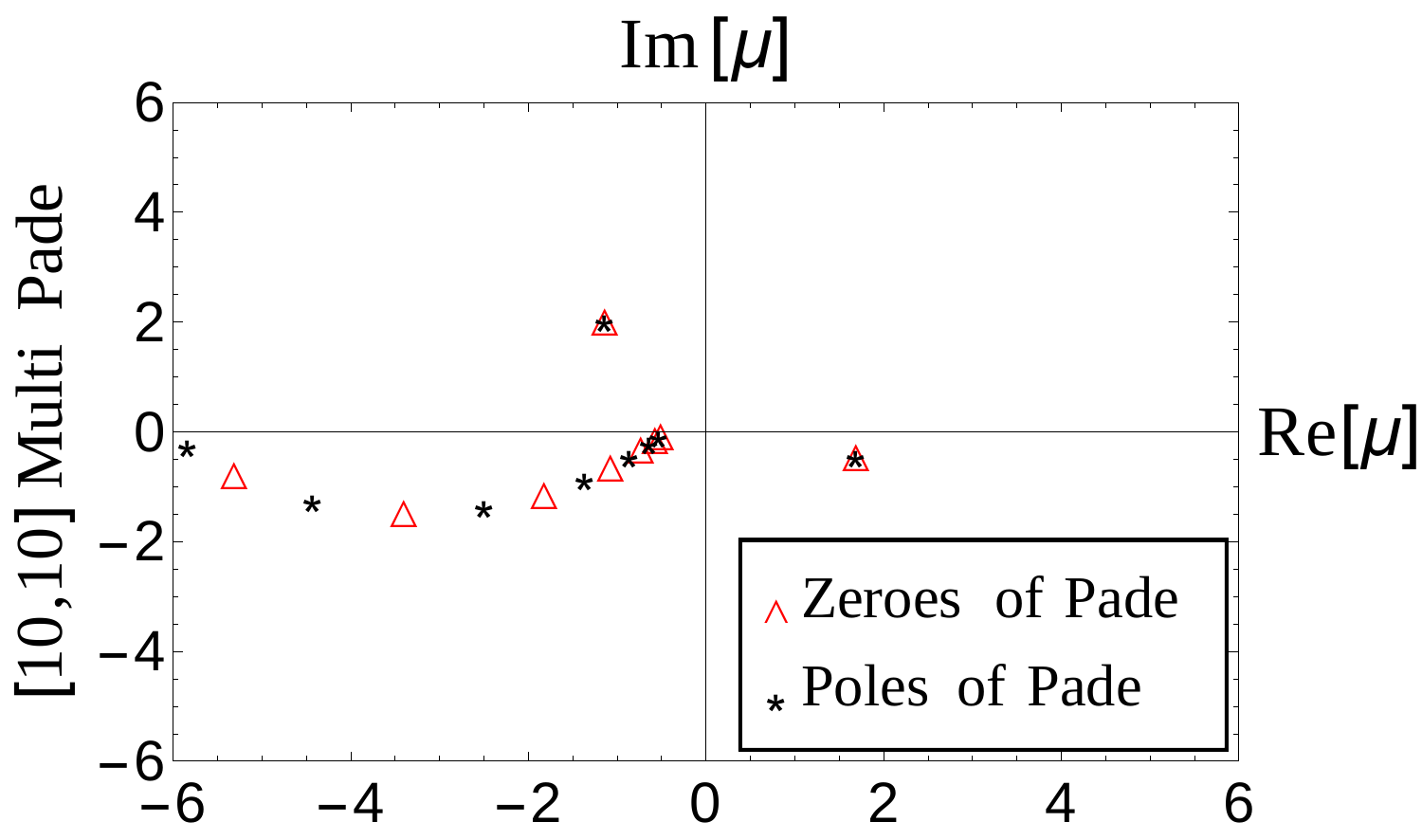}
        \hfill
        \includegraphics[scale=0.40]{multiPm331010.pdf}
        \caption{(Left) : Interval used Im$[\mu] \in [0,3]$, (top) Approximation is good but (bottom) signature of branch cut missed . (Right) : Interval used Im$[\mu] \in [-3,3]$,  (top) Approximation is good in the region considered and (choosing an interval which includes the branch cut symmetrically ensures) that the (bottom) branch cut is properly obtained.}
        \label{fig:appendixC_BranchZeroPole}
\end{figure*}

\noindent Another example where the interval dependence of a function is manifest is when the function has a branch cut. We would ask the reader to keep in mind that range dependence may not be apparent from the functional form always, whereas it is can be manifest in the structure of zeroes and poles. In Fig.~\ref{fig:appendixC_BranchZeroPole} is shown the example of such a function ($\sqrt{\frac{2\mu+1}{\mu+6}}$).

\section{MORE NUMERICAL EXPERIMENTS}
\label{sec:VariousSing}
Some more numerical experiments were performed to see how the Padé approximation treats different types of singularities (poles, branch points, essential singularities etc). While some literature exists on how the single point Padé treats these singularities with varying the order of the approximation \cite{yamada2014numerical} - not much exists on how multipoint Padé treats these singular points. This is also a nice way to test that the approximation works.

\noindent The reason we are focusing on a \enquote{cusp} like singularity is motivated from the periodicity properties of the partition function, which at and above the Roberge-Weiss temperature behaves like a cusp at the RW point and multiples of $\pi$. The free energy, by definition, also has this structure and hence it's first derivative becomes \emph{discontinuous}.

The first of the examples of known functions shown below mimics the above mentioned behaviour (corner function) while the second example is just to show how the multipoint Padé handles a genuine cusp~\footnote{Mathematically a cusp is different from a \enquote{corner function} - cusp functions are subsets of corner functions }.

\subsection{\enquote{Cusp} like function - and it's derivative:}
\begin{figure*}[h!] 
        \centering
        \includegraphics[scale=0.27]{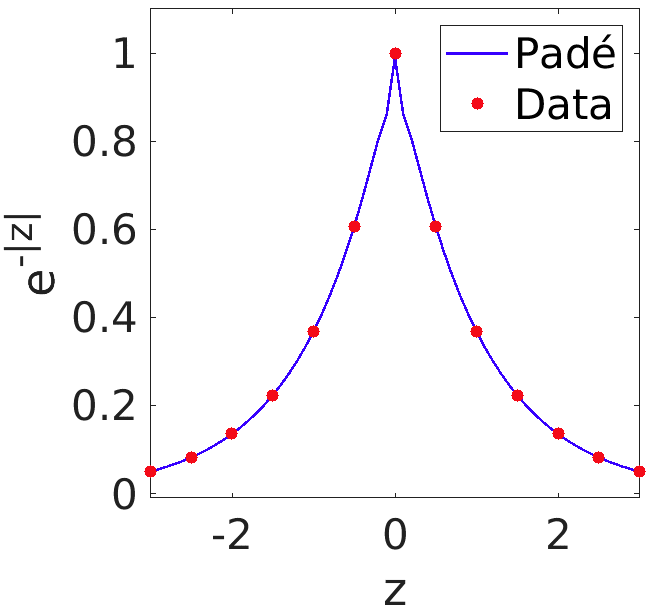}
 		\hfill
        \includegraphics[scale=0.27]{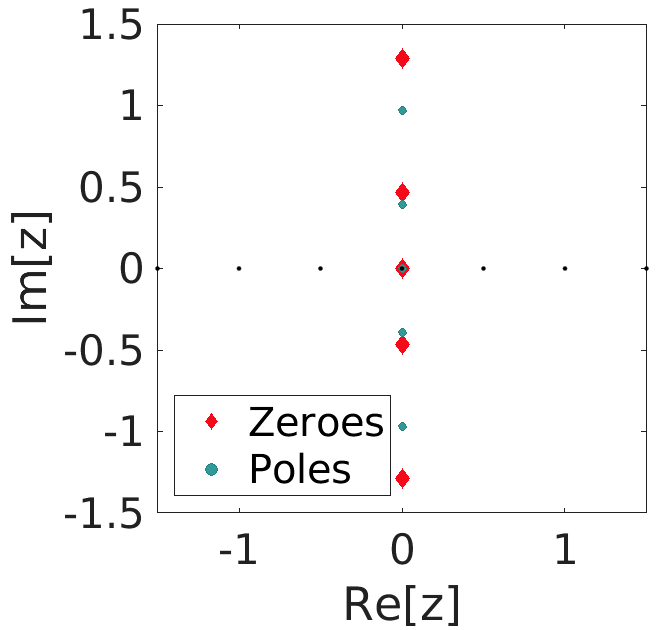}\\
        \includegraphics[scale=0.27]{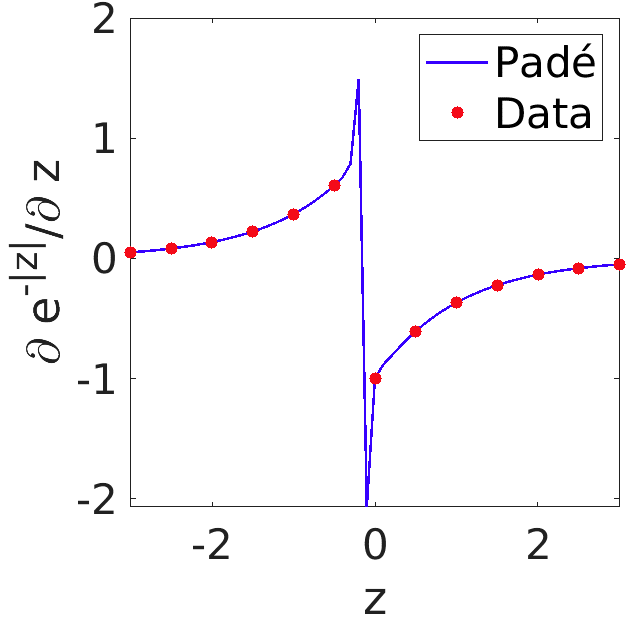}
        \hfill
        \includegraphics[scale=0.27]{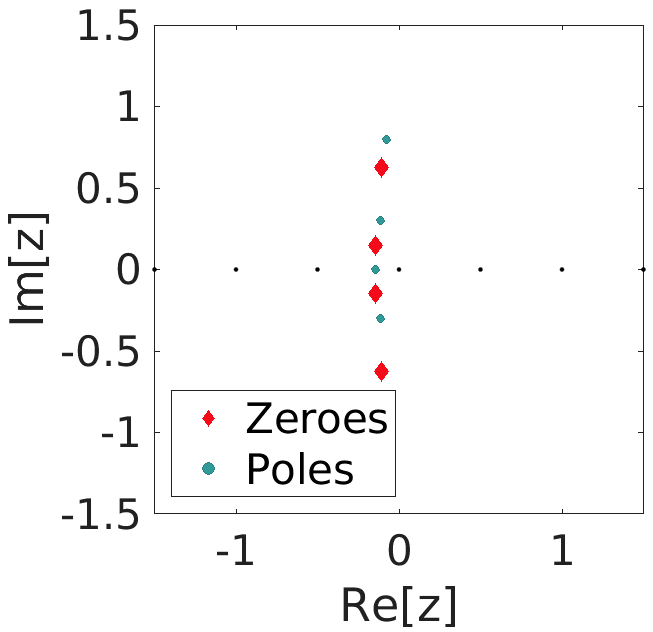}
        \caption{Multipoint Padé for : (Top) A cusp like function and it's singularity structure, (Bottom)  The derivative of the above function and its singularity structure.}
        \label{fig:appendixD_DerivCuspLike}
\end{figure*}

The exponent function with argument of negative absolute values and it's derivative showing a discontinuity similar to our case are shown in Fig. \ref{fig:appendixD_DerivCuspLike}.

Poles are strictly speaking the only singularity that a Padé approximation can have. When we demand a rational approximation of an irrational function, such as the square root function above, the only way rational function can mimic the branch cut is by placing a sequence of zeroes and poles alternately along the branch cut.

\subsection{Function with a genuine cusp singularity:}
Just as an extra example the Padé analysis of a genuine \emph{cusp} singularity is also presented below (Fig. \ref{fig:appendixD_Cusp}):
\begin{figure*}[h!] 
        \centering
        \includegraphics[scale=0.27]{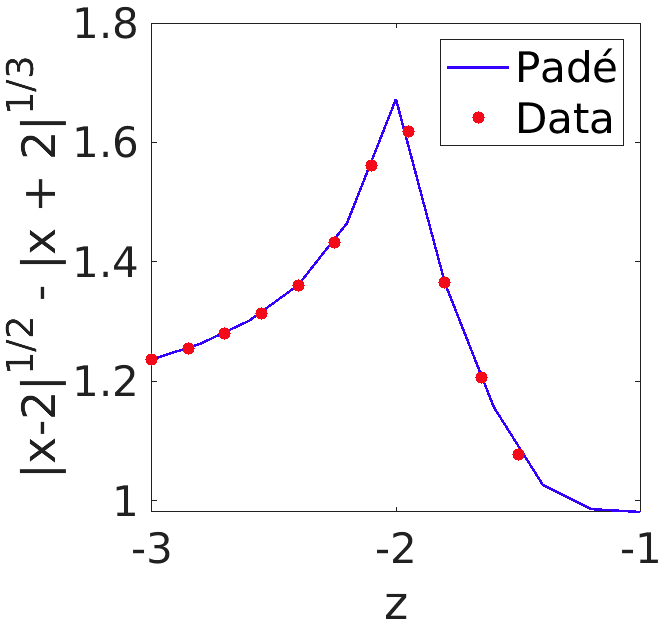}
        \hfill
        \includegraphics[scale=0.27]{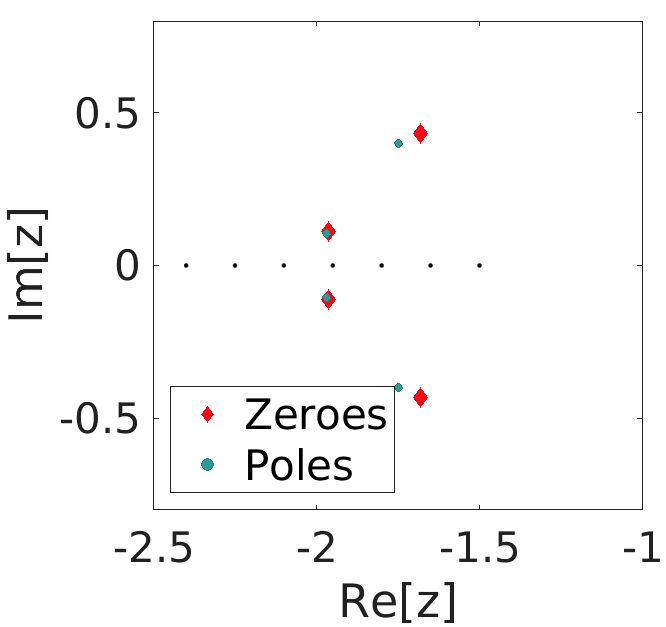}
        \caption{Multipoint Padé for a genuine cusp function and its Zeroes and Poles.}
        \label{fig:appendixD_Cusp}
\end{figure*}
\section{EFFECT OF NOISE ON POLE STABILITY}
\label{sec:LastApPADE}
\subsection{Statistical Error}
\begin{figure*}[!ht] 
        \centering
        \includegraphics[scale=0.54]{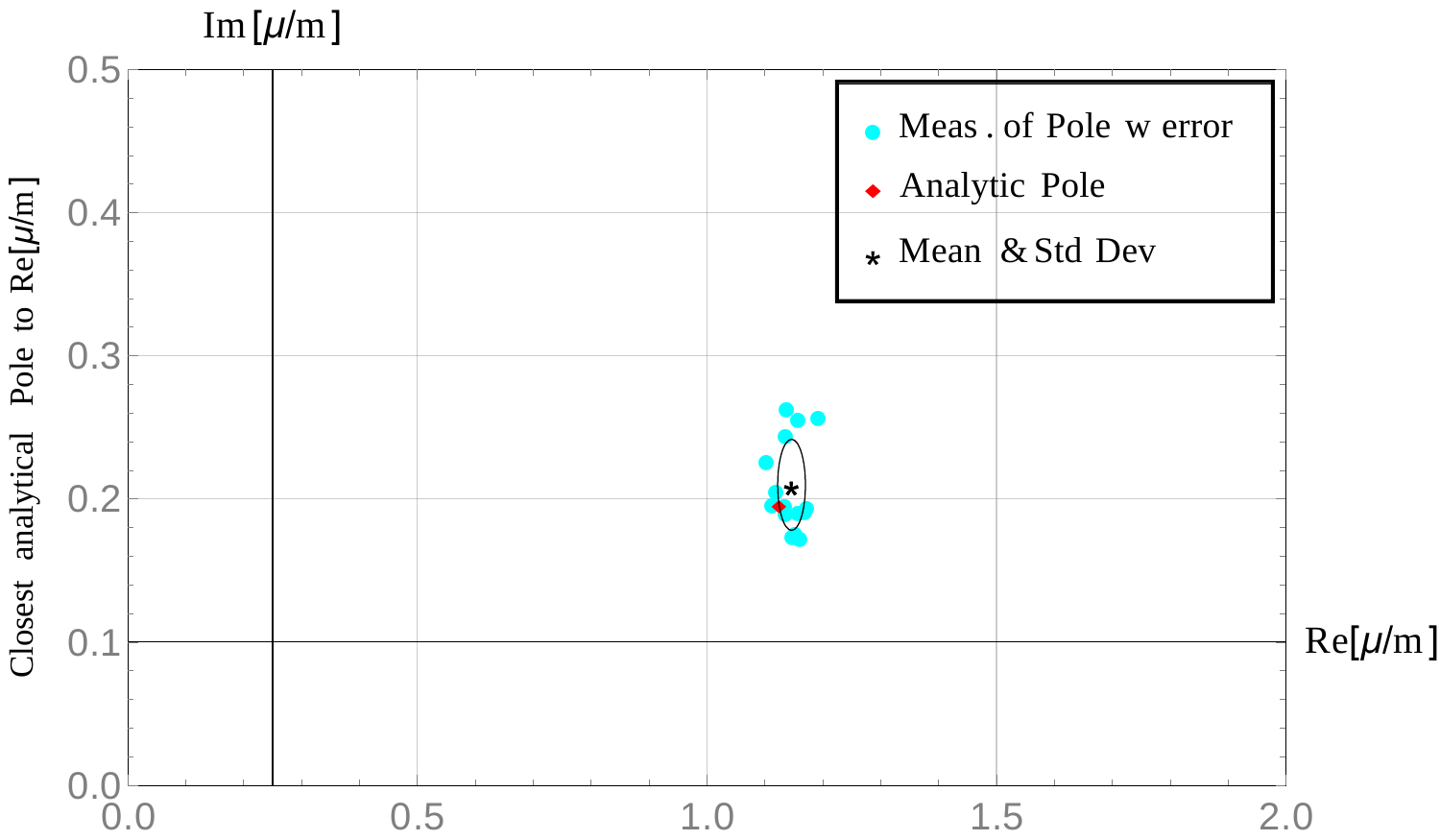}
        \hfill
        \includegraphics[scale=0.54]{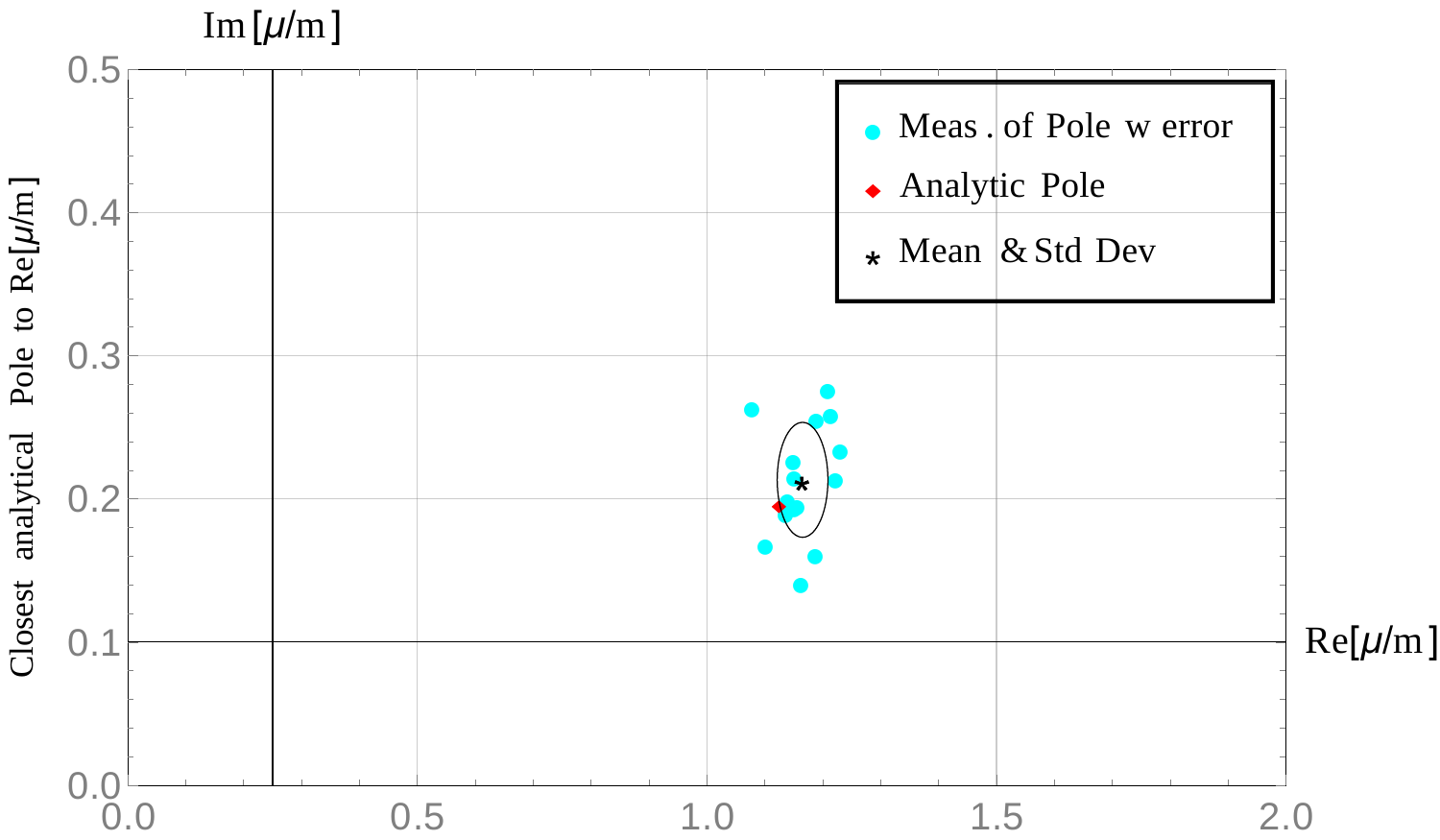}
         \caption{Closest singularity with a [4,4] Padé with (Left) 1\%  and (Right) 5\% errors on values and (Left) 10\% (Right) 15\%  on first derivatives respectively.}
         \label{fig:appendixE_Thirr44}
\end{figure*}

\begin{figure*}[!ht] 
        \centering
        \includegraphics[scale=0.54]{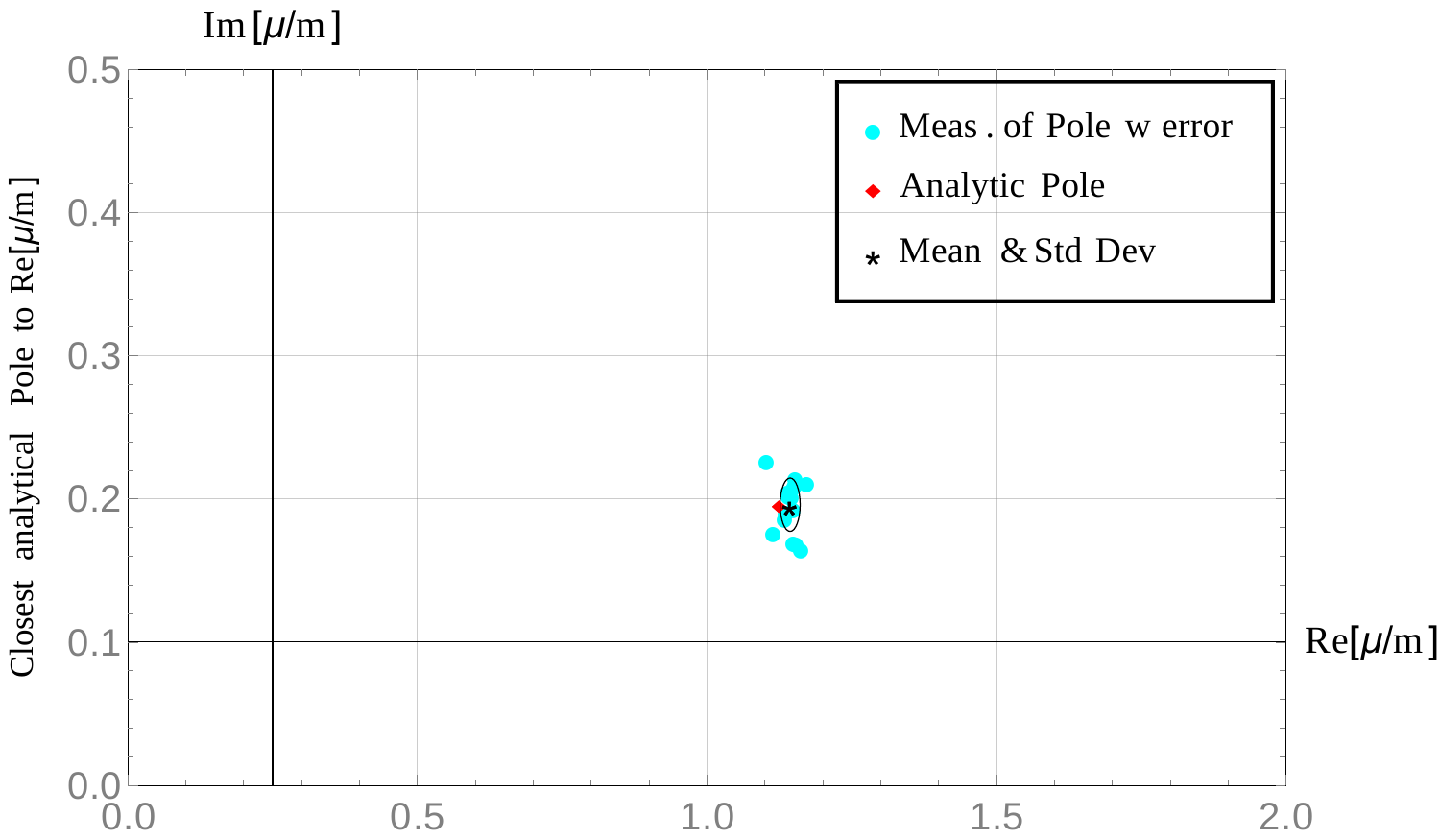}
        \hfill
        \includegraphics[scale=0.54]{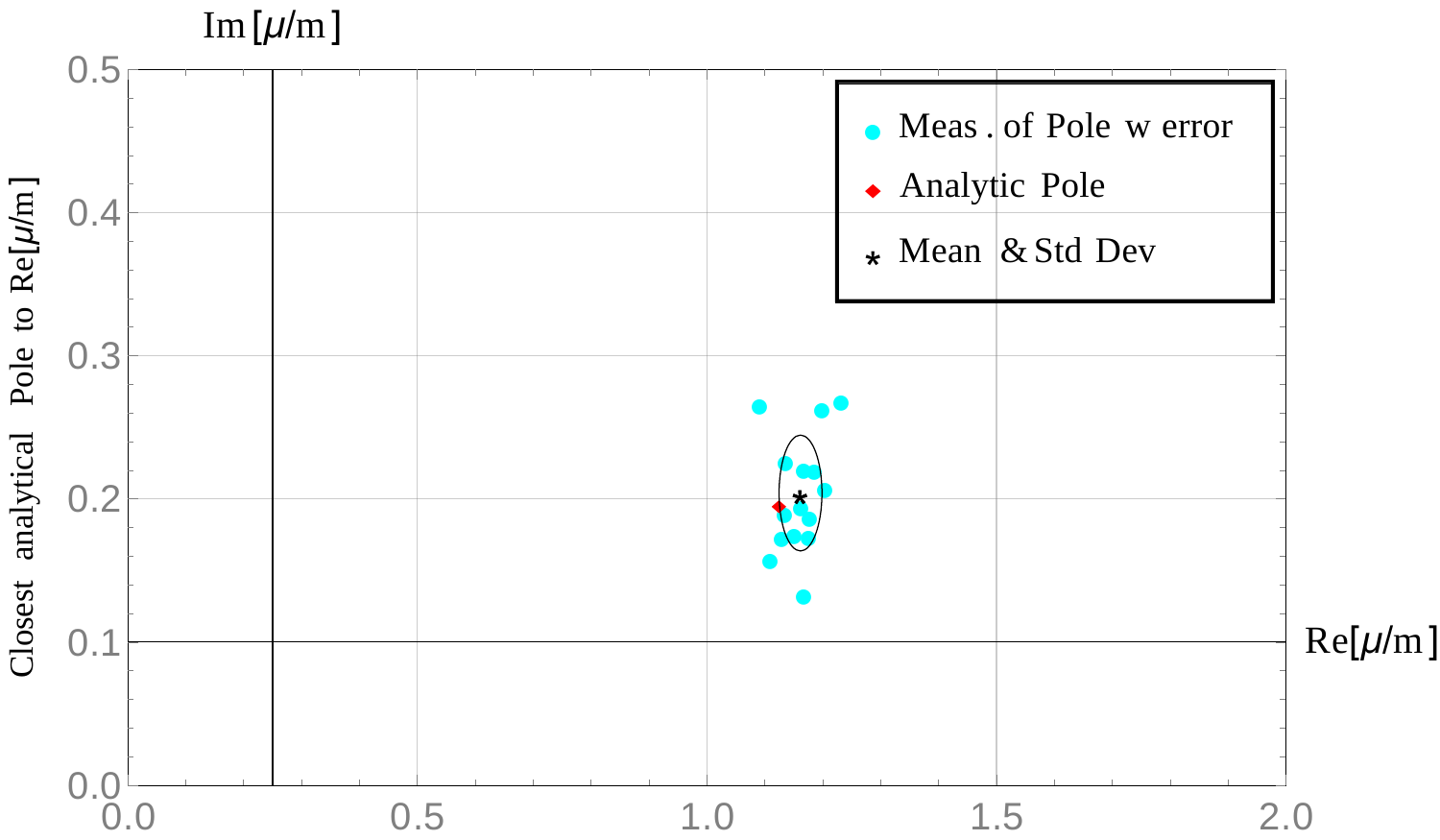}
         \caption{Closest singularity with a [5,5] Padé with (Left) 1\%  and (Right) 5\% errors on values and (Left) 10\% (Right) 15\%  on first derivatives respectively.}
         \label{fig:appendixE_Thirr55}
\end{figure*}

\begin{figure*}[!ht] 
        \centering
        \includegraphics[scale=0.54]{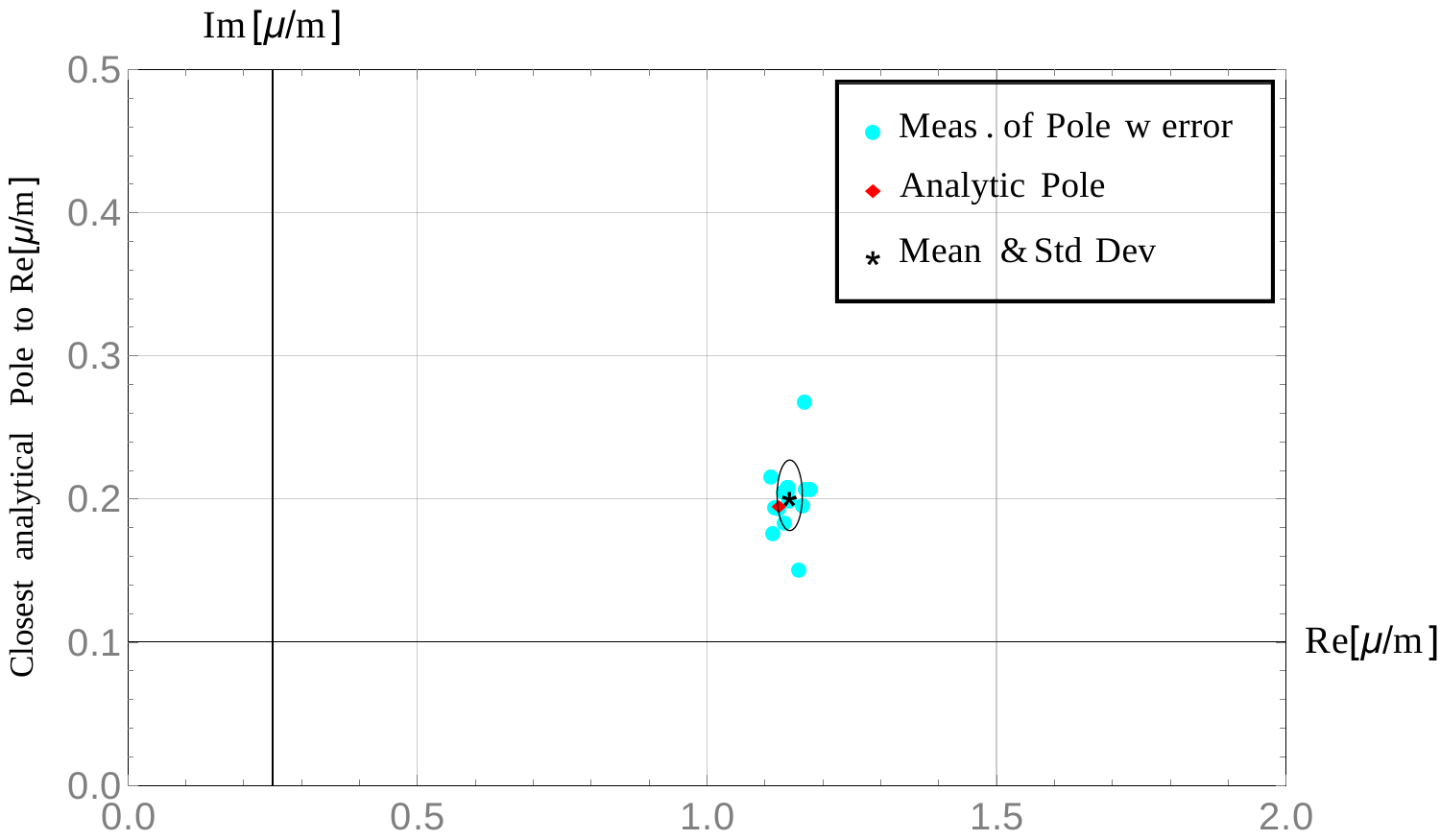}
        \hfill
        \includegraphics[scale=0.54]{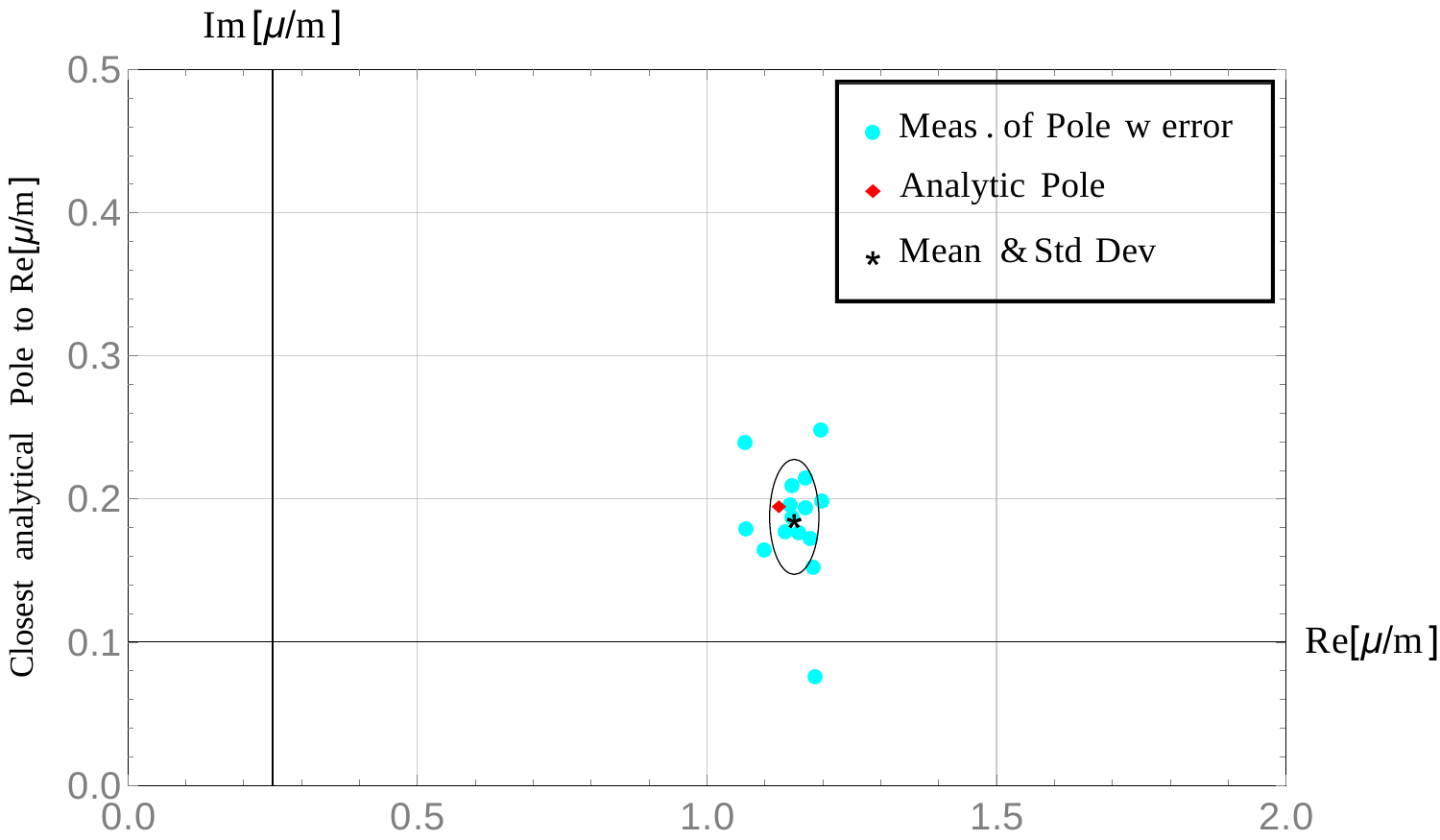}
         \caption{Closest singularity with a [6,6] Padé with (Left) 1\%  and (Right) 5\% errors on values and (Left) 10\% (Right) 15\%  on first derivatives respectively.}
         \label{fig:appendixE_Thirr66}
\end{figure*}

Since our data from the lattice simulations comes with noise, it is important to study the effect of noise on functions containing genuine singularities. It is already well known that in the presence of noise, poles move about around the true singularity. Also, the mean distance from the true pole increases with increasing the magnitude of error. This can be seen (again) with the help of the Thirring model (but the reader is free to choose their own test function). The figures below (Figs.~\ref{fig:appendixE_Thirr44},~\ref{fig:appendixE_Thirr55},~\ref{fig:appendixE_Thirr66}) are intended to mimic the lattice data at least where statistical errors are concerned. For instance, from our QCD data, we have around 1\% error on the $\chi_1^B$ and 10\% errors on $\chi_2^B$. In the figures, we show the effect of adding this combination of errors to the Thirring model. The take away message is that even though the poles move around - the signature of the singularity is present and consistent within errors with the true singularity.   

\subsection{Systematic error:}
\begin{figure*}[!ht] 
        \centering
        \includegraphics[scale=0.51]{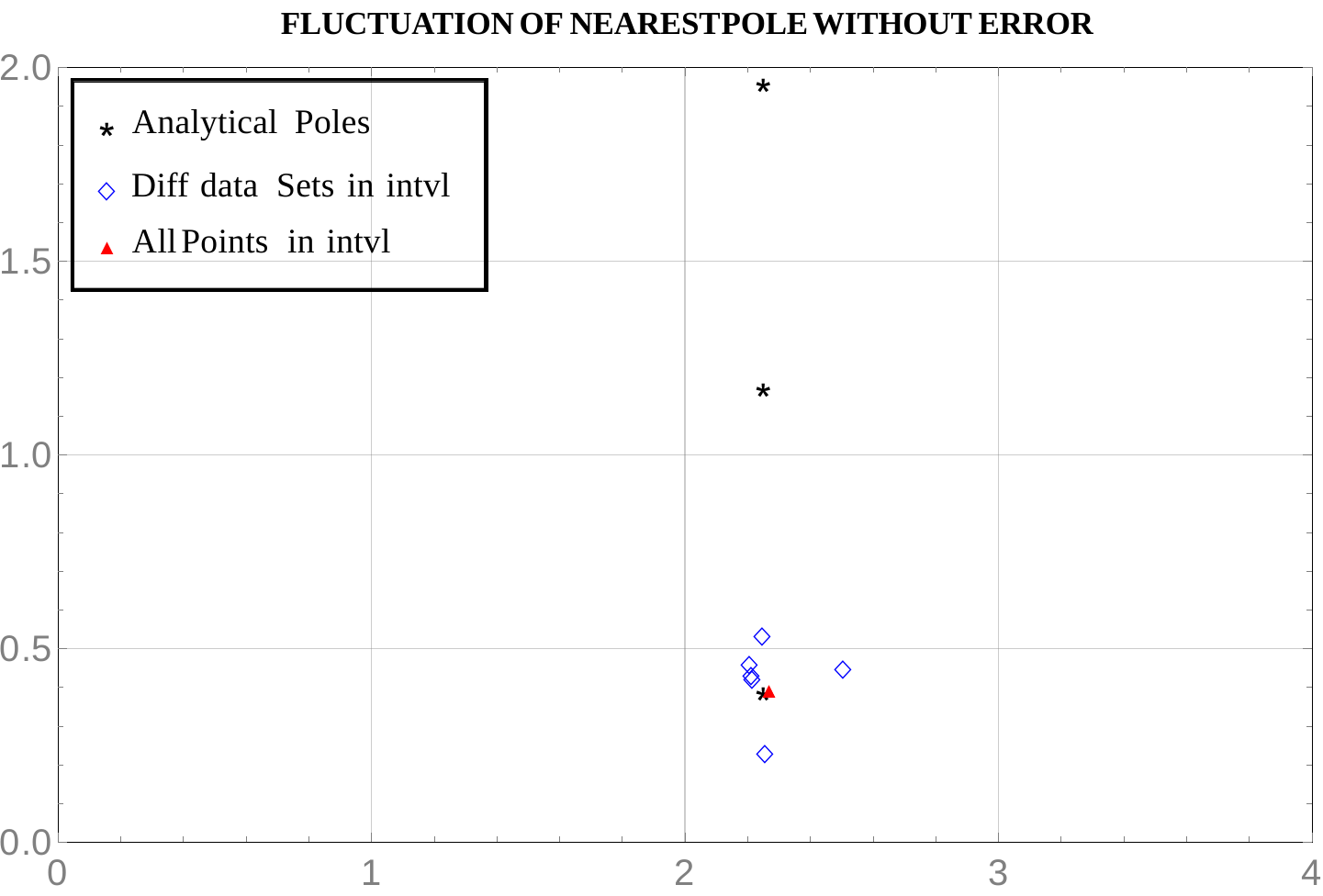}
        \hfill
        \includegraphics[scale=0.50]{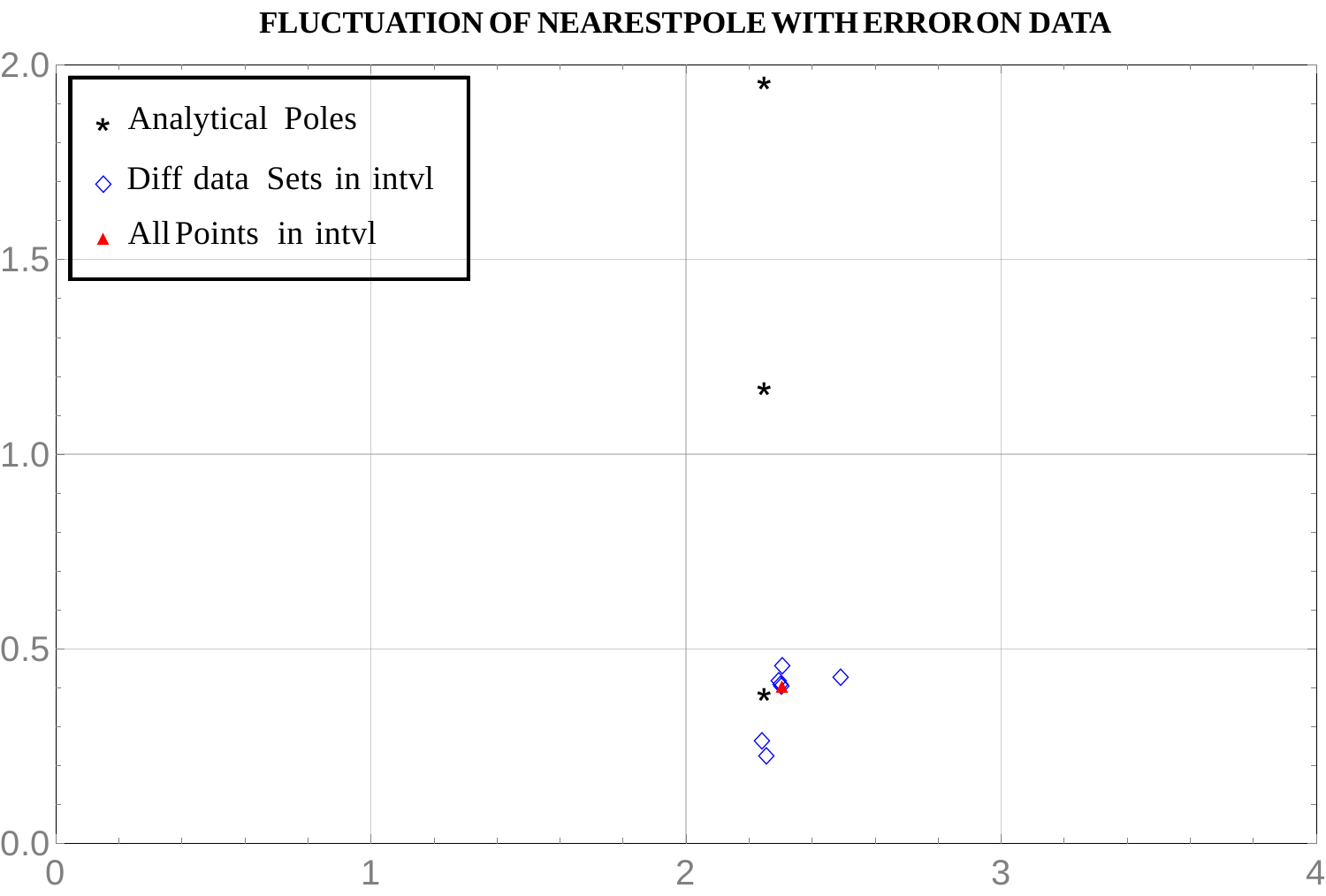}
         \caption{Padé built out of only values of the function (Thirring) at multipoints without noise (Left) and in the presence of noise (Right). As can be seen, the order of the Padé is low enough to observe movement of the pole with varying the set of points included to build the Padé even in the absence of noise.}
         \label{fig:appendixE_SystThirr1}
\end{figure*}

\begin{figure*}[!ht] 
        \centering
        \includegraphics[scale=0.55]{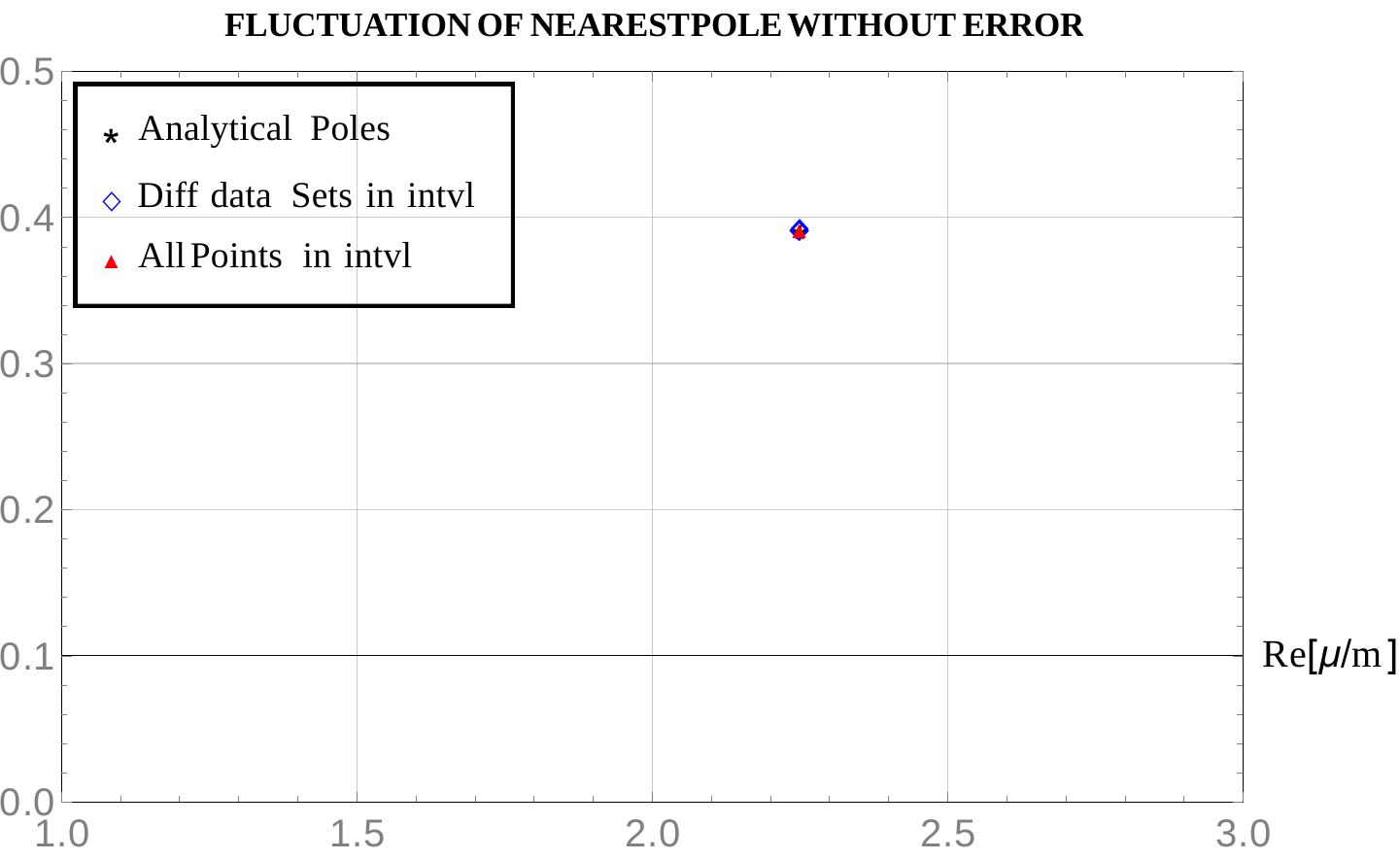}
        \hfill
        \includegraphics[scale=0.55]{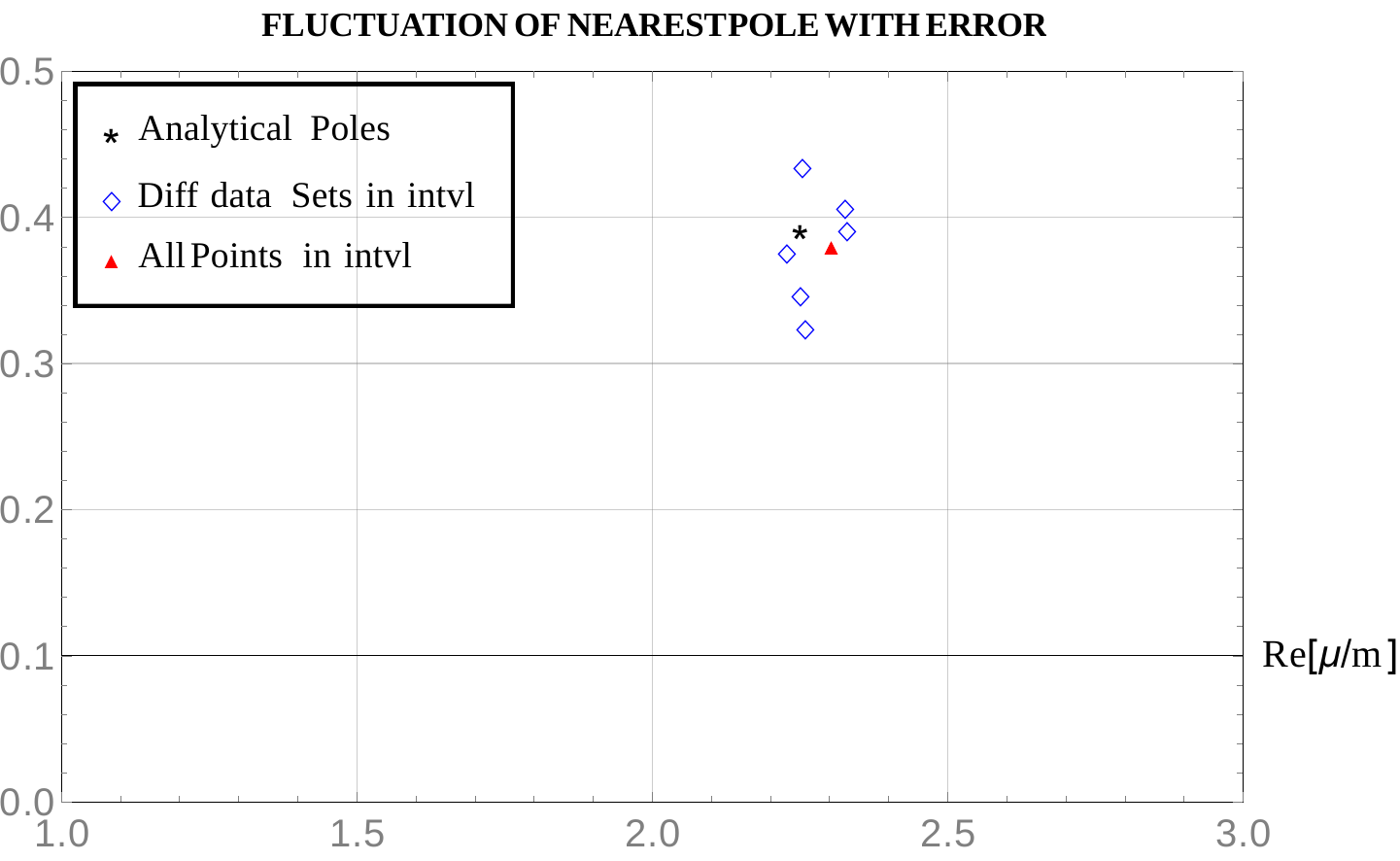}
         \caption{(Left) Padé built out of values and first derivatives of the function (Thirring) at multiple points. (Right) The same as (Left) but in the presence of errors. As can be seen, the order of the Padé is high enough to give a stable pole for the Padé built without errors (Left), whereas in (Right) we observe movement of the pole with varying the set of points included in the presence of noise.}
         \label{fig:appendixE_SystThirr2}
\end{figure*}

We have already seen the interval dependence of poles. Padé theory dictates that the true poles of a function remain fixed when changing orders of the Padé. While it is very clear in building single point Padé approximations what increasing or decreasing the order of a Padé means, it is not the case for multipoint. We can change the order in at least two distinct ways or a combination of them, by either increasing the number of points sampled or keeping the points fixed and increasing the derivatives at those points. The systematic errors that we want to highlight in this section are those that cause the pole to move around even at a fixed instance of statistical error while varying the order of the approximant as mentioned above (see Figs. \ref{fig:appendixE_SystThirr1} and \ref{fig:appendixE_SystThirr2}). 
\clearpage
\bibliography{PadeBib}
\end{document}